\documentclass[fleqn,usenatbib]{mnras}

\usepackage{newtxtext,newtxmath}

\usepackage[T1]{fontenc}
\usepackage{ae,aecompl}

\usepackage{graphicx}	
\usepackage{amsmath}	
\usepackage{amssymb}    
\usepackage{verbatim}
\usepackage{placeins}

\usepackage{etoolbox}
\makeatletter
\patchcmd\@combinedblfloats{\box\@outputbox}{\unvbox\@outputbox}{}{%
  \errmessage{\noexpand\@combinedblfloats could not be patched}%
}%
\makeatother

\defcitealias{Angles-Alcazar2017}{AA17}

\defcitealias{Ford2014}{F14}
\newcommand{\fordp}{\citetalias{Ford2014}}
\defcitealias{Peeples2014}{P14}
\newcommand{\peeplesp}{\citetalias{Peeples2014}}

\title[Origins of the CGM]{The Origins of the Circumgalactic Medium in the FIRE Simulations}

\author[Hafen et al.]{
\parbox{\textwidth}{
Zachary Hafen,$^{1}$\thanks{E-mail: zhafen@u.northwestern.edu}
Claude-Andr\'e Faucher-Gigu\`ere,$^{1}$
Daniel Angl\'es-Alc\'azar,$^2$
Jonathan Stern,$^1$
Du\v{s}an Kere\v{s}$,^3$
Cameron Hummels,$^4$
Clarke Esmerian,$^5$
Shea Garrison-Kimmel,$^4$
Kareem El-Badry,$^6$
Andrew Wetzel,$^7$
T. K. Chan,$^3$
Philip F. Hopkins,$^4$
Norman Murray$^8$
} \vspace{0.4cm}\\
\parbox{\textwidth}{
$^{1}$Department of Physics and Astronomy and Center for Interdisciplinary Exploration and Research in Astrophysics (CIERA),\\ Northwestern University, 2145 Sheridan Road, Evanston, IL 60208, USA \\
$^{2}$Center for Computational Astrophysics, Flatiron Institute, 162 Fifth Avenue, New York, NY 10010, USA\\
$^3$Department of Physics, Center for Astrophysics and Space Sciences, University of California, San Diego, 9500 Gilman Drive, \\ La Jolla, CA 9209, USA \\
$^4$TAPIR, Mailcode 350-17, California Institute of Technology, Pasadena, CA 91125, USA \\
$^5$Department of Astronomy and Astrophysics, The University of Chicago, Chicago, IL 60637, USA \\
$^6$Department of Astronomy and Theoretical Astrophysics Center, University of California, Berkeley, CA 94720-3411, USA \\
$^7$Department of Physics, University of California, Davis, CA, USA 95616 \\
$^8$Canadian Institute for Theoretical Astrophysics, 60 St George Street, University of Toronto, ON M5S 3H8, Canada
}
}

\date{Accepted XXX. Received YYY; in original form ZZZ}

\pubyear{2015}

\begin{document}
\label{firstpage}
\pagerange{\pageref{firstpage}--\pageref{lastpage}}
\maketitle
\newcommand{\evolutionMmaxsubMWq}{5\times10^{9}}
\newcommand{\evolutionMsatwindmaxdwarfy}{4\%}
\newcommand{\totmassgalaxysubMWmeanlow}{3\times10^{9}}
\newcommand{\metprofAllinnerlow}{3.4}
\newcommand{\totmetmassgalaxyMWmeanlow}{10^{9}}
\newcommand{\fracMWwindmeanhigh}{30\%}
\newcommand{\deltaMWsteeplow}{2.3}
\newcommand{\metMWwindmedhigh}{3\times10^{-1}}
\newcommand{\evolutionMminMWi}{8\times10^{9}}
\newcommand{\metdwarfwindmedlow}{3\times10^{-2}}
\newcommand{\fCGMsubMWmeanlow}{55}
\newcommand{\totmetmassCGMdwarfmeanlow}{9\times10^{5}}
\newcommand{\fracnepmetalmeanmdMWhigh}{15}
\newcommand{\evolutionMnepfinalsubMWq}{3\times10^{9}}
\newcommand{\meddenseCGMmasshigh}{3\%}
\newcommand{\totmetmassinterfaceMWmeanlow}{2\times10^{8}}
\newcommand{\mtenvmzlow}{1}
\newcommand{\evolutionMsatmaxdwarfy}{2\times10^{7}}
\newcommand{\totmetmassinterfacedwarfmeanlow}{2\times10^{6}}
\newcommand{\evolutionMsatwindminsubMWq}{3\%}
\newcommand{\metprofSatWindouterlow}{2.7}
\newcommand{\fracsatwindmetalmeanmdlow}{25}
\newcommand{\yboxstdhigh}{0.001}
\newcommand{\totmassinterfacesubMWmeanhigh}{9\times10^{8}}
\newcommand{\fCGMMWmeanlow}{40}
\newcommand{\fracnepmetalmedianMWhigh}{1\%}
\newcommand{\totmetmassCGMMWmeanlow}{2\times10^{8}}
\newcommand{\metprofIGMAccouterlow}{17}
\newcommand{\fracsatwindmetalmeanhigh}{10\%}
\newcommand{\deltaelevenmaxhigh}{0.3}
\newcommand{\metsubMWsatwindmedhigh}{3\times10^{-2}}
\newcommand{\evolutionMwindatpeakdwarfy}{10^{8}}
\newcommand{\fmetinterfaceMWmeanhigh}{15}
\newcommand{\profilelowMincreaseIGMAccretionlow}{6.6}
\newcommand{\evolutionMminsubMWq}{10^{9}}
\newcommand{\fracnepmetalmedianlowermasslow}{1\%}
\newcommand{\meanmhmdMWlow}{9\times10^{11}}
\newcommand{\metprofAllouterlow}{7}
\newcommand{\meanmhcoreMWhigh}{4\times10^{11}}
\newcommand{\fbaryonsubMWmeanlow}{0.4}
\newcommand{\deltaMWsteepminhigh}{1.4}
\newcommand{\evolutionMwindfinalMWi}{5\times10^{9}}
\newcommand{\meanstellarmasspercentingaplow}{10\%}
\newcommand{\metprofIGMAccouterhigh}{8.2}
\newcommand{\zstardivzstarcorrected}{1.4}
\newcommand{\fracnepdwarfmeanlow}{65}
\newcommand{\evolutionMnepfinalMWi}{4\times10^{10}}
\newcommand{\fracnepmetalmedianlowermasshigh}{4\%}
\newcommand{\fracsatmetalmeanhigh}{2\%}
\newcommand{\evolutionMwindfinalsubMWq}{2\times10^{9}}
\newcommand{\evolutionMsatwindmaxMWi}{30\%}
\newcommand{\fraclessmasswindmeanhigh}{40\%}
\newcommand{\meanmstarcoreMWlow}{8\times10^{10}}
\newcommand{\finterfacesubMWmeanlow}{15}
\newcommand{\fmetalMWmeanhigh}{0.8}
\newcommand{\totmassCGMsubMWmeanlow}{5\times10^{9}}
\newcommand{\fracwindmetalmedianmdlow}{60}
\newcommand{\totmetmassinterfacesubMWmeanhigh}{10^{6}}
\newcommand{\evolutionMsatwindatpeakMWi}{10^{10}}
\newcommand{\fracsubMWsatmeanhigh}{10^{-5}}
\newcommand{\fracMWsatmeanhigh}{10^{-5}}
\newcommand{\totmetmassCGMsubMWmeanhigh}{2\times10^{6}}
\newcommand{\fmetgalaxydwarfmeanhigh}{25}
\newcommand{\totmassCGMdwarfmeanlow}{8\times10^{8}}
\newcommand{\fracnepmetalmeanhigh}{4}
\newcommand{\fracsatwindmetalmeanfhigh}{15}
\newcommand{\mtwelvefIGMAccretionmedmassfrachigh}{0.77}
\newcommand{\evolutionMmaxdwarfy}{3\times10^{8}}
\newcommand{\fmetalMWmeanlow}{1}
\newcommand{\evolutionMnepatpeakMWi}{3\times10^{10}}
\newcommand{\meansubMWwindcoremasshigh}{80}
\newcommand{\totmassgalaxydwarfmeanlow}{3\times10^{8}}
\newcommand{\evolutionMsatmaxMWi}{3\times10^{9}}
\newcommand{\metprofIGMAccinnerlow}{3.3}
\newcommand{\evolutionMnepatpeakdwarfy}{2\times10^{8}}
\newcommand{\yboxmtenvlow}{0.034}
\newcommand{\fmetalsubMWmeanhigh}{0.7}
\newcommand{\metMWmdwindiqrlow}{0.4}
\newcommand{\fracwindmetalmeanflow}{90}
\newcommand{\metprofWindinnerlow}{1.5}
\newcommand{\fbaryonsubMWmeanhigh}{0.6}
\newcommand{\metMWwindmedlow}{4\times10^{-1}}
\newcommand{\deltaMWsteepmaxlow}{5.1}
\newcommand{\totmetmassgalaxysubMWmeanhigh}{10^{6}}
\newcommand{\totmassinterfaceMWmeanlow}{8\times10^{9}}
\newcommand{\fmetCGMdwarfmeanhigh}{60}
\newcommand{\metprofSatWindinnerlow}{1.1}
\newcommand{\fbaryonMWmeanlow}{0.7}
\newcommand{\fracsubMWsatmeanlow}{10^{-5}}
\newcommand{\metMWmdNEPmedhigh}{3\times10^{-2}}
\newcommand{\metsubMWwindmedlow}{2\times10^{-1}}
\newcommand{\fracnepmetalmedianlowermassmdhigh}{5\%}
\newcommand{\fmetgalaxysubMWmeanlow}{40}
\newcommand{\metMWsatwindboosthigh}{10^{0}}
\newcommand{\totmassgalaxyMWmeanhigh}{10\times10^{9}}
\newcommand{\fmetgalaxydwarfmeanlow}{20}
\newcommand{\metdwarfsatwindmedlow}{10^{-2}}
\newcommand{\fracnepMWmeanhigh}{45}
\newcommand{\rgalrviroutliervaluelow}{0.27}
\newcommand{\evolutionMsatwindatpeaksubMWq}{4\times10^{8}}
\newcommand{\fracMWsatwindmeanlow}{20\%}
\newcommand{\metMWsatwindboostlow}{3}
\newcommand{\fmetaldwarfmeanlow}{0.7}
\newcommand{\totmassinterfacesubMWmeanlow}{10^{9}}
\newcommand{\totmetmassgalaxyMWmeanhigh}{3\times10^{7}}
\newcommand{\fracwindmetalmeanmdlow}{55}
\newcommand{\meddenseCGMmasslow}{1\%}
\newcommand{\maxdenseinterfaceradiusratiohigh}{166\%}
\newcommand{\profilevarysubMWlow}{1.6}
\newcommand{\fgalaxyMWmeanlow}{45}
\newcommand{\totmassCGMsubMWmeanhigh}{3\times10^{9}}
\newcommand{\fracwindmetalmedianlow}{90\%}
\newcommand{\fracnepmetalmedianMWlow}{1\%}
\newcommand{\meansubMWsatellitewindpeakmasslow}{30}
\newcommand{\profilevarydwarfhigh}{1.5}
\newcommand{\metMWsatwindmedhigh}{5\times10^{-2}}
\newcommand{\totmassgalaxyMWmeanlow}{6\times10^{10}}
\newcommand{\fracnepmetalmeanmdMWlow}{25}
\newcommand{\totmetmassinterfaceMWmeanhigh}{10^{7}}
\newcommand{\fracMWsatmeanlow}{10^{-5}}
\newcommand{\fracwindmetalmedianflow}{95}
\newcommand{\totmetmassgalaxysubMWmeanlow}{2\times10^{7}}
\newcommand{\evolutionMsatwindmaxmassdwarfy}{2\times10^{7}}
\newcommand{\fracnepMWmeanlow}{60}
\newcommand{\evolutionMsatwindfinalsubMWq}{2\times10^{8}}
\newcommand{\fmetgalaxysubMWmeanhigh}{30}
\newcommand{\metprofSatWindouterhigh}{1.6}
\newcommand{\totmassgalaxysubMWmeanhigh}{10\times10^{8}}
\newcommand{\profilevarysubMWhigh}{1.7}
\newcommand{\totmassgalaxydwarfmeanhigh}{10^{8}}
\newcommand{\fCGMdwarfmeanlow}{70}
\newcommand{\fmetgalaxyMWmeanlow}{70}
\newcommand{\metwindspreadlow}{1}
\newcommand{\floormetalmassmaxhigh}{1\%}
\newcommand{\metsubMWsatwindmedlow}{7\times10^{-2}}
\newcommand{\metsatwindspreadhigh}{0.7}
\newcommand{\fmetinterfacesubMWmeanlow}{20}
\newcommand{\evolutionMmindwarfy}{2\times10^{8}}
\newcommand{\metdwarfsatwindmedhigh}{10^{-2}}
\newcommand{\totmassinterfacedwarfmeanlow}{4\times10^{8}}
\newcommand{\metprofWindinnerhigh}{0.93}
\newcommand{\fracwindmetalmeanfhigh}{80}
\newcommand{\evolutionMnepfinaldwarfy}{10^{8}}
\newcommand{\maxdenseinterfacemasshigh}{38\%}
\newcommand{\metdwarfwindmedhigh}{3\times10^{-2}}
\newcommand{\metprofWindouterlow}{2.5}
\newcommand{\fgalaxyMWmeanhigh}{30}
\newcommand{\totmetmassgalaxydwarfmeanhigh}{8\times10^{4}}
\newcommand{\evolutionMsatwindmaxmassMWi}{10^{10}}
\newcommand{\meanstellarpercentingaplow}{3\%}
\newcommand{\fmetaldwarfmeanhigh}{0.6}
\newcommand{\fracsatwindmetalmeanflow}{5}
\newcommand{\totmetmassCGMsubMWmeanlow}{10^{7}}
\newcommand{\meanMWsatellitewindpeakmasshigh}{30}
\newcommand{\fracnepmetalmeanlow}{1}
\newcommand{\fracnepmetalmedianMWmdlow}{20\%}
\newcommand{\fracsatwindmetalmeanlow}{15\%}
\newcommand{\deltaMWsmoothlow}{1.5}
\newcommand{\fgalaxysubMWmeanhigh}{20}
\newcommand{\finterfacesubMWmeanhigh}{20}
\newcommand{\fraclessmasssatwindmaxhigh}{20\%}
\newcommand{\fracwindmetalmeanlow}{75\%}
\newcommand{\evolutionMsatmaxsubMWq}{2\times10^{8}}
\newcommand{\fracwindmetalmedianfhigh}{85}
\newcommand{\metbudgdivmetbudgPeeples}{0.53}
\newcommand{\evolutionMsatwindmaxsubMWq}{10\%}
\newcommand{\maxpercentsplitCGMlow}{0.5}
\newcommand{\fmetCGMdwarfmeanlow}{50}
\newcommand{\maxdenseinterfacemasslow}{71\%}
\newcommand{\deltaMWsteepminlow}{1.3}
\newcommand{\fgalaxysubMWmeanlow}{30}
\newcommand{\fgalaxydwarfmeanhigh}{15}
\newcommand{\totmassinterfacedwarfmeanhigh}{8\times10^{7}}
\newcommand{\fretdivfretPeeples}{3.2}
\newcommand{\fmetinterfacesubMWmeanhigh}{25}
\newcommand{\fCGMdwarfmeanhigh}{75}
\newcommand{\fracnepsubMWmeanhigh}{55}
\newcommand{\meanMWsatellitewindpeakmasslow}{40}
\newcommand{\fracdwarfsatmeanhigh}{10^{-5}}
\newcommand{\metMWwindiqrlow}{1}
\newcommand{\fracnepmeanlow}{60\%}
\newcommand{\fmetCGMsubMWmeanhigh}{45}
\newcommand{\deltaMWsmoothhigh}{nan}
\newcommand{\fracsatmetalmeanmdhigh}{1}
\newcommand{\fracwindmetalmeanhigh}{75\%}
\newcommand{\fmetCGMMWmeanlow}{15}
\newcommand{\totmassinterfaceMWmeanhigh}{4\times10^{9}}
\newcommand{\fracsatmetalmeanlow}{3\%}
\newcommand{\fracnepmeanhigh}{50\%}
\newcommand{\finterfacemedianlow}{5\%}
\newcommand{\meandwarfwindcoremasshigh}{60}
\newcommand{\metsubMWwindmedhigh}{10^{-1}}
\newcommand{\profilelowMincreaseIGMAccretionhigh}{1.8}
\newcommand{\evolutionMsatwindminMWi}{9\%}
\newcommand{\meansubMWwindcoremasslow}{40}
\newcommand{\metmassdivmetmassPeeples}{1.7}
\newcommand{\evolutionMsatwindatpeakdwarfy}{10^{7}}
\newcommand{\totmetmassinterfacedwarfmeanhigh}{5\times10^{4}}
\newcommand{\meandwarfwindcoremasslow}{50}
\newcommand{\fracwindmetalmedianmdhigh}{80}
\newcommand{\fmetalsubMWmeanlow}{0.8}
\newcommand{\meanmstarcoreMWhigh}{5\times10^{9}}
\newcommand{\mtwelvefsatwindmedmassfrachigh}{0.8}
\newcommand{\fracMWwindmeanlow}{15\%}
\newcommand{\fCGMMWmeanhigh}{55}
\newcommand{\fmetinterfaceMWmeanlow}{10}
\newcommand{\meanstellarmasspercentingaphigh}{2\%}
\newcommand{\fracnepdwarfmeanhigh}{60}
\newcommand{\fracwindmetalmedianhigh}{85\%}
\newcommand{\rgalrviroutliernamehigh}{m11q}
\newcommand{\yboxhigh}{0.032}
\newcommand{\evolutionMsatwindmaxmasssubMWq}{10^{9}}
\newcommand{\metMWsatwindmedlow}{6\times10^{-2}}
\newcommand{\evolutionMsatwindfinaldwarfy}{5\times10^{6}}
\newcommand{\fracMWsatwindmeanhigh}{20\%}
\newcommand{\yboxlow}{0.036}
\newcommand{\fmetCGMMWmeanhigh}{45}
\newcommand{\metMWmdwindiqrhigh}{0.2}
\newcommand{\evolutionMnepinitialsubMWq}{8\times10^{8}}
\newcommand{\yboxstdlow}{0.001}
\newcommand{\fmetgalaxyMWmeanhigh}{35}
\newcommand{\metprofSatWindinnerhigh}{0.94}
\newcommand{\totmetmassinterfacesubMWmeanlow}{10^{7}}
\newcommand{\fracsatwindmetalmeanmdhigh}{10}
\newcommand{\fracnepmetalmedianlowermassmdlow}{2\%}
\newcommand{\mtwelvefIGMAccretionmedmassfraclow}{0.78}
\newcommand{\totmetmassCGMdwarfmeanhigh}{10^{5}}
\newcommand{\fracwindmetalmeanmdhigh}{75}
\newcommand{\deltatenmaxlow}{1.3}
\newcommand{\fracnepmetalmedianMWmdhigh}{20\%}
\newcommand{\meanstellarpercentingaphigh}{3\%}
\newcommand{\metMWwindiqrhigh}{0.5}
\newcommand{\finterfaceMWmeanlow}{5}
\newcommand{\fracsatmetalmeanmdlow}{5}
\newcommand{\fraclessmasswindmeanlow}{30\%}
\newcommand{\fracdwarfsatmeanlow}{10^{-5}}
\newcommand{\meanmhmdMWhigh}{2\times10^{11}}
\newcommand{\deltaMWsteepmaxhigh}{2.6}
\newcommand{\fbaryonMWmeanhigh}{0.7}
\newcommand{\meanmstarmdMWlow}{3\times10^{10}}
\newcommand{\fracsatmetalmeanflow}{1}
\newcommand{\fbaryondwarfmeanlow}{0.2}
\newcommand{\totmetmassgalaxydwarfmeanlow}{10^{6}}
\newcommand{\meanMWwindcoremasslow}{40}
\newcommand{\mtenvmstarlow}{7\times10^{5}}
\newcommand{\meddenseinterfacemasshigh}{3\%}
\newcommand{\evolutionMwindfinaldwarfy}{10^{8}}
\newcommand{\metprofAllinnerhigh}{2}
\newcommand{\fracsatmetalmeanfhigh}{2}
\newcommand{\fmetinterfacedwarfmeanlow}{30}
\newcommand{\deltaMWsteephigh}{1.7}
\newcommand{\finterfaceMWmeanhigh}{10}
\newcommand{\finterfacemedianhigh}{10\%}
\newcommand{\metMWmdNEPmedlow}{4\times10^{-2}}
\newcommand{\rgalrviroutliernamelow}{m11v}
\newcommand{\finterfacedwarfmeanlow}{15}
\newcommand{\totmassCGMMWmeanhigh}{2\times10^{10}}
\newcommand{\rgalrviroutliervaluehigh}{0.29}
\newcommand{\meanmhcoreMWlow}{10^{12}}
\newcommand{\metprofIGMAccinnerhigh}{1.4}
\newcommand{\evolutionMnepatpeaksubMWq}{2\times10^{9}}
\newcommand{\meanmstarmdMWhigh}{3\times10^{9}}
\newcommand{\meanMWwindcoremasshigh}{60}
\newcommand{\totmassCGMMWmeanlow}{5\times10^{10}}
\newcommand{\meddenseinterfacemasslow}{0\%}
\newcommand{\floormetalmassmaxlow}{1\%}
\newcommand{\deltaelevenmaxlow}{1.1}
\newcommand{\fmetCGMsubMWmeanlow}{35}
\newcommand{\fCGMsubMWmeanhigh}{60}
\newcommand{\totmetmassCGMMWmeanhigh}{4\times10^{7}}
\newcommand{\metprofWindouterhigh}{1.3}
\newcommand{\evolutionMnepinitialMWi}{3\times10^{9}}
\newcommand{\totmassCGMdwarfmeanhigh}{4\times10^{8}}
\newcommand{\fbaryondwarfmeanhigh}{0.3}
\newcommand{\fgalaxydwarfmeanlow}{10}
\newcommand{\mtwelvefsatwindmedmassfraclow}{0.78}
\newcommand{\fmetinterfacedwarfmeanhigh}{20}
\newcommand{\metprofAllouterhigh}{2.8}
\newcommand{\evolutionMsatwindmindwarfy}{1\%}
\newcommand{\fracnepsubMWmeanlow}{50}
\newcommand{\evolutionMnepinitialdwarfy}{2\times10^{8}}
\newcommand{\evolutionMwindatpeaksubMWq}{2\times10^{9}}
\newcommand{\profilevarydwarflow}{2.9}
\newcommand{\deltatenmaxhigh}{0.5}
\newcommand{\maxdenseinterfaceradiusratiolow}{31\%}
\newcommand{\evolutionMsatwindfinalMWi}{7\times10^{9}}
\newcommand{\meansubMWsatellitewindpeakmasshigh}{40}
\newcommand{\evolutionMmaxMWi}{6\times10^{10}}
\newcommand{\profilevaryMWlow}{1.5}
\newcommand{\finterfacedwarfmeanhigh}{10}
\newcommand{\metsatwindspreadlow}{0.6}
\newcommand{\profilevaryMWhigh}{1.3}
\newcommand{\fraclessmasssatwindmaxlow}{35\%}
\newcommand{\metwindspreadhigh}{1}
\newcommand{\evolutionMwindatpeakMWi}{2\times10^{10}}
\newcommand{\maxpercentsplitCGMhigh}{0.3}

\begin{abstract}

We use a particle tracking analysis to study the origins of the circumgalactic medium (CGM), separating it into (1) accretion from the intergalactic medium (IGM), (2) wind from the central galaxy, and (3) gas ejected from other galaxies.
Our sample consists of 21 FIRE-2 simulations, spanning the halo mass range $M_{\rm h} \sim 10^{10}-10^{12}$ M$_{\odot}$, and we focus on $z=0.25$ and $z=2$.
Owing to strong stellar feedback, only $\sim L^\star$ halos retain a baryon mass $\gtrsim50\%$ of their cosmic budget.
Metals are more efficiently retained by halos, with a retention fraction $\gtrsim50\%$.
Across all masses and redshifts analyzed $\gtrsim 60\%$ of the CGM mass originates as IGM accretion (some of which is associated with infalling halos).
Overall, the second most important contribution is wind from the central galaxy, though gas ejected or stripped from satellites can contribute a comparable mass in $\sim L^\star$ halos.
Gas can persist in the CGM for billions of years, resulting in well mixed-halo gas.
Sight lines through the CGM are therefore likely to intersect gas of multiple origins.
For low-redshift $\sim L^\star$ halos, cool gas ($T<10^{4.7}$ K) is distributed on average preferentially along the galaxy plane, however with strong halo-to-halo variability.
The metallicity of IGM accretion is systematically lower than the metallicity of winds (typically by $\gtrsim 1$ dex), although CGM and IGM metallicities depend significantly on the treatment of subgrid metal diffusion.
Our results highlight the multiple physical mechanisms that contribute to the CGM and will inform observational efforts to develop a cohesive picture.

\end{abstract}



\section{Introduction}

The circumgalactic medium (CGM) of galaxies is inferred to contain a baryonic mass comparable to or in excess of the galaxy mass~\citep[e.g.][]{Werk2014,Tumlinson2017}.
This large reservoir of gas, loosely defined as the gas immediately outside the galaxy but inside the dark matter halo, interfaces strongly with the galaxy:
accretion onto the galaxy is necessary to sustain galaxy growth over the age of the Universe \citep[e.g.][]{Prochaska2009, Bauermeister2010},
and in turn material from the galaxy returns to the CGM in the form of galactic winds driven by stellar and AGN feedback \citep[e.g.,][]{Heckman2000, Steidel2010, Jones2012, Rubin2014, Cicone2014}.
For recent reviews of the CGM see \cite{Putman2012}, \cite{Tumlinson2017}, and~\cite{Fox2017}.

Building a theoretical framework for the CGM requires accurately modeling both galaxies and their larger environment.
One way to approach this problem is through cosmological hydrodynamic simulations of galaxy formation~\citep[e.g.][]{Somerville2015}.
These simulations calculate the evolution of dark matter, gas, and stars according to the relevant physics (e.g. gravity, hydrodynamics, star formation, feedback, etc.). 
Cosmological galaxy formation simulations have been used to understand the CGM in a variety of ways, from analyses of the dynamics of the CGM in simulations~\citep[e.g.][]{Keres2005, Keres2009, Faucher-Giguere2011a, 2011MNRAS.414.2458V, Nelson2013, Oppenheimer2010, Wetzel2015, Oppenheimer2018} to those that use simulations to provide context to observations~\citep[e.g.][]{Faucher-Giguere2010, Faucher-Giguere2011,Hummels2013,Liang2015,Corlies2016, 2016MNRAS.462.2440T, Gutcke2017, Nelson2017, Roca-Fabrega2018}.

Our analysis makes use of high-resolution cosmological ``zoom-in'' simulations \citep[][]{1993ApJ...412..455K, 2014MNRAS.437.1894O} created as part of the FIRE project\footnote{\url{https://fire.northwestern.edu/}}~\citep[][]{Hopkins2014,Hopkins2017}.
It is particularly instructive to study CGM gas flows in the FIRE simulations because they generate galactic winds whose properties are not put in by hand, but rather emerge from energy/momentum injection by stellar feedback on the scale of star-forming regions. 
The FIRE simulations have been used to investigate the dynamics of gas flows in the CGM~\citep{Muratov2015,Muratov2016,Angles-Alcazar2017,Stewart2016} and the observability of the CGM through \ion{H}{i} and metal absorption, metal-line emission, the Sunyaev-Zel'dovich effect, and X-ray emission~\citep{Faucher-Giguere2015,Faucher-Giguere2016,Sravan2016,VandeVoort2016,Hafen2016}.
The observational comparisons are consistent with observations where constraints are available~\citep{Faucher-Giguere2015,Faucher-Giguere2016,Hafen2016}.

A powerful aspect of Lagrangian simulations is that they provide access to the full time history of gas resolution elements.
Following individual gas elements, or ``particle tracking'', has been used to study the accretion rate onto galaxies and halos,
recycling of galactic winds and their effects,
and the fate and origin of galaxy material~\citep[e.g.][]{Keres2005,Oppenheimer2010,Ubler2014,Nelson2015,Angles-Alcazar2017,DeFelippis2017,Oppenheimer2017,Crain2017,Brennan2018}.
The primary goal of this work is to use a particle tracking analysis to study the origin of gas and metals in the CGM. 
Our analysis builds on the analysis of \cite{Angles-Alcazar2017}, who developed a particle tracking algorithm suitable for use on FIRE simulations and who used it to investigate the origin of baryons in galaxies, linking baryon cycling processes to the mass assembly of galaxies.
\cite{Muratov2015} and \cite{Muratov2016} also studied the gas and metal content in the CGM of the FIRE simulations, but using instantaneous mass flow rates instead of particle tracking. 
Our analysis is similar to that of~\cite{Ford2014}, but also includes mass acquired through satellite galaxies and their winds.
Moreover, our zoom-in simulations have higher resolution and model galaxy formation physics using very different subgrid models than the full-volume simulations analyzed by \cite{Ford2014}.
We focus our analysis on $z=0.25$ and $z=2$ to connect to a wide variety of observations available at these redshifts~\citep[e.g.][]{Tumlinson2013,Steidel2010, Prochaska2017,Rudie2012,OMeara2013,Lehner2016,Chen2018}.

The primary observational method for studying CGM gas is through absorption line spectroscopy~\citep[e.g.,][]{Bahcall1969a}.
Observations with ground-based telescopes can measure ions in the CGM at both low and high redshifts~\citep[e.g.][]{2006ApJ...637..648S,2010ApJ...724L.176C, Steidel2010,Rudie2012, Kacprzak2012}.
In the last decade, new observations have been enabled at low redshift by the installation of the Cosmic Origins Spectrograph \citep{Green2012} to the \textit{Hubble Space Telescope} \citep[e.g.][]{Tumlinson2013,Werk2014,Liang2014,Johnson2017,Keeney2017}.

The interpretation of absorption line spectroscopy is complicated by its one-dimensional nature, which makes it difficult to distinguish CGM gas of  different physical origins.
One proposed diagnostic for distinguishing between origins is through the metallicity of the gas, assuming IGM accretion, satellite material, and winds are enriched by different amounts~\citep[e.g.][]{Lehner2013,Fox2013,Wotta2016,Fumagalli2016,Hafen2016,Lehner2016,Prochaska2017}.
Observations indicate a wide range of CGM metallicities (spanning $>2$ dex), which may reflect different origins~\citep[e.g.][though see Stern et al. 2016 who use multi-density ionization modeling and infer a significantly narrower metallicity distribution]{Lehner2013,Fumagalli2016,Wotta2016,Prochaska2017,Zahedy2018}.
Because observations alone do not unambiguously associate absorption systems with specific physical origins, simulation analyses like ours play a key role in developing observational diagnostics \citep[for a review of efforts in this area, see][]{2017ASSL..430..271F}.

We consider three main physically distinct origins for CGM gas:

\begin{enumerate}

\item The first origin we consider is accretion into the CGM from the IGM, hereafter referred to as \textit{IGM accretion}.
Some IGM accretion may fall spherically onto dark matter halos and generate a quasi-spherical shock~\citep{White1978,White1991}, while other accretion can penetrate into the halo in the form of cold streams~\citep[e.g.][]{Keres2005,Keres2009a,2009Natur.457..451D}.
IGM accretion is often assumed to be metal poor and provide a source of new gas to galaxies and their halos, as needed to sustain star formation.

\item The second origin is ejection into the CGM via winds from the central galaxy~\citep[e.g.][]{Heckman2017}, hereafter referred to simply as \textit{wind}.
Simulations show that winds regulate star formation and help produce galaxies that match the observed stellar mass-halo mass relation~\citep[e.g.][]{2007ApJ...655L..17B,2011MNRAS.416.1354D,Angles-Alcazar2014,Somerville2015}.
Material ejected in a wind may then recycle back onto the galaxy as ``wind recycling'' or a ``galactic fountain'' \citep[e.g.][]{Shapiro1976,Bregman1980,Oppenheimer2010,Ford2014,Angles-Alcazar2017}.

\item The third origin is arrival into the CGM via a galaxy other than the main galaxy.
For clarity, we divide this origin up into material that has since left the other galaxy (\textit{satellite wind}), and material that is within the virial radius of the central galaxy's halo but also in the ISM of another galaxy (\textit{satellite ISM}).
We include satellite ISM as a CGM component even though it is not diffuse halo gas because satellite ISM can contribute to absorption systems commonly used to probe the CGM observationally. 
It is therefore useful to quantify its contribution to galaxy halos. 
We will show that satellite ISM is a relatively minor component of the CGM by mass so including it does not significantly skew any of our results.

Because massive galaxies can host substantial populations of satellites, gas associated with satellite galaxies or their halos could be a significant contribution to the CGM. 
Gas ejected in winds from satellites can exist outside of the satellite for an extended period of time, and can also smoothly accrete onto the central galaxy and fuel in-situ star formation, which has been identified as an important growth mode for massive galaxies.
\cite{Angles-Alcazar2017} showed that this ``intergalactic transfer'' of gas via winds can account for $\sim 1/3$ of the $z=0$ stellar content of Milky-Way mass galaxies in FIRE-1 simulations.
In this paper, we quantify the contribution of ISM gas ejected from satellites to the CGM of central galaxies. 
Although we use the designation ``satellite wind,'' gas can leave a satellite not only via galactic outflows from the satellite~\citep[e.g.][]{Wang1993,Barger2016,McClure-Griffiths2018}, but also via ram pressure stripping~\citep[e.g.][]{Gunn1972,Tonnesen2010,Yun2018} or tidal stripping~\citep[e.g.][]{Connors2006}.
The Magellanic Stream in the halo of our own Galaxy is likely a result of one or multiple of these processes~\citep[e.g.][]{DOnghia2016,Bustard2018}.

\end{enumerate}

The plan of this paper is as follows. 
We describe our simulations in \S\ref{sec:simulations} and our analysis methodology in \S\ref{sec:analysis}.
We study the total baryon and metal mass contained in the CGM in our simulations in \S\ref{sec:mass_budgets}. 
We characterize the paths followed by CGM gas flows of different origins in \S\ref{sec:particle_pathlines}.
In \S\ref{sec:origin_by_mass} we investigate how much mass each origin contributes to the CGM as a function of halo mass, redshift, radial distance, and polar angle.
We address metallicity as a potential diagnostic of origin in \S\ref{sec:metallicity}.  
We discuss our results in \S\ref{sec:discussion} and summarize our conclusions in \S\ref{sec:conclusions}.

\section{Methods}

\subsection{Simulations}
\label{sec:simulations}

We use a sample of cosmological hydrodynamic ``zoom-in'' simulations created as part of the FIRE project.
The simulations were run with the multi-method gravity and hydrodynamics code \textsc{GIZMO}\footnote{\url{http://www.tapir.caltech.edu/\~phopkins/Site/GIZMO.html}}~\citep{Hopkins2015} in its meshless finite-mass (``MFM'') mode.
The MFM method is a Lagrangian method using Riemann solvers.
The method conserves mass, energy, and linear momentum to machine precision, and also conserves angular momentum. 
In MFM, the hydrodynamic solver is designed such that there is no mass flux between resolution elements; particle tracking can therefore be straightforwardly implemented by following resolution elements. 
\footnote{This is true modulo mass that can be added to gas resolution elements as a result of stellar mass loss, and particle splitting that can occur to prevent resolution elements from becoming too massive.
We discard particles that experience splitting, which we do not expect to significantly affect our results (e.g. for all simulations $<\maxpercentsplitCGMlow\%$ of gas resolution elements in the CGM at $z=0.25$ are split).} 
In a variety of tests, MFM has been shown to perform better than smoothed-particle hydrodynamics methods~\citep{Hopkins2015}.
\textsc{GIZMO} solves gravity using a heavily-modified version of the Tree-PM solver used in \textsc{GADGET-3}~\citep{Springel2005}, updated to use adaptive gravitational softening for gas. 
The particular sample of simulations used in our analysis was run with the FIRE-2 version of GIZMO~\citep{Hopkins2017}. 
In most cases, the main halo was chosen solely based on halo mass, however \texttt{m12r\_md} and \texttt{m12w\_md} were chosen to have an LMC-mass satellite at $z=0$.
We find that this selection does not significantly affect our analysis; 
throughout our analysis these halos have quantitative results consistent with other halos in the same mass range.
\texttt{m12r\_md} and \texttt{m12w\_md} will be presented in detail in Samuel et al., in prep, including the detailed selection methodology. 
The simulations we analyze are listed in Table~\ref{table:simulations_used}, and span the halo mass range $M_{\rm h}(z=0) \sim 10^{10} - 10^{12} M_\odot$. 
We identify and track the evolution of dark matter halos using the Amiga Halo Finder~\citep[AHF;][]{Gill2004,Knollmann2009} and adopt the virial overdensity definition of \cite{Bryan1998}.
In this paper, we define $M_{\star}$ as the stellar mass enclosed inside the galaxy radius as defined in \S\ref{sec:analysis}. 
Throughout, we assume a standard flat $\Lambda$CDM cosmology with $\Omega_{\rm m }\approx 0.32$, $\Omega_{\Lambda}=1-\Omega_{\rm m}$, $\Omega_{\rm b} \approx 0.049$, and $H_{0} \approx 67$ km s$^{-1}$ Mpc$^{-1}$ \citep[e.g.,][]{PlanckCollaboration2018}.\footnote{For consistency with previous work, some of our simulations were evolved with slightly different sets of cosmological parameters, but we do not expect this to significantly impact on any of our results given the small differences in the parameters.}

The full details of our simulation methods are provided in \cite{Hopkins2017}; we summarize key elements here. 
Radiative heating and cooling are tracked from  $T=10$ K to $10^{10}$ K,  including free-free emission, Compton scattering with the cosmic microwave background, photoelectric heating, high-temperature metal line cooling, and approximations for low-temperature cooling by molecules  and fine-structure metal lines.
Photo-heating and photo-ionization include effects from a cosmic UV background model \citep{Faucher-Giguere2009} and approximations for local sources and self-shielding of dense gas.
Star formation occurs only in self-gravitating gas (under the \citealt{Hopkins2013b} criteria), but also with a requirement that it is molecular, self-shielding, and has a density of at least $n_{\rm H} = 1000$ cm$^{-3}$.
Stellar feedback includes momentum from radiation pressure;  energy, momentum, mass, and metals from Type Ia and II supernovae and stellar winds; and photo-ionization and photo-electric heating.
Star particles are treated as independent stellar populations, and the feedback quantities are drawn from the \textsc{STARBURST99} stellar evolution models~\citep{Leitherer1999} assuming the IMF of \cite{Kroupa2001}.
We track the evolution of 9 independent metal species.

A subset of our simulations utilize a sub-grid model for metal transport between neighboring resolution elements (see Table~\ref{table:simulations_used}).
The purpose of the model is to capture sub-resolution-level metal diffusion that is not by default accounted for by the MFM hydrodynamic solver. 
Our metal transport model is described in detail in \cite{Hopkins2017a} and \cite{Hopkins2017}, but we summarize the most important aspects here.
The model assumes that the transport of metals between neighboring resolution elements follows a diffusion equation, $\partial_t ( \rho Z ) = \nabla \cdot ( \kappa \rho \nabla Z ) $, where $\rho$ is the density, $Z$ is the metallicity, and $\kappa$ is the relevant diffusion coefficient.
The primary assumption for the metal diffusion model is that the diffusion timescale scales with the eddy turnover time of the largest unresolved turbulent eddies, i.e. that unresolved turbulence exchanges metals between resolution elements.
This means $\kappa \sim \kappa_{\rm turb} \sim \lambda_{\rm eddy} v_{\rm eddy}$, where $\kappa_{\rm turb}$ is the ``eddy diffusivity'' calculated via the \cite{Smagorinsky1963} approximation~\citep[e.g.][]{Shen2010}.
This approximation is appropriate under certain conditions usually met in our simulations, but if these conditions are violated then the diffusivity can be significantly overestimated~\citep[e.g.][]{Colbrook2017}.
Our subgrid prescription caps the diffusivity at the equivalent value allowed in the meshless finite-volume (``MFV'') mode of GIZMO, which explicitly allow mass-flux between elements. 
This diffusivity (and the diffusivity that occurs in other grid-based codes) generally overestimates the true physical diffusivity. 
For galaxies up to $\sim L^{\star}$ simulated in the cosmological context, previous tests suggest that including subgrid turbulent metal diffusion has mostly negligible effects on the dynamical evolution of galaxies \citep[e.g.,][]{Su2016}, but the dynamical effects could be more important for more massive galaxies which develop hot halos whose cooling can be modified substantially by metal mixing. 
Unsurprisingly, turbulent metal diffusion can have large effects on metallicity distributions of gas and stars in both the ISM and CGM~\citep[e.g.,][]{Escala2018, Rennehan2018}, and so it is important to investigate its effects on CGM properties.

\begin{table*}
\caption{Simulation parameters.}
\begin{tabular}{ccccccccccc}
\hline
Name   & $M_{\rm vir}(z=0)$     & $M_\star(z=0)$     & $R_{\textrm{vir}}(z=0)$   &  $m_\textrm{b}$           & $m_\textrm{dm}$              & $\epsilon_{\textrm{b}}$          & $\epsilon_\star$           & $\epsilon_{\textrm{dm}}$           &    Metal Diffusion?   & Reference        \\
               & ($M_\odot$)                   & ($M_\odot$)                 & (kpc)                  & ($M_\odot$)       & ($M_\odot$)           &  (pc)      & (pc)         &  (pc)          &                                      &                            \\
 \hline
\texttt{m10q}    & 8.0e9                 & 2.0e6                   &  52                        &  250           &  1300         & 0.4          & 1.4           &      28.5   &               No                       &     A              \\  
\texttt{m10y}    & 1.4e10               & 1.1e7                   &  64                        &  260           &  1250         & 0.2          & 2.1           &      21      &                No                      &   A              \\ 
\texttt{m10v}    & 2.7e10                  & 1.2e6                  &  78                         &  250           &  1300         & 0.4          & 1.4           &      28.5   &              No                        &     A             \\  
\texttt{m10z}    & 3.7e10               & 3.9e7                   &  86                        &  260           &  1250         & 0.2          & 2.1           &      21      &                No                     &    B               \\ 
\texttt{m11a}    &  4.1e10              & 1.2e8                   &  90                        &  2100           &  1e4         & 0.4          & 4.3           &      43      &                No                     &     B               \\ 
\texttt{m11b}    &  4.4e10              & 1.2e8                   &  92                        &  2100           &  1e4         & 0.4          & 4.3           &      43      &                No                     &     B              \\ 
\texttt{m11c}    & 1.4e11               & 8.9e8                   &  140                        &  2100           &  1040         & 0.4          & 4.3          &      43      &                No                     &    B               \\ 
\texttt{m11q}    & 1.5e11               & 2.0e9                   &  140                        &  880           &  4400         & 0.5          & 2.0           &      20      &                No                     &    A              \\ 
\texttt{m11v}    & 2.6e11               & 2.6e9                   &  170                        &  7070           &  3.52e4         & 0.4          & 1.4           &      28.5      &                No                     &    A            \\ 
\texttt{m12i}    & 1.1e12                & 7.4e10                   &  270                        &  7070           &  3.52e4         & 1.0         & 4.0           &      20       &                No                     &    C              \\ 
\texttt{m12m}    & 1.6e12             & 1.4e11                   &  300                        &  7070           &  3.52e4         & 1.0         & 4.0           &      20       &                No                     &   A               \\ 
\texttt{m12f}    & 1.6e12               & 9.4e10                   &  300                        &  7070           &  3.52e4         & 1.0         & 4.0           &      20       &                No                     &   D            \\
\texttt{m11i\_md}    & 7.2e10               & 1e9                   &  110                        &  7070           &  3.52e4         & 1.0         & 4.0           &      40       &                Yes                     &    E              \\ 
\texttt{m11e\_md}    & 1.6e11               & 1.5e9                  &  140                        &  7070           &  3.52e4         & 1.0         & 4.0           &      40       &                Yes                     &    E              \\ 
\texttt{m11h\_md}    & 1.9e11               & 3.9e9                   &  150                        &  7070           &  3.52e4         & 1.0         & 4.0           &      40       &                Yes                     &    E              \\ 
\texttt{m11d\_md}    & 3.0e11               & 4.5e9                   &  170                        &  7070           &  3.52e4         & 1.0         & 4.0           &      40       &                Yes                     &    E              \\ 
\texttt{m12z\_md}    & 8.0e11                 & 2.3e10                  &  240                        &  4170           &  2.14e4         & 0.4         & 3.2           &      33       &                Yes                     &    F              \\ 
\texttt{m12w\_md}    & 9.8e11                 & 6.2e10                  &  260                        &  7070           &  3.52e4         & 0.5         & 4.0           &      40       &                Yes                     &    G              \\ 
\texttt{m12r\_md}    & 1.0e12                  & 1.9e10                 &  270                        &  7070           &  3.52e4         & 0.5         & 4.0           &      40       &                Yes                     &    G             \\
\texttt{m12b\_md}    & 1.3e12                  & 9.0e10                  &  290                        &  7070           &  3.52e4         & 0.5         & 4.0           &      40       &                Yes                     &    F             \\ 
\texttt{m12c\_md}    & 1.3e12                  & 6.4e10                  &  280                        &  7070           &  3.52e4         & 0.5         & 4.0           &      40       &                Yes                     &   F                \\ 
\hline
\end{tabular}
\\
\begin{flushleft}
$M_{\rm vir}(z=0)$ is the total mass contained within $R_{\rm vir}$ and $M_\star(z=0)$ is the total stellar mass contained within the galaxy radius~(\S\ref{sec:analysis}). 
$m_{\textrm{dm}}$ and $m_\textrm{b}$ are the dark matter and initial gas particle masses.
The simulations use adaptive gravitational softening lengths for the gas but fixed softening lengths for the dark matter and stars. 
$\epsilon_\textrm{b}$ is the minimum force softening length for the gas and $\epsilon_{\star}$ and $\epsilon_{\textrm{dm}}$ are Plummer-equivalent gravitational softening lengths for stars and dark matter.
 $R_{\rm vir}$ and the softening lengths are in proper units.
The references in the final column are:
A: \cite{Hopkins2017},
B: \cite{Chan2018},
C: \cite{Wetzel2016},
D: \cite{Garrison-Kimmel2017a},
E: \cite{El-Badry2017},
F: \cite{Garrison-Kimmel2018},
G: Samuel et al., in prep.
\end{flushleft}
\label{table:simulations_used}
\end{table*}

\subsection{Particle Tracking Analysis}
\label{sec:analysis}

Our analysis builds on the work of \cite{Angles-Alcazar2017} but extends the particle-tracking analysis to the origins of the CGM, as opposed to galactic baryons. 
The code used for our analysis is a greatly-extended version of the code used in~\cite{Angles-Alcazar2017}.
Our code is written in Python, and has been updated to be parallelized through \textsc{Jug}~\citep{Coelho2017}.
The computation of many cosmological quantities is done using Colossus~\citep{Diemer2017}, and we use \textsc{unyt}~\citep{Goldbaum2018} to maintain consistent units

\subsubsection{Galaxy and CGM Definitions}
\label{sec:galdef}

Within each halo or subhalo we identify baryons belonging to that halo's primary galaxy as all gas and stars inside $R_{\rm gal} = 4R_{\star,0.5}$, where $R_{\star,0.5}$ is the stellar half-mass radius. 
We found that this radius provided a good compromise between including the vast majority of stars belonging to galaxies while minimizing artifacts arising from going too far out.
If a particle is inside $R_{\rm gal}$ of multiple galaxies, we associate that particle with the least massive galaxy.
By choosing the least massive galaxy, we preferentially associate particles with satellite galaxies.
This is important for satellites that may pass within $R_{\rm gal}$ of the main galaxy, either as part of a fly-by or in the process of merging. 
For a halo to contain a galaxy we require that it contains greater than 10 star particles.
We apply an additional cut on gas and only count gas with $n_{\rm H} > 0.13$ cm$^{-3}$ as part of the halo's central galaxy ISM, where $n_{\rm H}$ is the baryon number density.
This is similar to a widely-used density threshold to identify ISM gas in subgrid models used in large-volume cosmological simulations \citep[e.g.,][]{2003MNRAS.339..289S,Ford2014}. 
This method of associating particles with galaxies is different from that of \cite{Angles-Alcazar2017}, who used the galaxy finder~\textsc{SKID}~\citep{Stadel2001}. 
We switched from SKID to AHF because the computational cost of \textsc{SKID}, making it computationally prohibitive for the high-resolution FIRE-2 simulations with a large number of particles per halo.

To avoid increasing $R_{\star, 0.5}$ spuriously due to other galaxies in the halo or halo stars, $R_{\star,0.5}$ is evaluated as the stellar half-mass radius only for stars within $0.15 R_{\rm vir}$.
This alone is insufficient to prevent significant sudden increases in $R_{\star,0.5}$ during galaxy mergers and close encounters, so for our main galaxy we smooth $R_{\star,0.5}$ over $\approx 500$ Myr using a top-hat window.

The CGM of the main galaxy is defined as all gas with a radial distance from the main galaxy $ R_{\rm CGM, inner} < r < R_{\rm vir}$, where $R_{\rm CGM,~inner} \equiv \max( 1.2 R_{\rm gal}, 0.1 R_{\rm vir})$.
Our choice of $R_{\rm CGM, inner}$ is made such that when $ R_{\rm gal} > 0.1 R_{\rm vir}$ (which can happen at high redshift) we do not include in the CGM gas that is part of the galaxy.
In our analysis, wind  from the main galaxy is defined as all material that is currently inside the CGM, but was previously inside the main galaxy (see \S\ref{sec:classifications}).
To avoid spuriously classifying material that crossed $R_{\rm gal}$ due to small-scale fluctuations as wind, we ensure that the there is always a distance $\Delta_{\rm gap}$ between the edge of the galaxy and the inner boundary of the CGM. 
We define the ``galaxy-halo interface'' as all gas within $R_{\rm gal}$ with $n_{\rm H} < 0.13$ cm$^{-3}$ plus all gas with $R_{\rm gal} < r < R_{\rm CGM,inner}$, i.e. gas that is inside the halo but outside both the galaxy and the CGM. 
The galaxy-halo interface constitutes a small fraction of the total halo baryonic mass (a median $\sim\finterfacemedianlow$ at $z=0.25$), and quantifying its detailed properties is beyond the scope of this paper.

Throughout the paper we divide the simulations into mass bins based on their $z=0$ halo mass.
For brevity, we will refer to main halos that are the progenitors of halos with $M_{\rm h}(z=0) \sim 10^{10}, 10^{11}, 10^{12} M_\odot$ simply as  $10^{10} M_\odot$ progenitors,  $10^{11} M_\odot$ progenitors, and $10^{12} M_\odot$ progenitors.

\subsubsection{Particle Classification}
\label{sec:classifications}

\begin{figure}
\centering
\includegraphics[width=0.75\columnwidth]{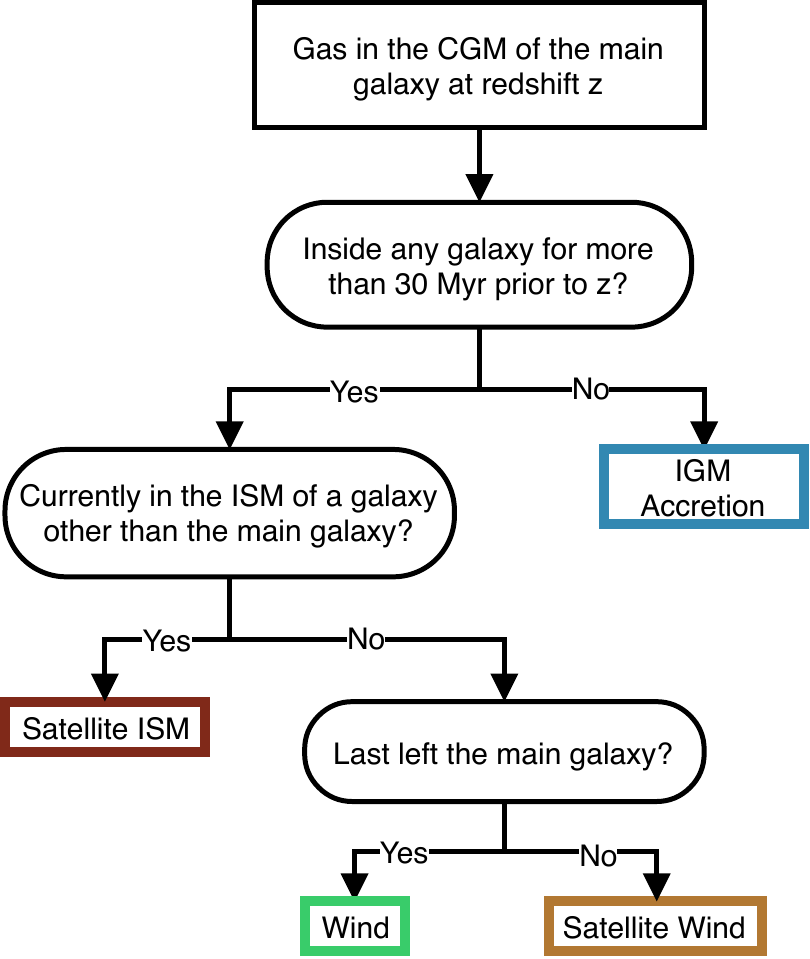}
\caption{
Flow chart summarizing how we classify the origin of gas elements in the CGM of a galaxy. 
\textit{IGM accretion} gas flows into the CGM from the intergalactic medium.
\textit{Winds} and \textit{satellite winds} are ejected from the main galaxy or another galaxy respectively.
\textit{Satellite ISM} is currently part of a satellite galaxy located inside the CGM.
}
\label{fig:CGM_origins}
\end{figure}

For each simulation in our sample, we analyze the CGM of that simulation's main halo and its origins at two main cosmic times: one at $z=0.25$ and one at $z=2$.
At low redshift, $z=0.25$ is a redshift representative of many observational studies of the CGM with the Hubble Space Telescope ~\citep[e.g.][]{Lehner2013, Tumlinson2013,Werk2014,Stocke2014,Prochaska2017,Chen2018}.
Meanwhile, $z=2$ is the peak of cosmic star formation, a critical point for understanding galaxy growth, and is a regime where we can probe the CGM with ground-based observatories~\citep[e.g.][]{2006ApJ...637..648S,Steidel2010,Rudie2012,OMeara2013}.
For each targeted redshift snapshot we identify all gas particles that have a radial distance from the center of the main galaxy of $(0.1-1) R_{\rm vir}$, and then randomly sample $10^5$ of those particles (in our simulations the CGM contains $\sim 10^6 - 10^7$ particles, so downsampling reduces memory usage while maintaining a statistically robust sample).
Note that when $R_{\rm gal} > 0.1 R_{\rm vir}$ particles within $R_{\rm gal}$ are not actually classified as part the CGM in our analysis, as explained in~\S\ref{sec:galdef}. 
Because the gas particles have approximately constant mass (there are some variations owing to mass loss from stellar evolution, which can increase the mass of gas neighbors), a random sample of particles is nearly equivalent to a mass-weighted sample.
For each particle we compile the full history of the particle's temperature, density, metallicity, position, velocity, particle type (i.e. star or gas), mass, and star formation rate at each snapshot of the simulation.
The time difference between snapshots is $\sim 10-25$ Myr.
We combine this information with the halo catalog produced by AHF to identify each particle's proximity to the main galaxy as well as the nearest galaxy other than the main galaxy.
In addition to analyzing snapshots at $z=0.25$ and $z=2$ for all simulations in our sample, we study the redshift evolution of the CGM by analyzing 25 additional snapshots from $z=0$ to $z=3$ for a subset of our simulations.
Specifically, we analyze in more detail simulations \texttt{m12i}, \texttt{m11q}, and \texttt{m10y} as representatives of $10^{12}$, $10^{11}$ and $10^{10} M_\odot$ progenitors respectively.
We repeat the same sampling and tracking procedure for each of these snapshots.

Figure~\ref{fig:CGM_origins} summarizes how we classify the origin of CGM gas in our analysis.
Consider a Lagrangian particle in the CGM of the main halo at a redshift $z$. 
We first consider whether or not this particle has spent more than a pre-processing time $t_{\rm pro}$ in any galaxy prior to $z$.
Following \cite{Angles-Alcazar2017}, if the particle has not spent $t_{\rm pro}$ in other galaxies, then its origin is \textit{IGM accretion}.
The pre-processing time $t_{\rm pro}$ is taken to be slightly larger than the maximum time interval between snapshots, i.e. $t_{\rm pro}=30$ Myr.
This takes advantage of the higher output frequency in our simulations compared to \cite{Angles-Alcazar2017}~(where $t_{\rm pro}$ was set to $100$ Myr).
Note that gas that has been in or is in the halo of another galaxy can still be classified as IGM accretion, provided it did not enter or spend significant time inside the galaxy itself.
While some very small halos exist below our resolution limit, star formation in these halos should be strongly suppressed by the UV background, so these halos should not process the gas. 
If the particle is processed, we determine whether or not it is currently, at $z$, inside another galaxy.
If it is inside another galaxy at $z$, then we classify it as belonging to \textit{satellite ISM}.
If not then we find whether the particle last left the main galaxy or a satellite galaxy.
If it last left the main galaxy the particle is classified as \textit{wind}.
If it last left a satellite galaxy the particle is classified as \textit{satellite wind}.
Note that gas classified as satellite wind could also have arrived in the CGM through tidal or ram pressure stripping from a satellite galaxy, or could have arrived in the CGM after being ejected by another galaxy not currently in the CGM and that may not become a satellite. 
We use the term satellite wind because the majority of this material arrives in the CGM after having been ejected from a satellite galaxy in a wind (see \S\ref{sec:particle_pathlines}).
The classifications described here are for a particular moment in time, and a gas particle can change classifications over time, e.g. gas that is first classified as IGM accretion before later being ejected in a galactic wind.

Throughout this work we focus on understanding the properties of gas according to its \emph{most recent} origin, as for many purposes this is most informative. 
For example, winds may return into the CGM gas that accreted onto the central galaxy as IGM accretion or satellite wind.
However, to understand the effects of this gas on CGM dynamics and enrichment it is important to recognize that this wind gas carries energy and metals recently output by the central galaxy.

\section{Results}

\subsection{Baryon and Metal Census}
\label{sec:mass_budgets}

In this section we summarize the mass content of the main halos in our simulations.
Unlike the rest of this work, the calculations done in this section use the simulation data only at $z=0.25$ and $z=2$, i.e. these are instantaneous measurements that do not employ particle tracking.
   
\begin{figure}
\includegraphics[width=\columnwidth]{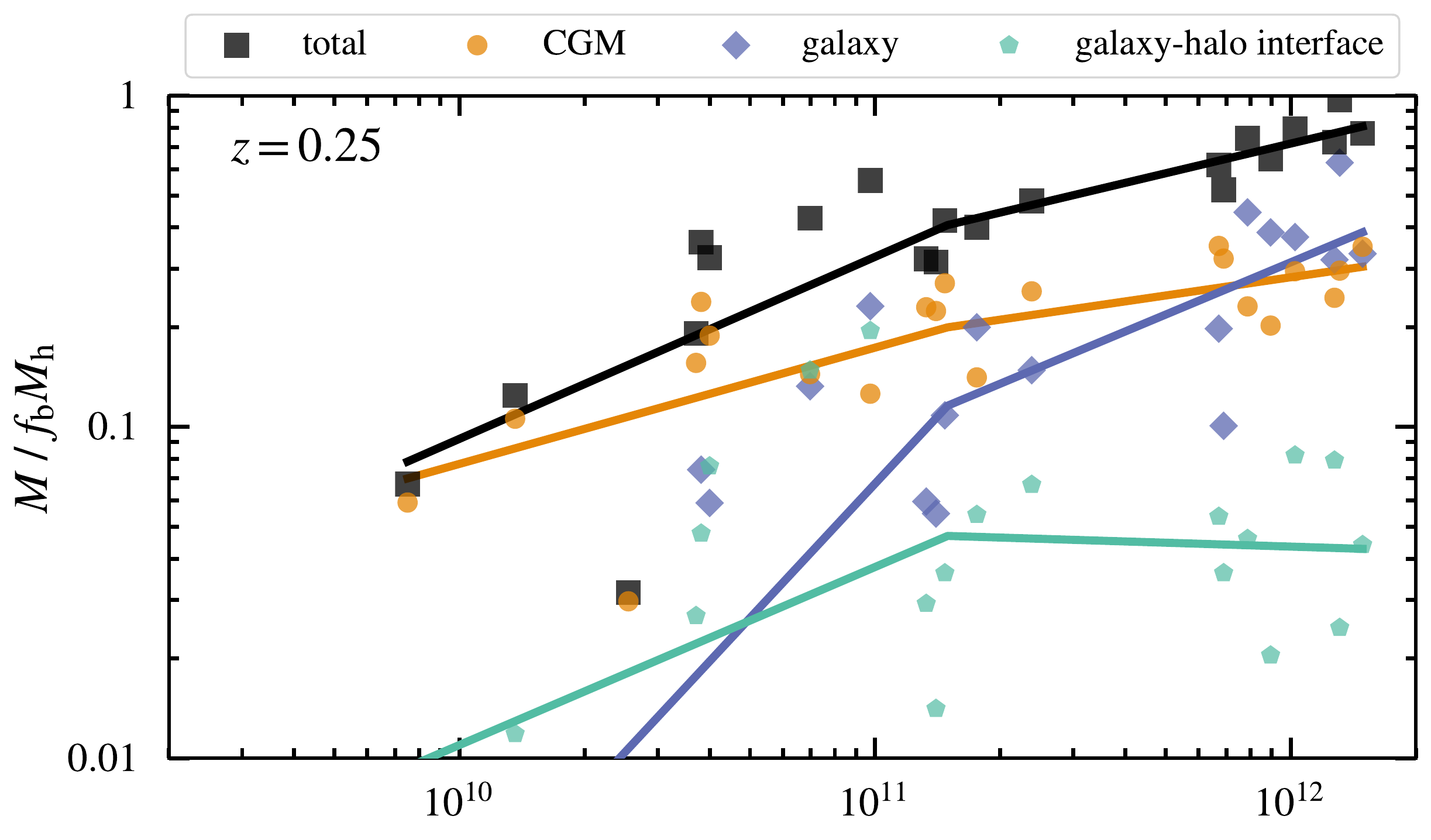}
\includegraphics[width=\columnwidth]{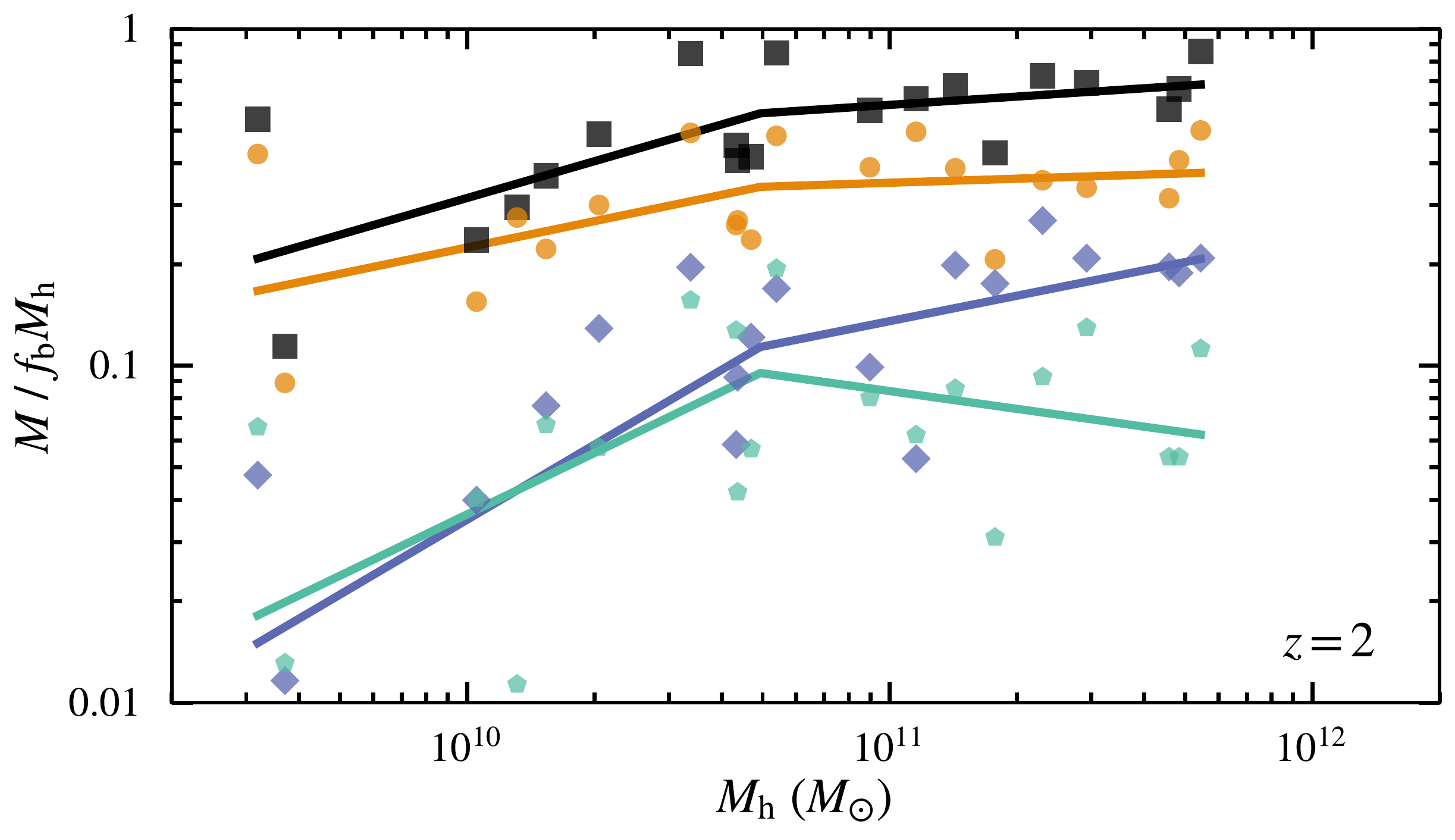}
\caption{
Mass composition of main halos in units of the halo's cosmological baryon budget,  $f_{\rm b} M_{\rm h}\equiv (\Omega_{\rm b}/\Omega_{\rm m} ) M_{\rm h}$, broken down into different categories at $z=0.25$ (top) and $z=2$ (bottom).
Black squares and purple diamonds are the total baryonic masses (including gas and stars) in the halo and the main galaxy respectively.
Orange circles  and turquoise pentagons are the gas masses of the CGM and the galaxy-halo interface respectively. 
The lines are meant to guide the eye, and connect the medians of each mass bin.
The x-axis shows the halo mass of the main halo in each zoom-in simulations.
Due to strong feedback especially effective in lower mass galaxies, our halos are consistently below the baryon budget, except for $10^{12} M_\odot$ progenitors, which have a baryon mass on average $\gtrsim 80\%$ of the baryon budget. 
The CGM provides $\gtrsim \fCGMMWmeanlow\%$ of the mass that remains in the halo across our entire halo mass range. 
}
\label{fig:mass_budget}
\end{figure}

Figure~\ref{fig:mass_budget} shows the total baryonic mass contained inside the main halo ($<R_{\rm vir}$) for each simulation in our sample at $z=0.25$ and $z=2$. 
The mass is displayed in units of the cosmological baryon budget, $f_{\rm b} M_{\rm h} \equiv ( \Omega_{\rm b} / \Omega_{\rm m} ) M_{\rm h}$.
The mass is divided up into four categories, according to our definitions of galaxies and the CGM outlined in \S\ref{sec:analysis}.
The total baryonic masses of the halo are the total masses of all gas and stars inside $R_{\rm vir}$. 

More massive halos retain a greater fraction of their baryons. 
The progenitors of $10^{(10,11,12)} M_\odot$ halos have a mean total halo baryon mass $M_{\rm baryon} \sim(\fbaryondwarfmeanlow,\fbaryonsubMWmeanlow,\fbaryonMWmeanlow) f_{\rm b} M_{\rm h}$ at $z=0.25$ and  $\sim(\fbaryondwarfmeanhigh,\fbaryonsubMWmeanhigh,\fbaryonMWmeanhigh)f_{\rm b} M_{\rm h}$ at $z=2$.
The baryonic mass of $10^{10} M_\odot$ halos is well below the universal baryon fraction because their potential can be too shallow to accrete photo-heated gas effectively~\citep[e.g.][]{Thoul1996, Gnedin2000, Okamoto2008, Faucher-Giguere2011,El-Badry2017} and to hold onto any wind, which is expelled from these halos more efficiently than in higher mass halos~\citep[e.g.][]{Muratov2015}.
The baryon budgets also suggest that winds can expel some gas from the halos of $10^{11} M_\odot$ and $10^{12} M_\odot$ progenitors.
Note that we do not display simulation \texttt{m10v} in any of our $z\sim2$ results because it forms late and has a very small stellar mass at $z>1$ ($M_\star (z=2) < 10^4 M_\odot$).

At $z=0.25$ the fractional CGM contribution to the baryon mass increases with decreasing halo mass: from $\sim$\fCGMMWmeanlow\% of the halo's baryon mass in $10^{12}$ M$_\odot$ progenitors to $\sim$\fCGMdwarfmeanlow\% in $10^{10}$ M$_\odot$ progenitors.
At $z=2$ the CGM provides most of the baryonic mass in the halo, contributing at minimum $\sim$\fCGMMWmeanhigh\% of the halo's baryon mass for $10^{12} M_\odot$ progenitors.
Galaxies contribute most of the remaining mass to the halo, and the contribution to the halo's baryon mass by the galaxy-halo interface is small (at most $\sim\finterfacedwarfmeanlow\%$ in the case of $10^{10} M_\odot$ progenitors at $z=0.25$).

\begin{figure}
\includegraphics[width=\columnwidth]{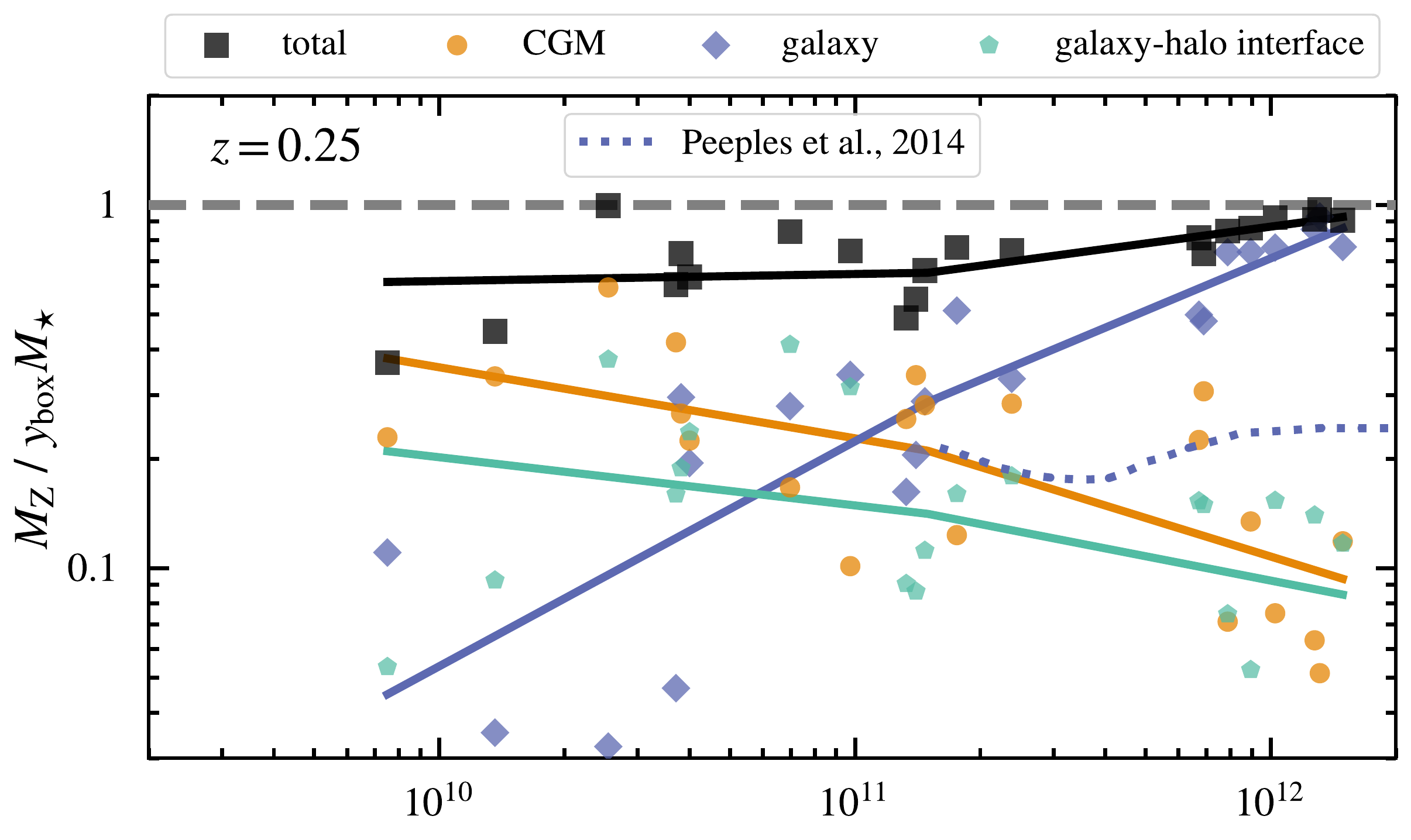}
\includegraphics[width=\columnwidth]{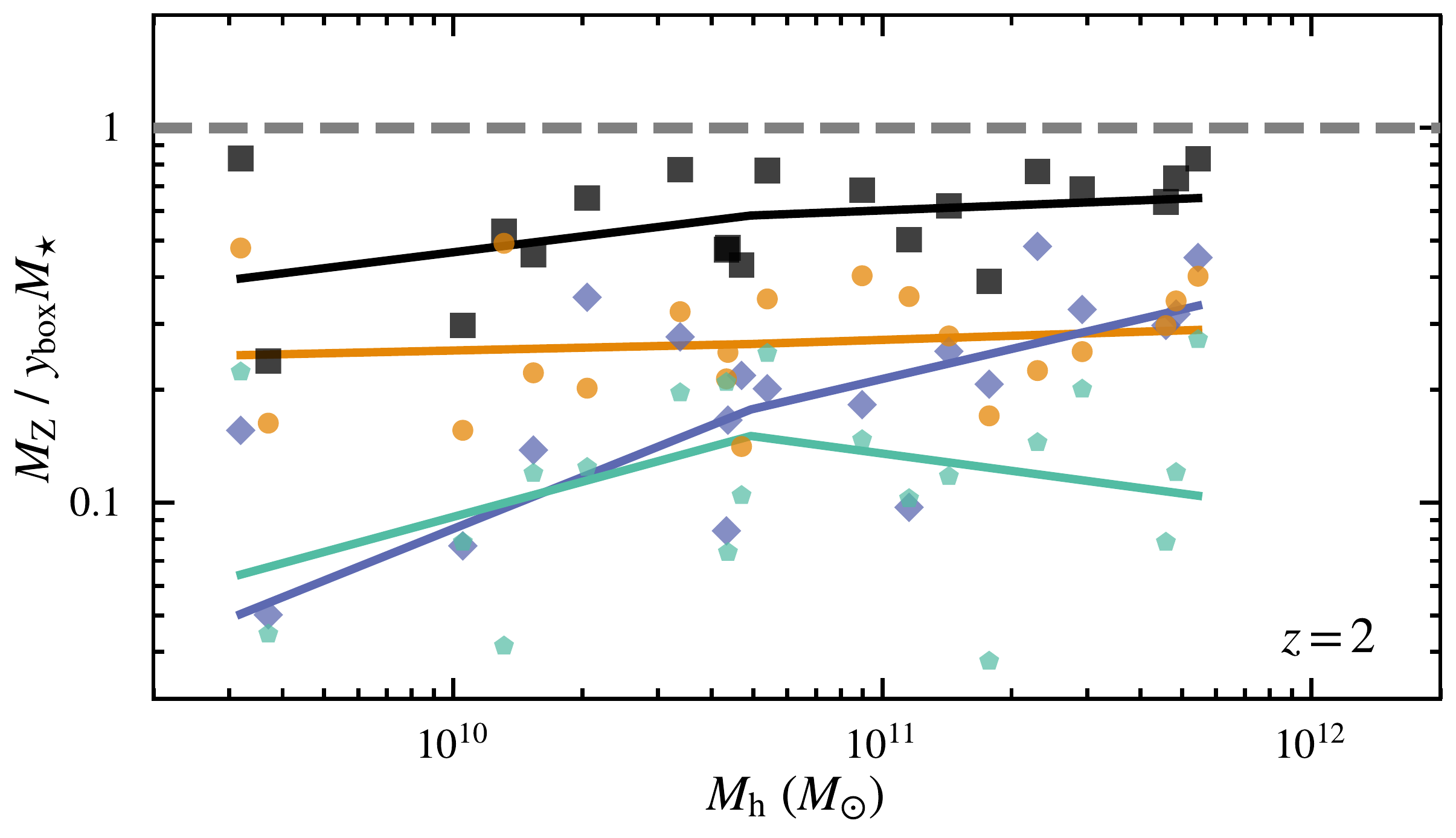}
\caption{
Similar to Figure~\ref{fig:mass_budget}, but for metals in the simulations, in units of the available metal budget $y_{\rm box} M_\star$, where the estimated yield is $y_{\rm box} \equiv M_{\rm Z, box}/M_{\rm \star, box}$, i.e. the total mass of metals in the simulation volume divided by the total stellar mass in the simulation volume.
The dotted purple line is an estimate of the galaxy metal mass by \protect\cite{Peeples2014}, in units of their estimate of the available metal budget (see discussion in \S\ref{sec:discussion_metal_content}).
At low halo masses and at $z=2$, where winds are highly mass loaded, most of the metals in our halos are in the CGM.
}
\label{fig:metal_mass_budget}
\end{figure}

Figure~\ref{fig:metal_mass_budget} is similar to Figure~\ref{fig:mass_budget}, but for the metal mass contained inside the halo, in units of our estimated metal budget, $y_{\rm box} M_\star$.
Here $y_{\rm box} \equiv (M_{\rm Z, box} - Z_{\rm floor} M_{\rm box} )/ M_{\star, \rm box}$ is the empirical yield of the simulation,
where $M_{\rm Z,box}$ is the total metal mass in the simulation volume (from both gas and stars),
$Z_{\rm floor} = 10^{-4}Z_\odot$ is the metallicity floor of our simulations,
$M_{\rm box}$ is the total baryonic mass in the simulation volume,
and $M_{\rm\star,box}$ is the total stellar mass in the simulation volume.
In other words, the metal budget is simply the metal mass produced by stars in the central galaxy assuming the yield for the central galaxy is the same as the yield for the simulation. 
The empirical yield of the simulation is consistent between different simulations:
the mean and standard deviation of $y_{\rm box}$ across the simulations is $(\yboxlow\pm\yboxstdlow,\yboxhigh\pm\yboxstdhigh) $ at $z=(0.25, 2)$.
To account for contribution from the metallicity floor, when we plot the metal mass in Figure~\ref{fig:metal_mass_budget}, we actually plot $M_{\rm Z} - Z_{\rm floor} M_{\rm baryon}$

In a few situations the main galaxy is about to undergo a major merger with another galaxy. 
In that case, the CGM of the two galaxies can overlap before the galaxies merge, causing the total gas mass inside $R_{\rm vir}$ to increase by order unity while the stellar mass of the main central remains unchanged. 
To avoid a CGM metal mass exceeding the metal budget estimated from the central galaxy, we use the total stellar mass contained in $R_{\rm vir}$ when calculating the metal budget for the CGM and the total halo metal mass.
For the galaxy and galaxy-halo interface metal budget we only use the stellar mass of the main galaxy.
In most cases, the total stellar mass and the galaxy stellar mass are very similar and this has no significant effect.

The progenitors of $10^{(10,11,12)} M_\odot$ halos have a mean total halo metal mass (i.e., the total metal mass retained inside $R_{\rm vir}$) $M_{\rm Z} \sim(\fmetaldwarfmeanlow,\fmetalsubMWmeanlow,\fmetalMWmeanlow) ~ y_{\rm box} M_\star$ at $z=0.25$ and  $\sim(\fmetaldwarfmeanhigh,\fmetalsubMWmeanhigh,\fmetalMWmeanhigh) ~ y_{\rm box} M_\star$ at $z=2$. 
The halo metal mass is a larger fraction of its metal budget, $y_{\rm box} M_\star$, than is the halo baryon mass relative to the cosmic baryon budget. 
This suggests that winds, which carry much of the metals, preferentially remain in the halo. 
As discussed by \cite{Muratov2016}, this is because galactic winds sweep up halo gas (most of which is contributed by IGM accretion), preferentially ejecting relatively metal-poor gas from halos. 
Further, metal-enriched winds cool more rapidly, which may increase the recycling of winds onto galaxies. 

Compared to the baryon mass, the CGM typically contains a smaller fraction of the halo metal mass: as little as $\sim$\fmetCGMMWmeanlow\% of the halo metal mass (again for $10^{12} M_\odot$ progenitors at $z=0.25$), but up to $\sim$\fmetCGMdwarfmeanhigh\% for $10^{10} M_\odot$ progenitors at $z=2$.
This is explained by the fact that a large fraction of the CGM mass comes from metal-poor IGM accretion (\S\ref{sec:origin_by_mass}), while metals are produced in galaxies and then potentially expelled as winds.
The contribution of the galaxy-halo interface to the halo metal mass is on average higher than its contribution to the halo baryon mass, and can contribute up to $\sim$\fmetinterfacedwarfmeanhigh\% in the case of  $10^{10} M_\odot$ progenitors at $z=2$.
This is because the metallicity of the galaxy-halo interface is higher than the metallicity of the CGM, having been enriched by the metals in the adjacent galaxy.

Our total halo metal masses versus stellar mass agree well with the results from \cite{Ma2015}, who studied the metal budget for simulations run with the FIRE-1 version of the FIRE code (using a P-SPH hydrodynamic solver instead of the MFM solver), similar to \cite{Muratov2016}. 
However the CGM metallicities in FIRE-2 simulations can nevertheless differ significantly from those of FIRE-1 simulations.
We discuss this in Appendix~\ref{sec:supplementary_material}. 
In \S\ref{sec:discussion} we compare our halo and galaxy metal masses to those from~\cite{Peeples2014} (purple dashed line in the top panel of Figure~\ref{fig:metal_mass_budget}) and \cite{Christensen2018}.

The simulations in our sample that use a subgrid prescription for turbulent metal diffusion are simulations with $10^{11}$ or $10^{12} M_\odot$ progenitor main halos. 
The halo baryon and total metal masses for simulations with and without metal diffusion are consistent with one another (Appendix~\ref{sec:supplementary_material}).
This is also the case for most quantities analyzed in this paper, with the exception of quantities directly dependent on metal diffusion (e.g., the metallicity distributions in \S\ref{sec:metallicity}).
As such, unless we explicitly state otherwise, we simultaneously analyze results from both simulations with and without metal diffusion. 
Note, however, that there may be more significant changes owing to subgrid metal diffusion for more massive halos or for other halo properties not studied here.

\subsection{Pathlines for Different CGM Origins}
\label{sec:particle_pathlines}

\begin{figure*}
\centering
\includegraphics[height=0.29\textheight]{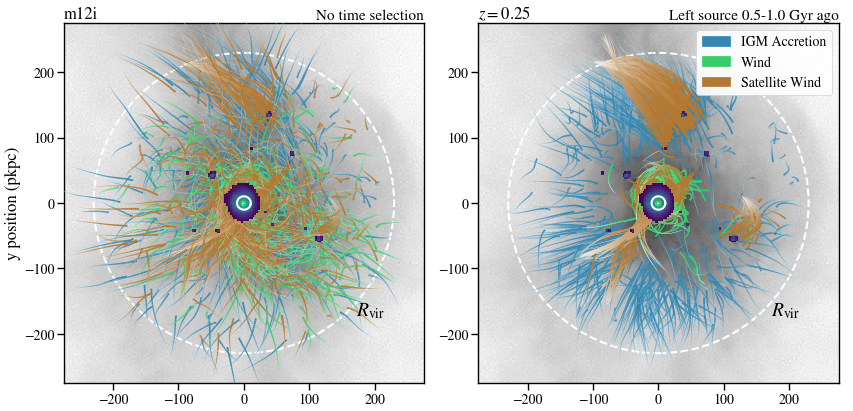}
\includegraphics[height=0.29\textheight]{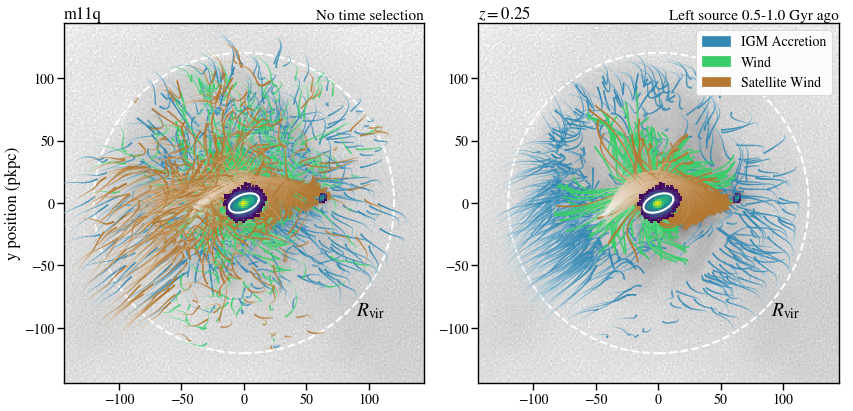}
\includegraphics[height=0.305\textheight]{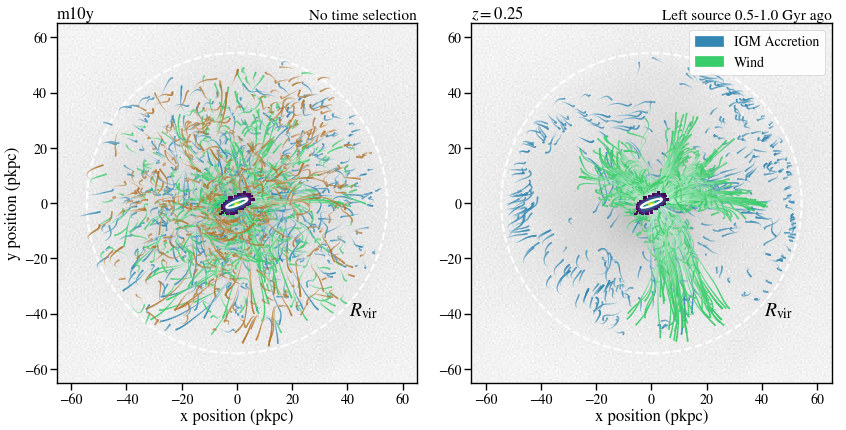}
\caption{
Paths traced over the course of 1 Gyr by 500 randomly-selected particles (per origin) found in the CGM of \texttt{m12i} (top), \texttt{m11q} (center), and \texttt{m10y} (bottom) at $z=0.25$.
The color of the path is chosen based upon its classification at $z=0.25$.
The line becomes thinner and whiter the farther back in time.
The stellar mass surface density  is plotted as a viridis (blue-green-yellow) histogram. 
\textbf{Left:} Trajectories for a random CGM particle sample, with no filtering based on time. 
Trajectories of particles from different origins overlap in complex patterns, producing well-mixed halos. 
\textbf{Right:} Trajectories for CGM particles filtered to select particles that left their source 0.5-1 Gyr ago. 
These trajectories reveal the more coherent recent behavior of gas elements from different origins.
They also show the effect of winds and/or photoheating on gas accreting from the IGM, particularly in \texttt{m10y}.
}
\label{fig:pathlines_CGM_snum465}
\end{figure*}

\begin{figure*}
\centering
\includegraphics[height=0.29\textheight]{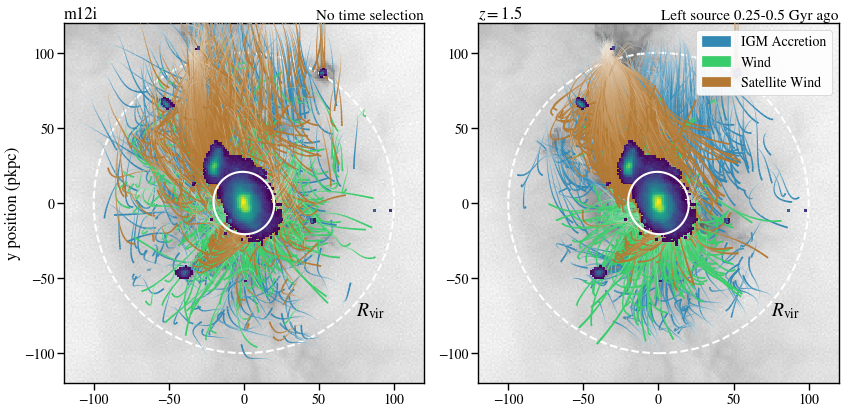}
\includegraphics[height=0.29\textheight]{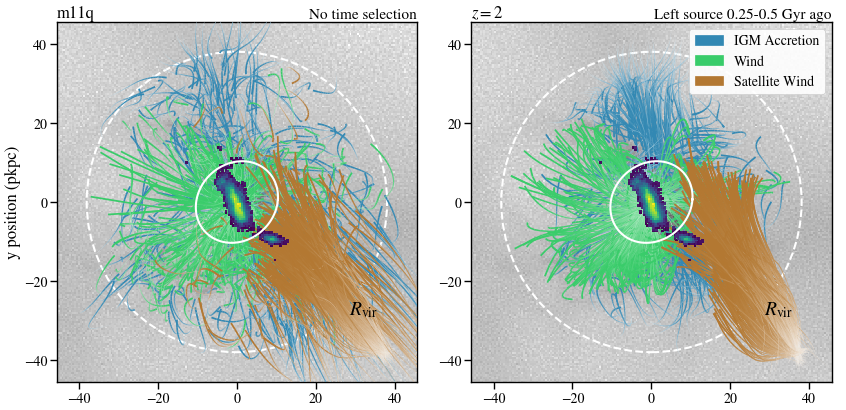}
\includegraphics[height=0.305\textheight]{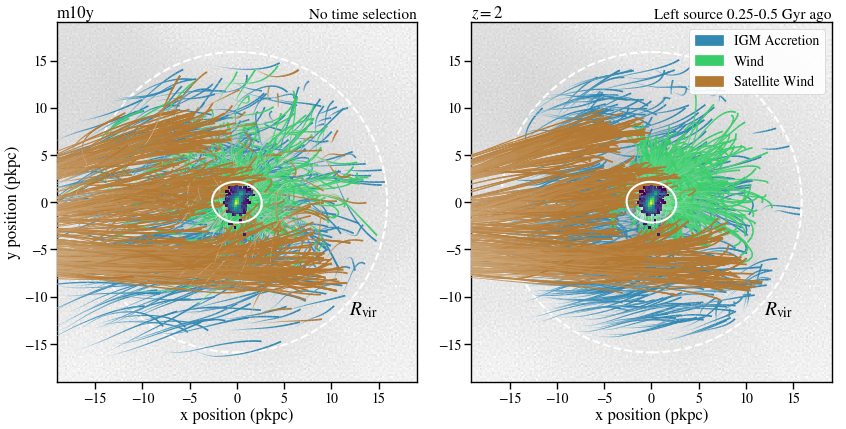}
\caption{
Same as Figure~\ref{fig:pathlines_CGM_snum465}, but for $z\sim2$ (for \texttt{m12i}, we show results at $z=1.5$ to avoid a major merger taking place at $z=2$).
To account for shorter halo dynamical times at this cosmic time, particle trajectories are plotted over the course of 0.5 Gyr, as opposed to 1 Gyr as in Figure~\ref{fig:pathlines_CGM_snum465}. 
Relative to $z\sim 0.25$, the full CGM at $z=2$ is better described by the properties of gas which left its source in the recent past (250-500 Myr before). 
This reflects the more dynamic nature of the high-redshift CGM, which has not yet accumulated a well-mixed distribution of halo gas.
}
\label{fig:pathlines_CGM_snum172}
\end{figure*}

\begin{figure}
\includegraphics[width=\columnwidth]{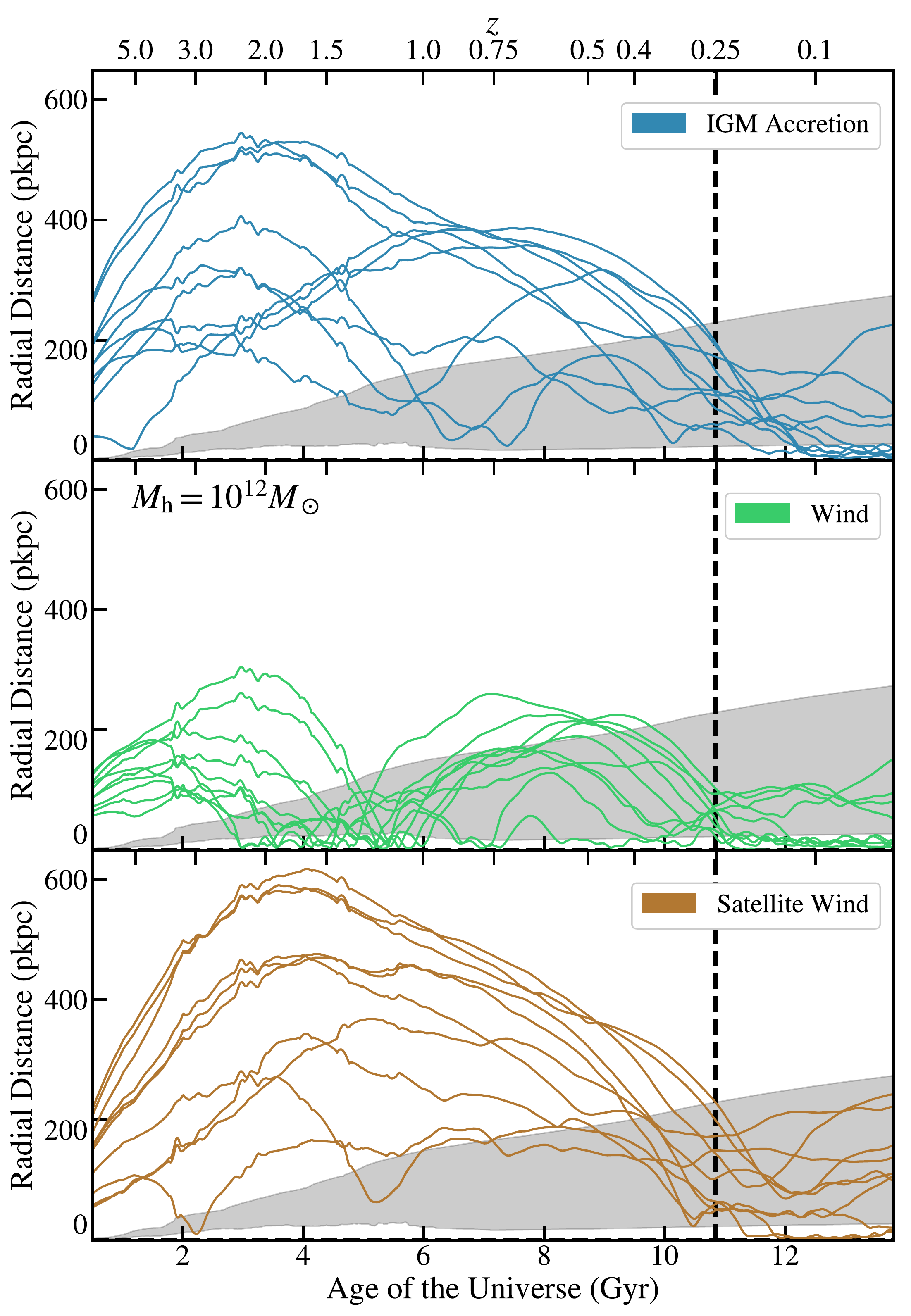}
\caption{
Radial distance (in proper kpc) vs. time over the course of the simulation for particles found in the CGM of \texttt{m12i} at $z=0.25$  (the vertical dashed line).
Ten randomly selected particles are shown for each classification at $z=0.25$. 
IGM accretion is shown in the top panel, winds from central galaxies in the middle panel, and satellite wind particles in the bottom panel.
The shaded grey region shows the CGM (the white band underneath indicates the extent of the galaxy+galaxy-halo interface).
Each origin has characteristic dynamics: 
IGM accretion accretes onto the halo as a broad flow over a wide range of times,
wind from the central galaxy is regularly recycled on time scales ranging from tens of Myrs to Gyrs, and the paths followed by satellite wind are similar to the paths taken by IGM accretion but with stronger correlation among particles.
}
\label{fig:r_vs_time_m12i_CGM_snum465}
\end{figure}

\begin{figure}
\includegraphics[width=\columnwidth]{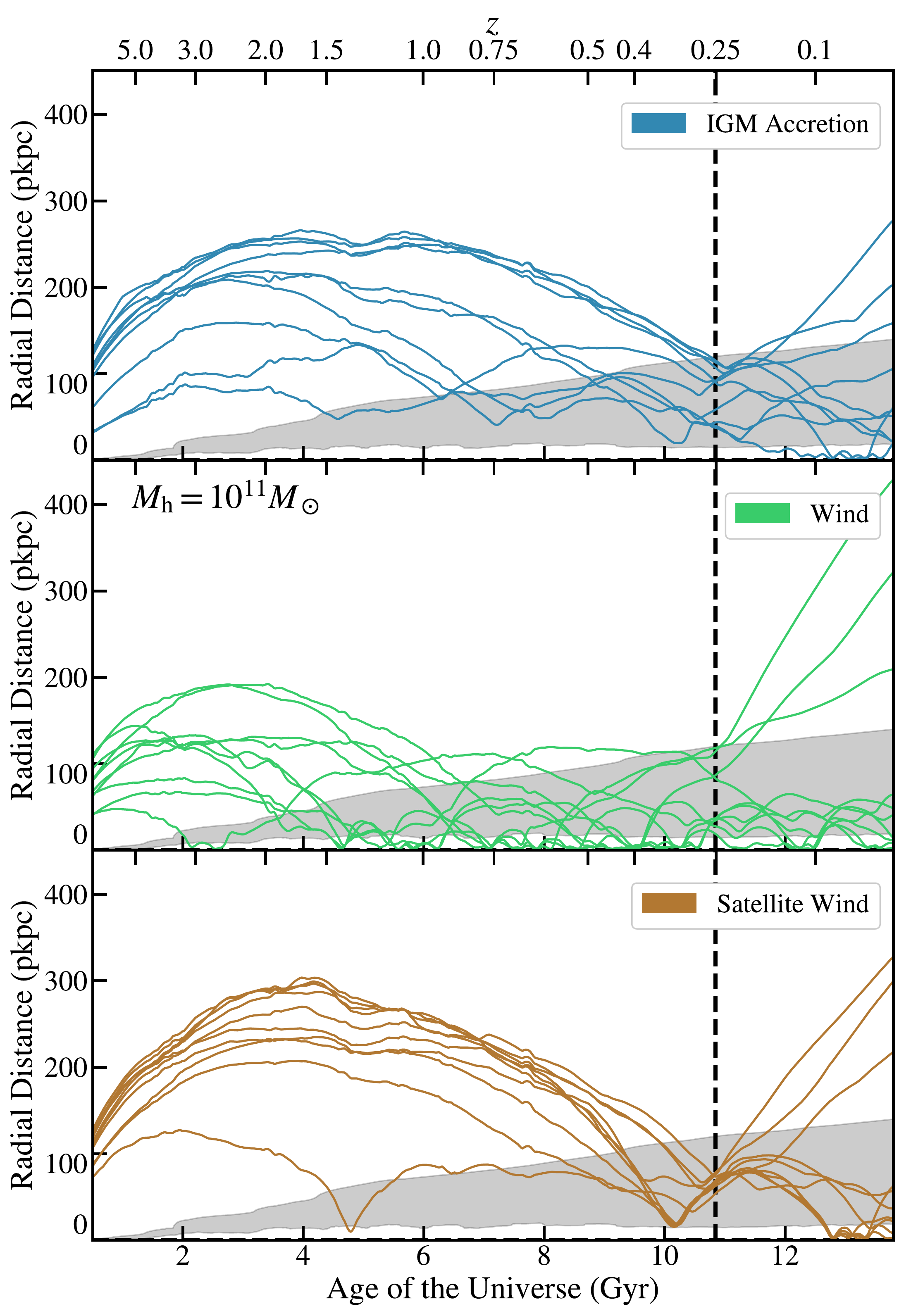}
\caption{
Same as Figure~\ref{fig:r_vs_time_m12i_CGM_snum465}, but for \texttt{m11q}.
As illustrated in Figure \ref{fig:pathlines_CGM_snum465}, winds from the central galaxy can prevent IGM accretion and satellite wind from accreting onto the central galaxy (preventative feedback). 
Gas of all origins can remain in the CGM for multiple Gyrs.
}
\label{fig:r_vs_time_m11q_CGM_snum465}
\end{figure}

\begin{figure}
\includegraphics[width=\columnwidth]{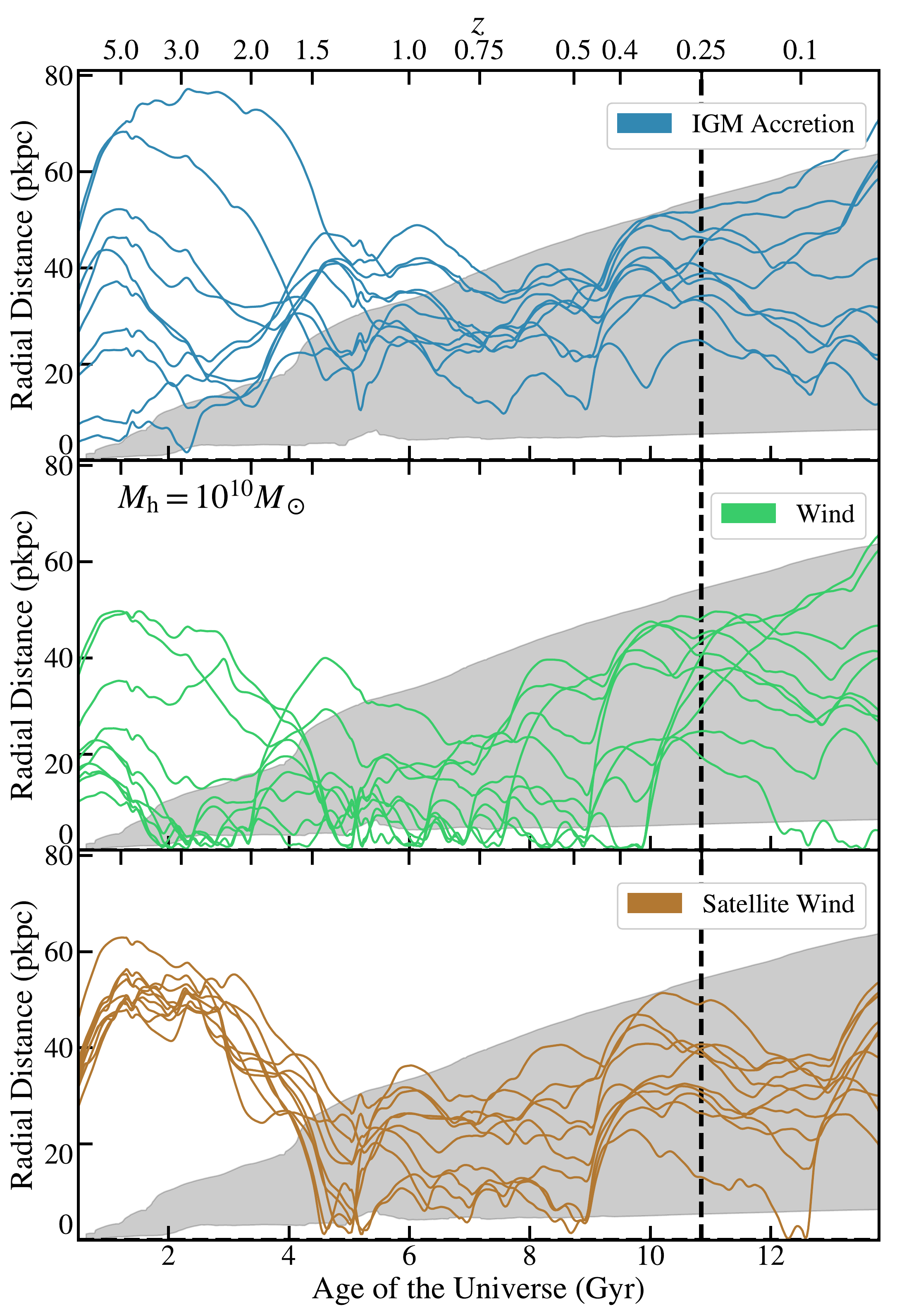}
\caption{
Same as Figure~\ref{fig:r_vs_time_m12i_CGM_snum465}, but for \texttt{m10y}.
Much of the gas in the CGM of $10^{10} M_\odot$ progenitors accretes into the halo early and is loosely bound in the halo, hovering at a broad range of distances from the central galaxy. 
This is most likely owing to photoheating of halo gas by the cosmic UV background and wind from the central galaxy.
}
\label{fig:r_vs_time_m10y_CGM_snum465}
\end{figure}

In this section we analyze the trajectories for different particles in the CGM of our simulations, according to their origin as classified in \S\ref{sec:classifications}. 
We do this in two ways:
by considering their projected trajectory relative to a galaxy and its environment over a limited period of time, and also by considering pathlines of particles over the full history of the simulations.
In addition, we use the 3D interactive visualization tool \textsc{Firefly}\footnote{See the Firefly homepage at \url{https://galaxies.northwestern.edu/firefly}.
Interactive visualizations of the datasets presented in this paper can be accessed on the web at \url{zhafen.github.io/CGM-origins} and \url{zhafen.github.io/CGM-origins-pathlines}.
\label{foot:firefly}
}~\citep{Geller2018} to gain more insight into the full pathlines and distribution of CGM gas in the simulated halos.

The left columns of Figures~\ref{fig:pathlines_CGM_snum465} and~\ref{fig:pathlines_CGM_snum172} shows the paths traced by 500 gas particles per classification for 1 Gyr prior to $z=0.25$ and 0.5 Gyr prior to $z\sim2$ respectively, for three representative simulations corresponding to different mass bins.
These simulations were chosen at random, and other simulations show similar behavior.
We plot the pathlines of \texttt{m12i} at $z=1.5$ instead of $z=2$ because \texttt{m12i} is undergoing a major merger at $z=2$ which is not representative of most times and which introduces some artifacts in our analysis.
The virial radius of the main galaxy is plotted as a dashed white circle, and a circle with radius $R_{\rm gal}$ is centered on the main galaxies. The circle is rotated such that the normal is parallel to the total angular momentum of the galaxy's stars, indicating the orientation of the galactic disk.
The location of stars is plotted as a histogram using a log-scale viridis (i.e. blue-yellow-green) colormap.
The gas density distribution at $z=0.25$ or $z=2$ is shown as a gray background (logarithmic stretch). 

As can be seen from Figure~\ref{fig:pathlines_CGM_snum465}, at $z=0.25$ most of the volume of the three halos shown is filled with particle trajectories that overlap in complex patterns.
These particles come from different origins, and are mixed together relatively well.
These particles are part of the CGM's diffuse halo gas, i.e. they are not part of any substructure or satellite galaxy, which is $\gtrsim 60\%$ of the CGM by mass by $z=0.25$ for all analyzed halos (Appendix \ref{sec:mass_by_phase}).

The right columns of Figures~\ref{fig:pathlines_CGM_snum465} and~\ref{fig:pathlines_CGM_snum172} are similar to the left column, but we select our sample from particles that have accreted onto the CGM (in the case of IGM accretion) or left a galaxy (in the case of wind or satellite wind) between 0.5 and 1 Gyr ago (in the case of Figure~\ref{fig:pathlines_CGM_snum465}) or 0.25 and 0.5 Gyr ago (in the case of Figure~\ref{fig:pathlines_CGM_snum172}). 
This excludes gas accreted onto the CGM more than 1 Gyr or 0.5 Gyr ago, respectively.
For the $M_{\rm h} \sim 10^{12}$ M$_{\odot}$ halo, this has a similar effect as excluding most hot virialized gas.
Overall, this this selection reveals more coherent gas flows before they have time to become well-mixed with the CGM.

For all the simulations shown, recent IGM accretion is highly anisotropic at $z \approx 2$, consistent with the expected filamentary accretion at high redshift \citep[e.g.,][]{Keres2005, Dekel2006, 2011MNRAS.412L.118F}. 
By $z=0.25$, IGM accretion remains significantly anisotropic but overall much less so.
In the case of \texttt{m12i} at $z=1.5$, the dominant direction of IGM accretion is associated with other infalling structures, as can be seen by the co-aligned gas overdensity and satellite wind.
This is not surprising, because filamentary structures that build halos bring in both IGM accretion and satellite galaxies.
For \texttt{m10y} at $z=0.25$, the location of the gas changes only slightly over the course of 1 Gyr, as can be seen by the very short length of the pathlines.
This is because the potentials of the $10^{10}M_\odot$ halos are shallow enough that photo-heated gas in the IGM does not accrete strongly onto the halo.

To complement the 2D projections, Figures~\ref{fig:r_vs_time_m12i_CGM_snum465}, \ref{fig:r_vs_time_m11q_CGM_snum465}, and \ref{fig:r_vs_time_m10y_CGM_snum465} show the radial distance from the main galaxy of \texttt{m12i}, \texttt{m11q}, \texttt{m10y}, respectively, as a function of time for 10 randomly-selected particles per classification found in the CGM at $z=0.25$.
For these plots we do not show the paths traced by particles that are classified as satellite ISM, because such gas makes up a small fraction of the total halo gas mass.
These figures show that particles initially follow the cosmological expansion (moving radially outward), before reaching a turnaround radius.
This is reminiscent of the spherical collapse of halo formation \citep[][]{1972ApJ...176....1G}. 
For \texttt{m12i}, IGM accretion can fall in from over 500 proper kpc away; this is a measure of the Lagrangian volume of the Universe contributing to the assembly of the halo.
We now highlight some characteristics of the different modes of CGM assembly illustrated by particle trajectories.

As expected, satellite wind is associated with galaxies other than the main central galaxies. 
Note that the renderings in Figures~\ref{fig:pathlines_CGM_snum465} and \ref{fig:pathlines_CGM_snum172} show the location of stars at the current (target) redshift but the particle trajectories show past behavior, and therefore in the right column the white parts of the particle trajectories converge where the galaxy was located in the past. 
This is why the points of origin of satellite wind do not in general coincide with the location of stellar clumps. 
Interestingly, the images show that in many cases (especially at $z\approx 2$) satellite wind gas was actually ejected before the source entered the main halo, i.e. the source galaxy only later becomes a satellite of the main central galaxy. 
An extreme case is \texttt{m10y} at $z=2$, where the majority of the satellite wind in the halo was forcibly ejected from the halo of another galaxy $\sim 50$ kpc away.
On the other hand, gas removed from other galaxies primarily through stripping appears to be less common, thus the name ``satellite wind.''
Also as expected, the radial distance plots in Figures~\ref{fig:r_vs_time_m12i_CGM_snum465} and~\ref{fig:r_vs_time_m11q_CGM_snum465} show that the histories of particles are more highly correlated in the case of satellite wind compared to IGM accretion (which can originate from directions other than from other galaxies). 

Wind from the main galaxy often undergoes frequent recycling.
Some of these recycling periods are short, $\lesssim 100$ Myr, while others can be as long as 10 Gyrs (see \citealt{Angles-Alcazar2017}).
Once out of the galaxy, wind can remain in the halo for billions of years at a time.
This occurs for other modes of CGM growth as well.
As can be seen in Figures~\ref{fig:r_vs_time_m12i_CGM_snum465} and~\ref{fig:r_vs_time_m11q_CGM_snum465}, some of the IGM accretion gas in the CGM of \texttt{m12i} and \texttt{m11q} at $z=0.25$ was in the CGM previously, but was pushed out by winds before eventually reentering $R_{\rm vir}$.
This is most clearly seen for gas particles that start at a radii $r_0$ from the central galaxy, begin to accrete, and instead turn around and reach radii $r > r_0$, suggesting energy injection.
This ``preventative feedback'' is especially visible in the visualizations of \texttt{m11q} near $z=0.25$ because the halo is undergoing a strong outflow event at that redshift.
In this case, wind from the central galaxy is pushing out and reversing some of the recent IGM accretion, and may also interact significantly with wind from the satellite. 
Wind is also found on average closer to the central galaxy compared to the other origins, as discussed in \S\ref{sec:origin_by_radius}.
As found in previous analyses of the FIRE simulations ~\citep[e.g.][]{Muratov2015,Faucher-Giguere2015}, strong outflows can suppress infall onto galaxies and cause temporary decreases in the star formation rate.

\subsection{Origins of the CGM}
\label{sec:origin_by_mass}

In this section we analyze how gas from different origins contributes to the CGM as a function of halo mass (\S\ref{sec:origin_by_halo_mass}), time (\S\ref{sec:origin_by_time}), radius (\S\ref{sec:origin_by_radius}), and polar angle relative to the stellar disk (\S\ref{sec:origin_by_angle}) .

\subsubsection{Origin by Halo Mass}
\label{sec:origin_by_halo_mass}
\begin{figure}
\includegraphics[width=\columnwidth]{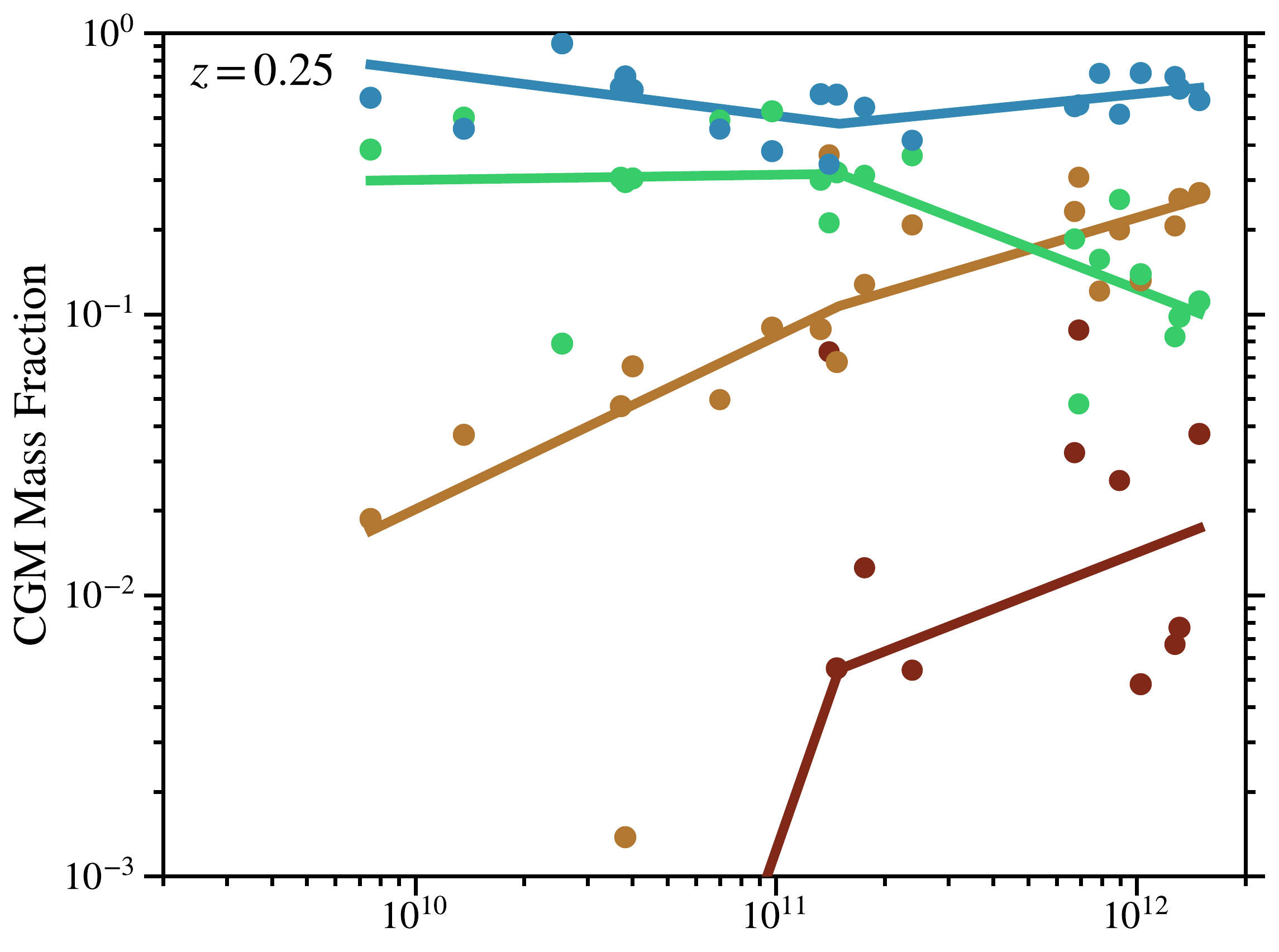}
\includegraphics[width=\columnwidth]{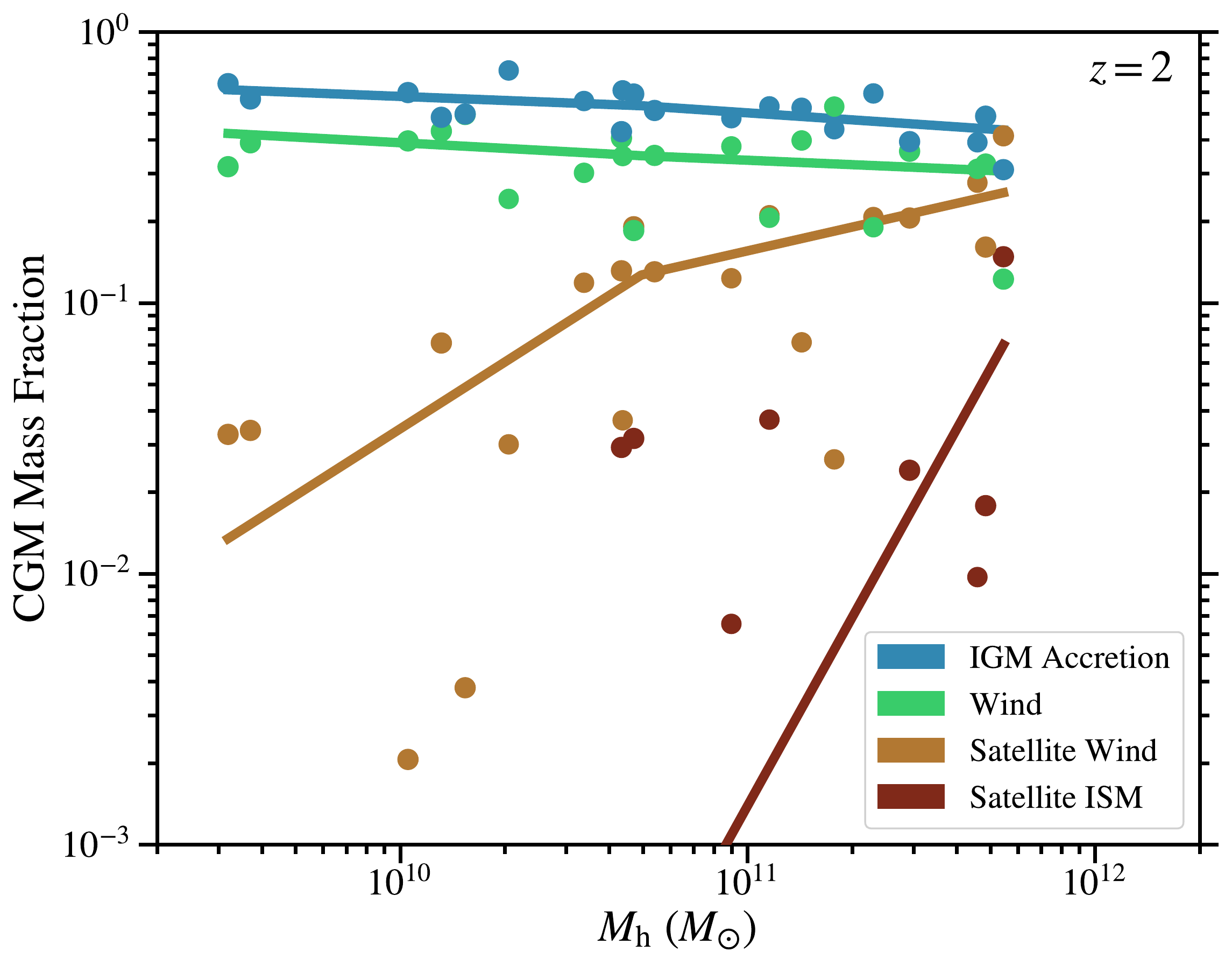}
\caption{
\textbf{Top:}
Fraction of the total CGM mass contributed by different origins at $z=0.25$.
Each point is a value from a simulation, and the color represents origin.
The lines are meant to guide the eye, and connect the medians for each mass bin.
IGM accretion is on average the dominant component at all halo masses.
Wind from central galaxies is generally the second most important component by mass, except at low redshift in $M_{\rm h}\sim 10^{12} M_{\odot}$ halos, for which winds from satellite galaxies can be more important. 
\textbf{Bottom:}
Same but at $z=2$.
The same general trends remain as at $z=0.25$, except that wind from central galaxies contributes more than wind from satellites at all halo masses considered. 
This is consistent with $M_{\rm h}\sim 10^{12} M_{\odot}$  progenitors driving strong winds at high redshift, but much weaker winds as they settle into well-ordered disks at low redshift \citep[e.g.][]{Muratov2015}. 
The increasing importance of satellite winds with increasing halo mass is consistent with more galaxy mass contributed by intergalactic transfer at higher halo mass~\citep[][]{Angles-Alcazar2017}.
}
\label{fig:CGM_mass_frac_vs_Mh_CGM}
\end{figure}

\begin{figure}
\includegraphics[width=\columnwidth]{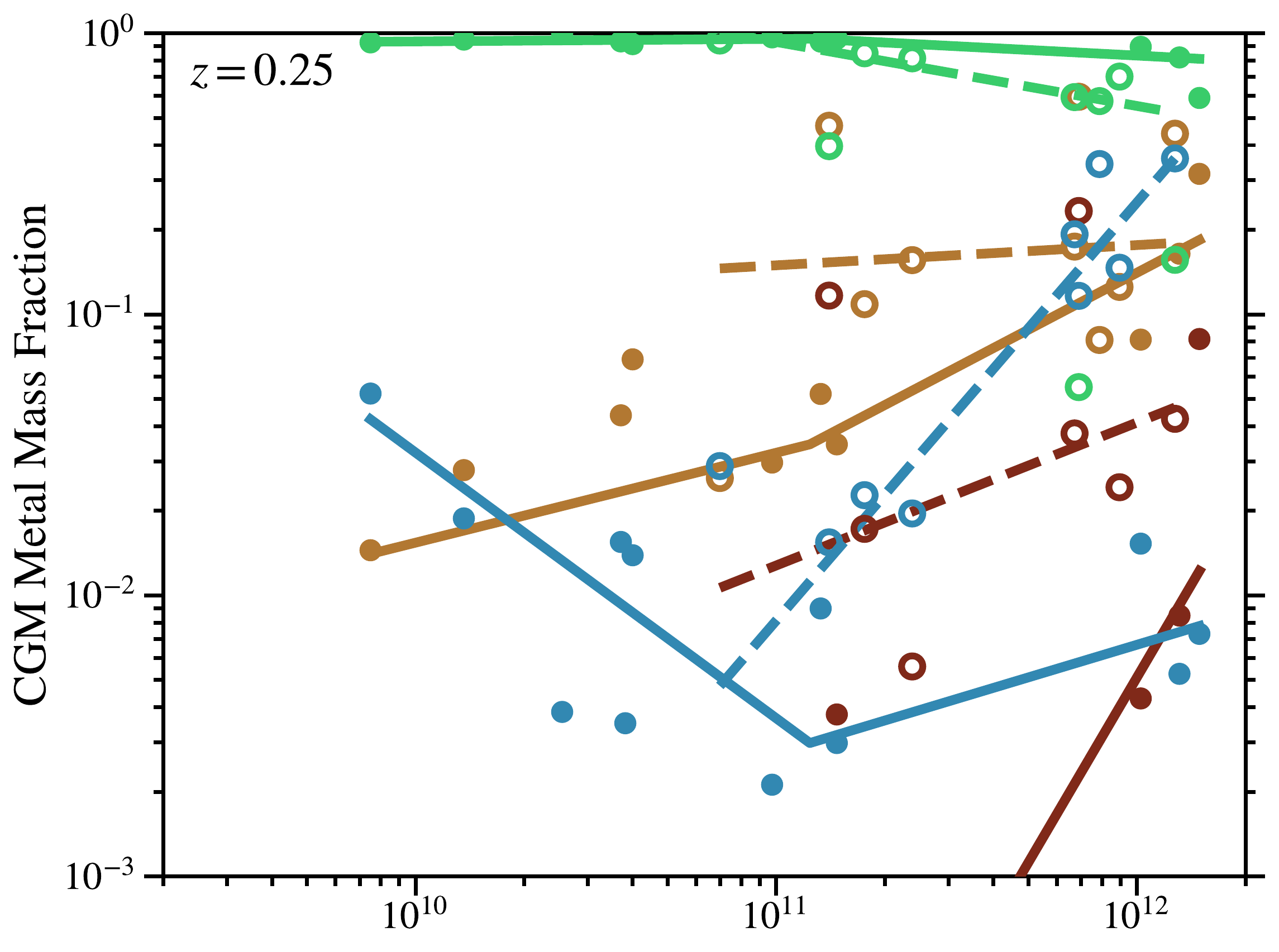}
\includegraphics[width=\columnwidth]{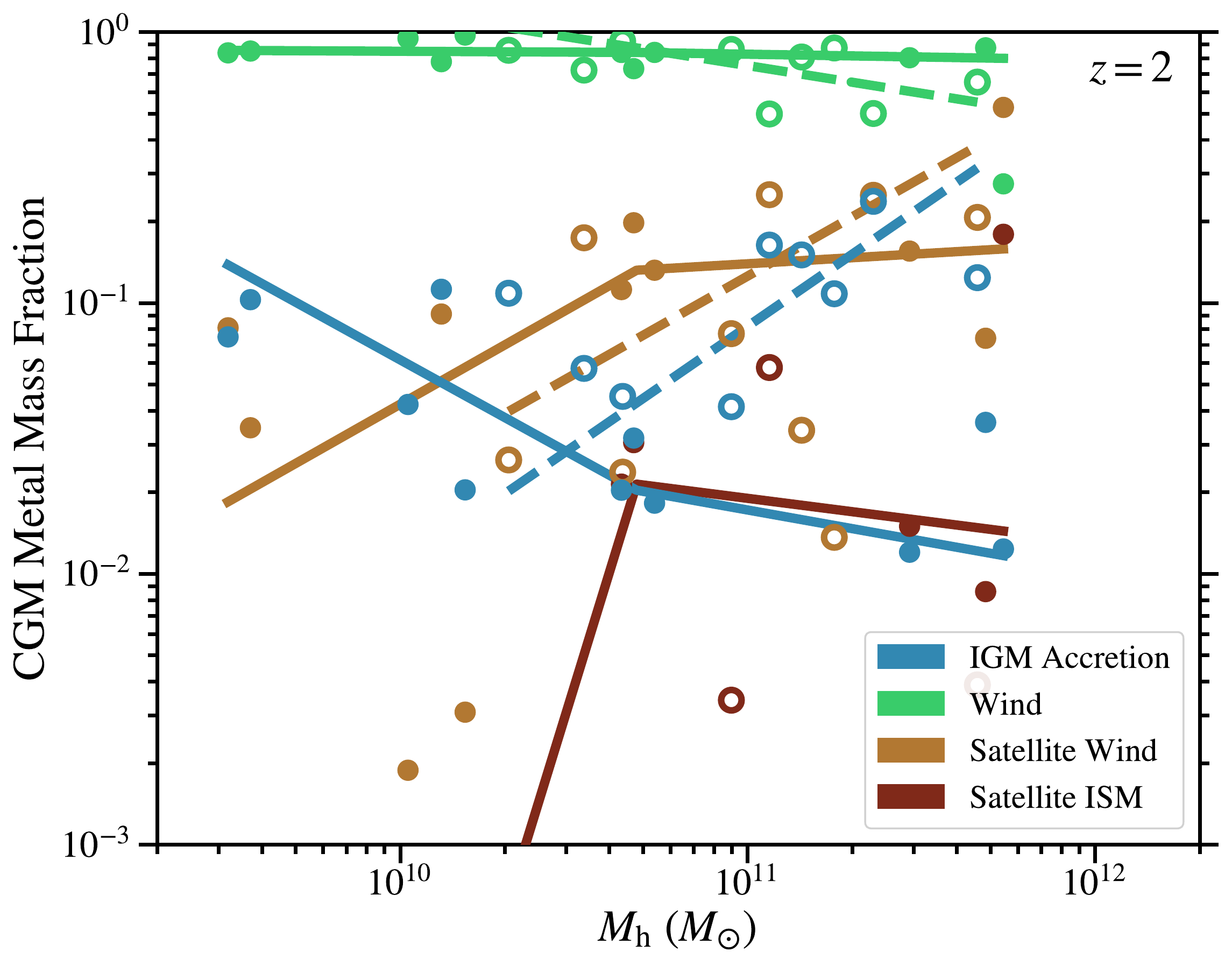}
\caption{
Fraction of CGM metal mass in gas from different origins.
Solid circles are simulations without subgrid metal diffusion and open circles are simulations evolved with identical physics but including a subgrid model for unresolved turbulent metal diffusion.
The solid (dashed) lines use data from the simulations without (with) subgrid metal diffusion and connect the medians in each mass bin.
For metal diffusion simulations we omit the line connecting the median for the $10^{10} M_\odot$ progenitors because only one such simulation is run with our metal diffusion prescription.
Additionally, at $z=2$ the medians for satellite ISM are below the x-axis for simulations with metal diffusion, so the line is not displayed.
At all masses, most of the metals are found in wind from the central galaxy. 
We discuss in \S \ref{sec:origin_by_halo_mass} the different interpretations of results with and without subgrid metal diffusion (where metals are at the target redshift vs. where metals were produced). 
Even in simulations without subgrid metal diffusion, IGM accretion can be enriched by interactions with halo stars or by brief interactions with other galaxies.}
\label{fig:CGM_metal_mass_frac_vs_Mh_CGM}
\end{figure}

Figure~\ref{fig:CGM_mass_frac_vs_Mh_CGM} shows the origin of mass in the CGM at $z=0.25$ and $z=2$ (within $R_{\rm vir}$, as before). 
Note that we include gas belonging to satellite galaxies (satellite ISM) as part of the CGM in this plot, even though it is not diffuse halo gas in a strict sense. 
We do this because satellite ISM can contribute to absorption systems commonly used to probe the CGM observationally, and so it is useful to know how much of it is present in galaxy halos. 
As the figure shows, satellite ISM is a relatively minor component by mass so including it in the total CGM mass does not significantly skew any of our results. 

Gas that arrives in the CGM through accretion from the IGM makes up $\gtrsim \fracnepmeanlow$ of the mass on average, across all halo masses and redshifts considered.
At $z=0.25$, wind from central galaxies is generally the second most important component by mass, except in $M_{\rm h}\sim 10^{12} M_{\odot}$ halos, for which winds from satellite galaxies can be more important. 
We find the same general trends at $z=2$, except that wind from central galaxies contributes more than wind from satellites at all halo masses considered. 
This is consistent with the fact that $M_{\rm h}\sim 10^{12} M_{\odot}$  progenitors drive strong winds in the FIRE simulations at high redshift, but much weaker winds as they settle into well-ordered disks with steady star formation rates at low redshift \citep[e.g.][]{Muratov2015}. 
The increasing importance of satellite winds with increasing halo mass is consistent with the trend with increasing halo mass found for the contribution of intergalactic transfer to galaxy growth found in the FIRE simulations \citep[][]{Angles-Alcazar2017}.

Figure~\ref{fig:CGM_metal_mass_frac_vs_Mh_CGM} shows the fraction of the total CGM metal mass contributed by different origins.
As in  Figure~\ref{fig:metal_mass_budget}, we exclude the metal mass contributed by the metallicity floor in order to focus on the metals produced by stars. 
In this plot, we distinguish between simulations that were evolved with the subgrid model for turbulent metal diffusion (open circles) and those that did not include subgrid metal diffusion. 
The solid (dashed) trend lines are for simulations without (with) metal diffusion.
For subgrid metal diffusion simulations we do not extend the trend lines to the $10^{10}$ M$_{\odot}$ halo mass bin because we only analyze one simulation including subgrid metal diffusion in this mass bin. 

We separately show trends with halo mass for the CGM metal mass fractions for the simulations with and without subgrid metal diffusion. 
We stress that the interpretation differs between the two sets of results.
For the simulations without subgrid metal diffusion, the CGM metal mass fractions most directly measure \emph{where metals in the CGM were produced}. 
This is because in the simulations without subgrid diffusion particles can only accumulate metals by direct contact with stellar processes that produce metals (stellar winds and supernovae).
In this case the mass fractions are dominated by winds from the central galaxy, winds from satellites, and satellite ISM. 
As we will show more directly in \S \ref{sec:metallicity}, even in the simulations without subgrid metal diffusion gas classified as IGM accretion can have a metallicity greater than the metallicity floor. 
This enrichment can occur via halo stars and brief interactions with other galaxies (too short for our algorithm to classify the particles as having been processed by other galaxies).

The CGM metal mass fraction in subgrid metal diffusion simulations includes both metals directly accumulated by interactions with star particles as well as through diffusion of metals from neighbor gas particles, which can have a different origin. 
The subgrid metal diffusion process does not change the CGM origin categories in our particle tracking analysis. 
Thus, when subgrid metal diffusion is include, the CGM origin category does not necessarily reflect where the metal mass carried by a gas particle were produced.

Although simulations with subgrid metal diffusion ``lose'' some information on the history of a particle's metals, they are very useful to include in our analysis, for a few reasons.
First, the subgrid metal diffusion prescription in FIRE-2 simulations has been shown to be necessary in order to produce realistic stellar metallicity distributions \citep[][]{Escala2018}, and the same may be true for the CGM. To connect with metallicities inferred with observations, it is thus important to explore the effects of subgrid metal diffusion on our results. 
Second, our model for subgrid metal diffusion is approximate (as are all such models), which introduces significant uncertainties in our predictions \citep[e.g.][]{Rennehan2018}. 
Including both simulations with and without this subgrid model in our analysis allows us to gauge these uncertainties. For example, the difference in a given origin's CGM metal mass between simulations of similar mass halos with/without metal diffusion is a measure of the metal fraction due to metal return by stellar processes vs. subgrid mixing with neighbors.\footnote{We note, however, that this measure is imperfect due to variance between different halos and the fact that subgrid metal diffusion can also introduce some dynamical differences through cooling.}

Subgrid metal diffusion primarily affects the metal mass content of gas accreted from the IGM. 
This is because these particles experience no or little direct interactions with star particles. 
When subgrid metal diffusion is included, the diffusion enrichment process rapidly becomes important.
At $z=0.25$, the metal mass fraction in IGM accretion in the $M_{\rm h}\sim 10^{12}$ M$_{\odot}$ halos can be $\sim 1-2$ dex higher than in similar runs without subgrid metal diffusion.
Despite large fractional changes in metallicity, this enhancement generally occurs through the diffusion of a relatively small absolute metal mass.

Subgrid metal diffusion also appears to have a major effect on the halo mass dependence of the IGM accretion metal mass fraction, which has strong positive slope with increasing halo mass at $z=0.25$ in the simulations with subgrid metal diffusion. 
The fact that the effects of subgrid metal diffusion are strongest for the most massive halos and at late times suggests that the magnitude of the effect is enhanced by the presence of a virialized hot halo into which metals can diffuse efficiently from winds (the more massive halos at low redshifts are the ones which sustain the most substantial hot halos).
This is because metals in well-mixed, volume-filling hot halos naturally come into contact with IGM accretion flows.

\subsubsection{Origin by Time}
\label{sec:origin_by_time}
\begin{figure}
\includegraphics[width=\columnwidth]{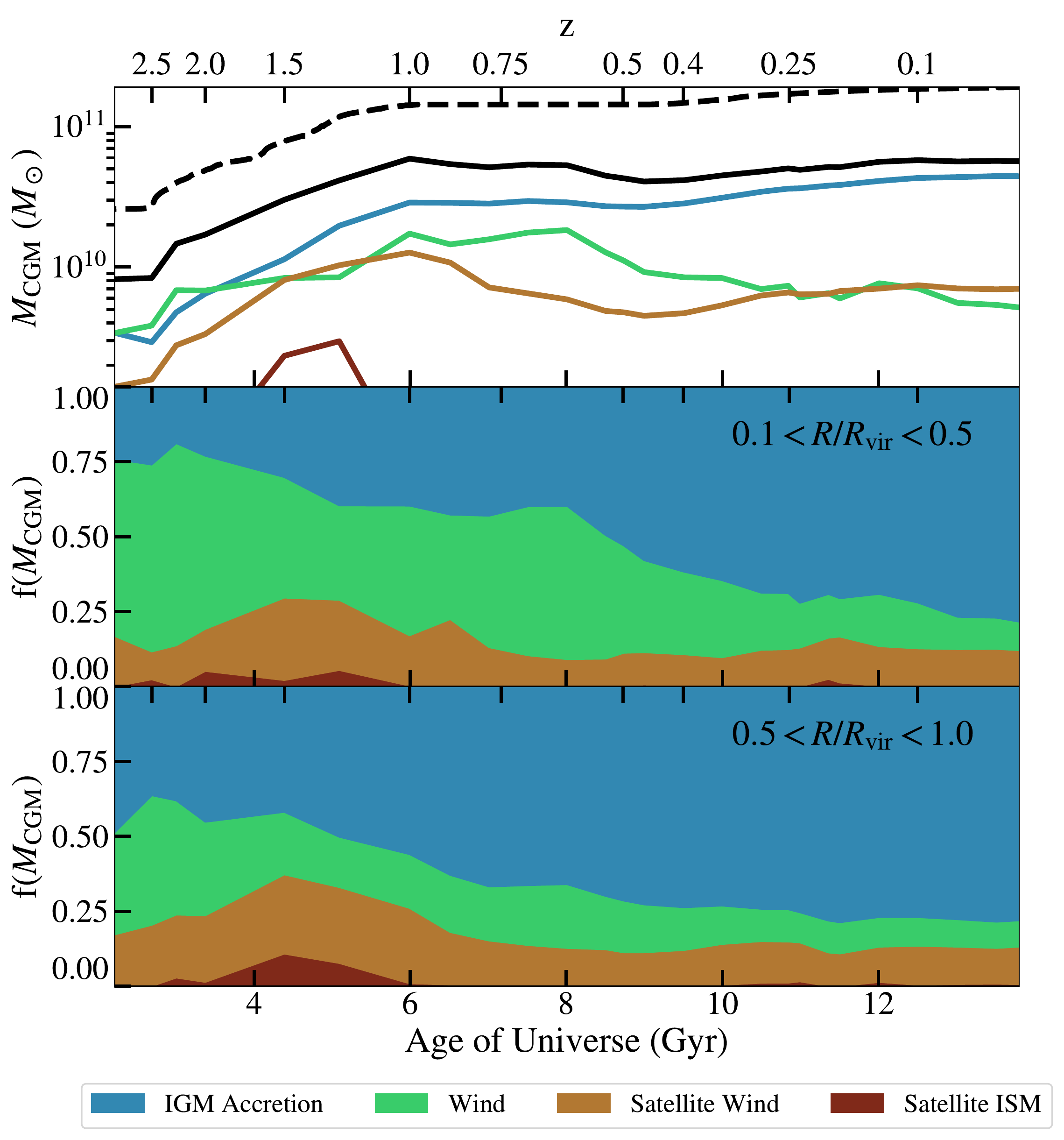}
\caption{
Evolution of the CGM mass for the main halo of simulation \texttt{m12i}, which has  $M_{\rm h} ( z=0 )\sim 10^{12} M_\odot$.
The top panel shows the evolution of the total mass in the CGM (solid black line).
The colored lines represent the CGM mass from different origins.
The black dashed line shows the cosmic baryon budget, $(\Omega_{\rm b}/\Omega_{\rm m}) M_{\rm h}$.
The bottom two panels show the evolution of $f(M_{\rm CGM})$, the fraction of the total CGM mass contributed by different origins in the inner halo ($0.1 < R / R_{\rm vir} < 0.5$) and the outer ($0.5 < R/R_{\rm vir} < 1$) halo. 
As the galaxy grows, it develops a virialized hot halo while the efficiency of galactic wind driving declines. 
As a result, the fractional contribution of IGM accretion to the CGM mass becomes increasingly dominant.
}
\label{fig:m12i_CGM_redshift_mass_fraction}
\end{figure}
\begin{figure}
\includegraphics[width=\columnwidth]{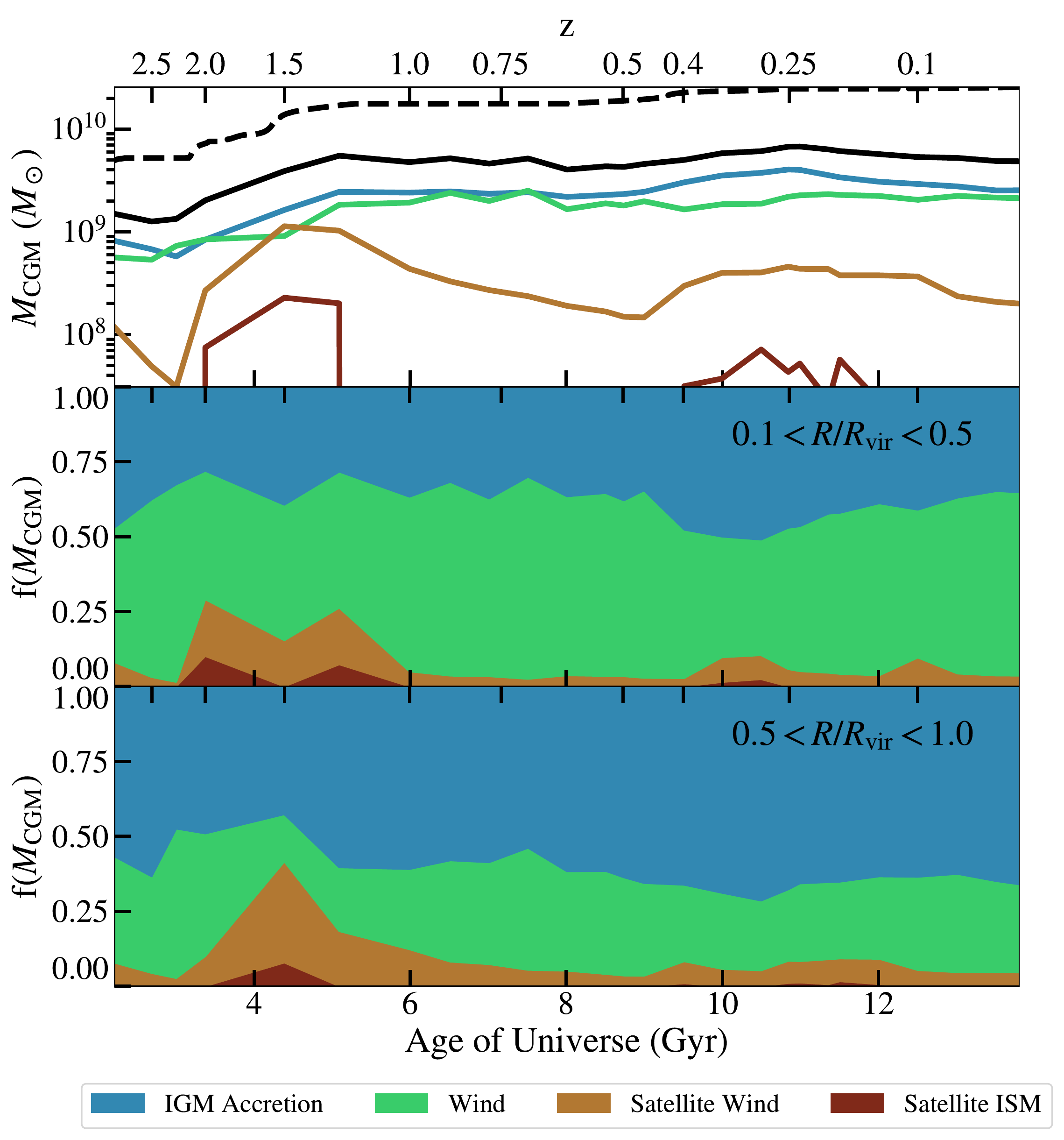}
\caption{
Same as Figure~\ref{fig:m12i_CGM_redshift_mass_fraction}, but for the main halo of simulation \texttt{m11q}, which has  $M_{\rm h} ( z=0 )\sim 10^{11} M_\odot$.
For this halo, the relative contributions to the CGM mass from IGM accretion and wind from the central galaxy are approximately constant with time. 
This reflects the fact that FIRE galaxies drive significant galactic winds all the way to $z=0$ at this mass scale.
}
\label{fig:m11q_CGM_redshift_mass_fraction}
\end{figure}
\begin{figure}
\includegraphics[width=\columnwidth]{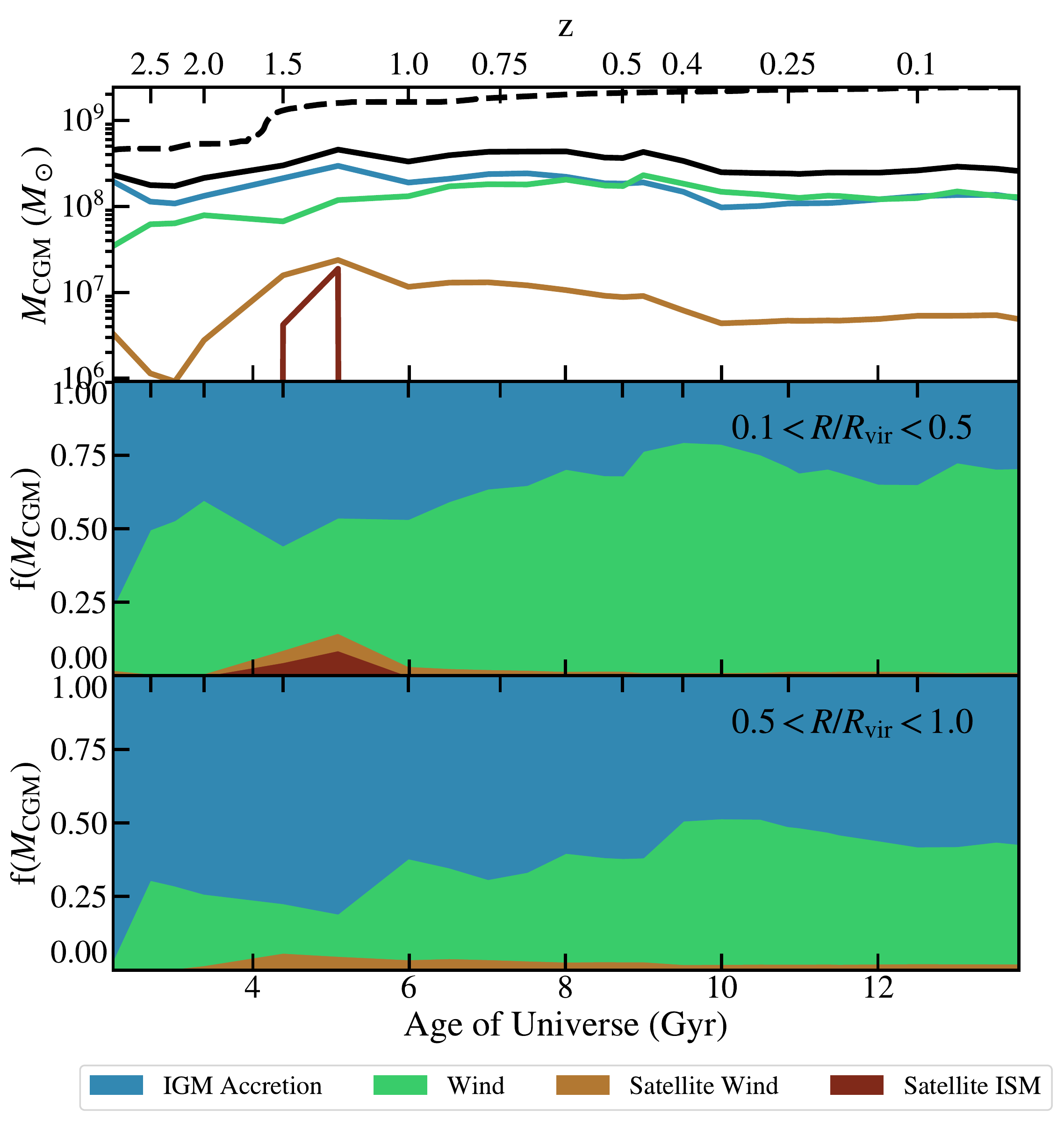}
\caption{
Same as Figure~\ref{fig:m12i_CGM_redshift_mass_fraction}, but for the main halo of simulation \texttt{m10y}, which has  $M_{\rm h} ( z=0 )\sim 10^{10} M_\odot$.
The redshift evolution of the IGM accretion fraction relative to the wind fraction is inverse that of \texttt{m12i}, with wind from the central galaxy contributing an increasingly large fraction of the CGM mass with decreasing redshift. 
Satellites contribute little to the CGM of this dwarf galaxy at all redshifts.
}
\label{fig:m10y_CGM_redshift_mass_fraction_galdefv2}
\end{figure}

In this section we summarize in more detail the redshift evolution of the CGM origins for one representative simulation per progenitor mass bin.
For each simulation, we consider separately the mass fractions in the inner halo ($0.1 < R / R_{\rm vir} < 0.5$) and the outer halo ($0.5 < R/R_{\rm vir} < 1$). 

The evolution of CGM mass composition for \texttt{m12i} is shown in Figure~\ref{fig:m12i_CGM_redshift_mass_fraction}.
The total CGM mass (black line in the top panel) stays below the baryon budget for the entirety of the redshift interval shown ($z<2.5$).
Over $0\lesssim z\lesssim 1$, a time interval during which the total mass in the CGM plateaus, the CGM mass fraction from IGM accretion increases by a factor $\sim 2$ in the outer halo.  
This occurs primarily as a result of the decreasing contribution from wind from the central galaxy. 
This decrease in wind mass is driven by a transition in the FIRE simulations from bursty, powerful winds at $z\gtrsim1$ to more steady galaxies with weaker outflows in $\sim L^{\star}$ galaxies at low redshift~\citep[e.g.][]{Muratov2015,Sparre2017,2017MNRAS.465.1682H,Angles-Alcazar2017,Angles-Alcazar2017a,Faucher-Giguere2017}. 
The drop in wind contribution is very apparent in the inner halo, where more of the wind comes from recently ejected material as opposed to long-lived halo gas.
For the same reasons at all redshifts wind from the central galaxy contributes a larger CGM mass fraction in the inner halo than in the outer halo.
Pseudo-evolution of the virial radius~\citep[e.g.][]{Liang2015} also plays a role in the increasing IGM accretion mass, since at larger radii IGM accretion is more prevalent and the mass in radial shells is approximately constant with increasing radius (\S\ref{sec:origin_by_radius}). 
On the other hand, the CGM mass fraction from satellite wind is more similar between the inner halo and the outer halo. 
We quantify radial profiles of different CGM origins in more detail in the next section.

The redshift evolution of \texttt{m11q} is shown in Figure~\ref{fig:m11q_CGM_redshift_mass_fraction}.
The increased contributions from satellite wind and satellite ISM at $z\sim 1.5$ correspond to an infalling intergalactic filament containing multiple dwarf galaxies. 
Another minor merger is apparent at $z \sim 0.25$. 
Unlike for \texttt{m12i}, the fractional contribution of wind from the central galaxy to the CGM is approximately constant throughout the redshift interval considered. 
This is because galaxies at this mass scale in FIRE continue to drive significant galactic winds all the way to $z=0$.

The redshift evolution of our representative $M_{\rm h}(z=0) \sim 10^{10} M_\odot$ simulation, \texttt{m10y}, is shown in Figure~\ref{fig:m10y_CGM_redshift_mass_fraction_galdefv2}.
Interestingly, the redshift evolution of the IGM accretion fraction relative to the wind fraction is inverse that of \texttt{m12i}, with wind from the central galaxy contributing an increasingly large fraction of the CGM mass with decreasing redshift. 
This difference likely owes to the presence of winds that efficiently remove accreted gas from the halo outskirts in small systems~\citep[e.g.][]{Muratov2015}, and inefficient accretion owing to photo-heating of the IGM~\citep[the maximum halo mass affected by the UV background increases with deacreasing redshift, e.g.][]{Faucher-Giguere2011,El-Badry2017}.
The effects of winds on IGM accretion in dwarfs is visible in Figure~\ref{fig:r_vs_time_m10y_CGM_snum465} as sharp, coherent upticks in radial distance of IGM accretion paths. 
Satellites contribute little to the CGM at all redshifts, consistent with the relative paucity of satellite galaxies around dwarfs.

\subsubsection{Origin by Radius}
\label{sec:origin_by_radius}

\begin{figure*}
\centering
\begin{minipage}{0.495\textwidth}
\centering
\includegraphics[width=\textwidth]{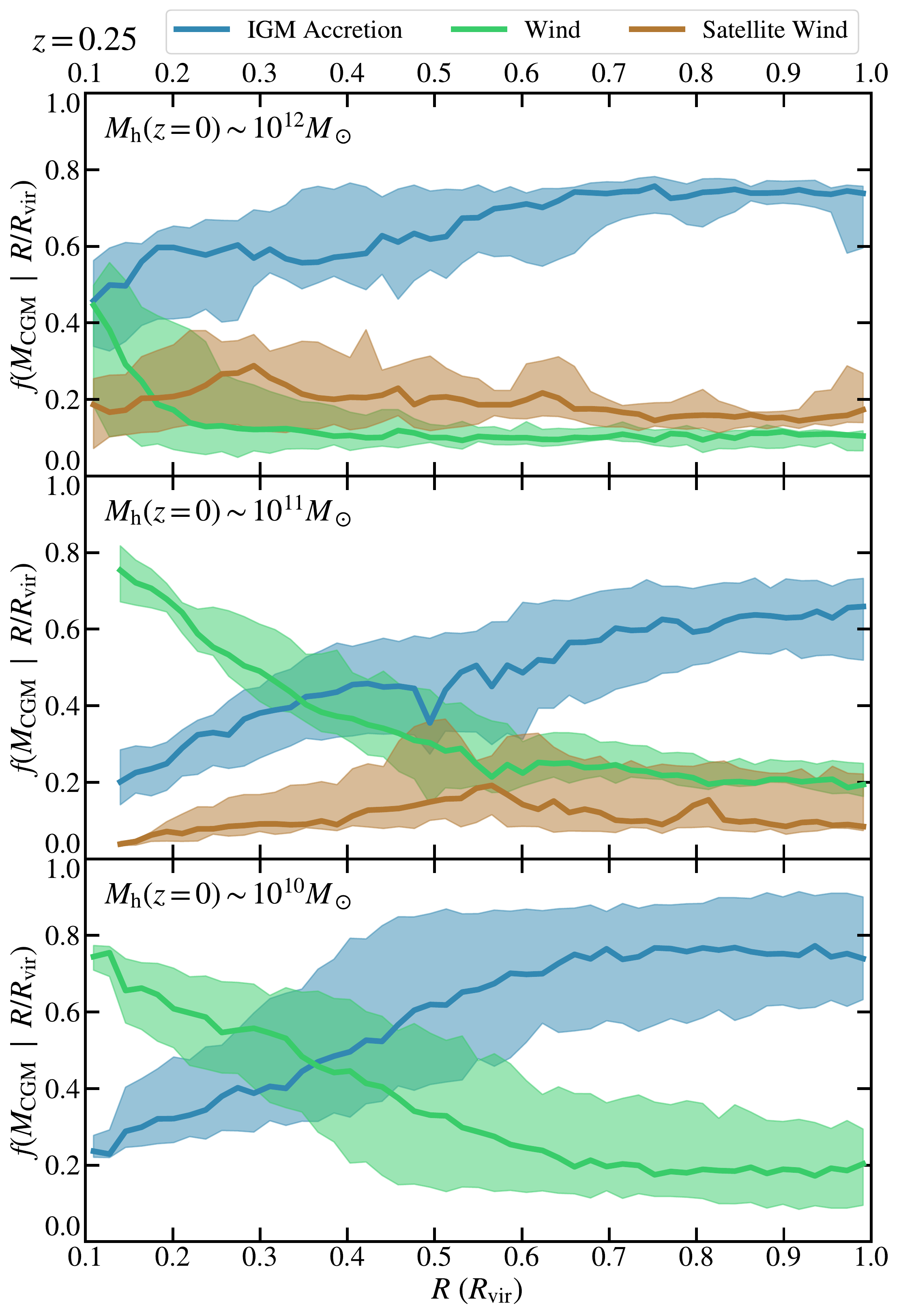}
\end{minipage} \hfill
\begin{minipage}{0.495\textwidth}
\centering
\includegraphics[width=\textwidth]{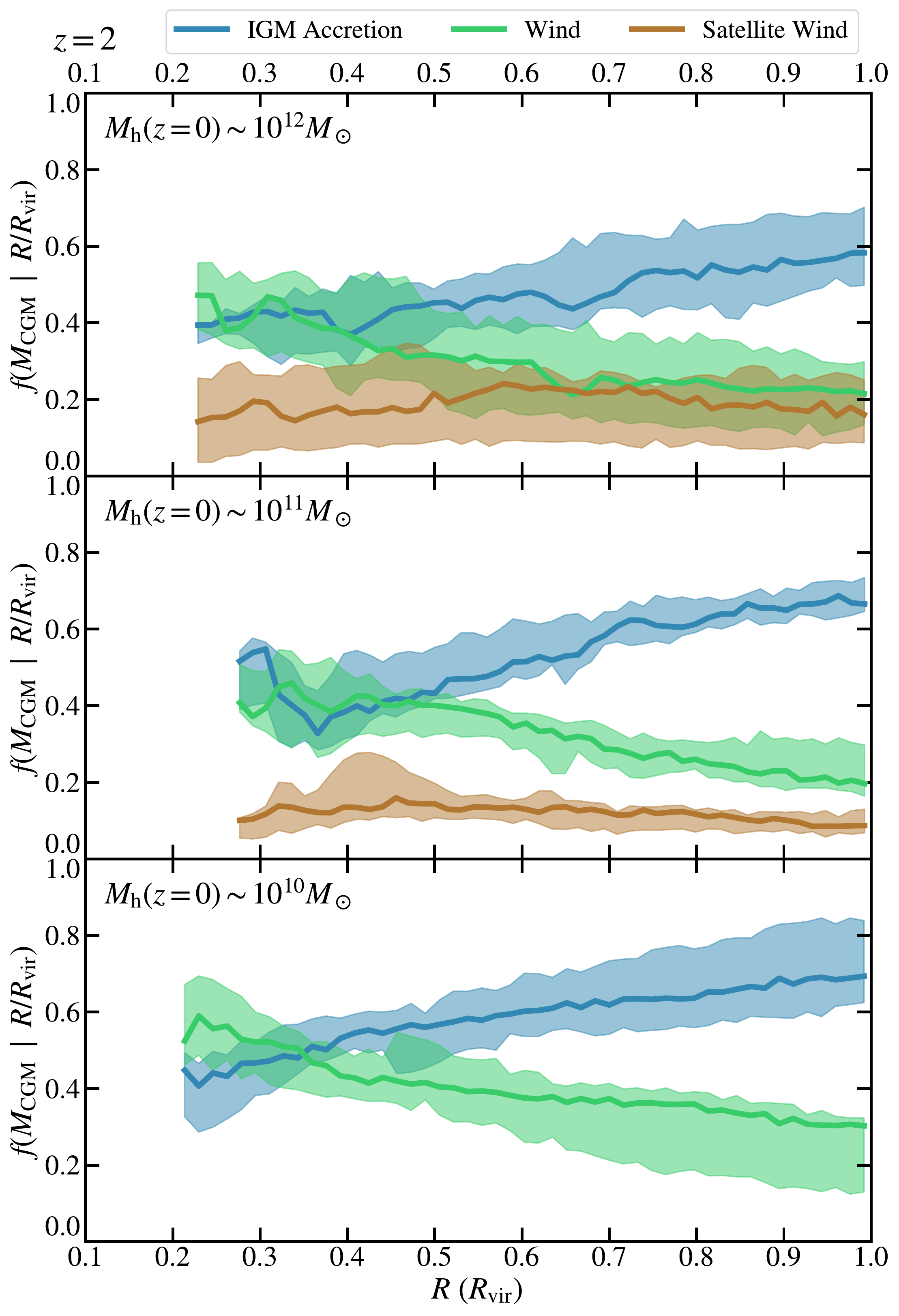}
\end{minipage} \hfill
\caption{
Fraction of the total CGM mass vs. radius contributed by different origins for different halo mass ranges at $z=0.25$ (left) and $z=2$ (right).
The solid lines indicate median mass fractions and the shaded regions are the 16th and 84th percentiles. 
Across all halo masses and redshifts considered, IGM accretion provides $\sim 60-80\%$ of the CGM gas mass at $R \sim R_{\rm vir}$, but only $\sim 20-60\%$ of the gas mass in the inner CGM ($0.1R_{\rm vir}<R<0.5R_{\rm vir}$).
}
\label{fig:CGM_mass_frac_profile}
\end{figure*}

\begin{figure*}
\centering
\begin{minipage}{0.495\textwidth}
\centering
\includegraphics[width=\textwidth]{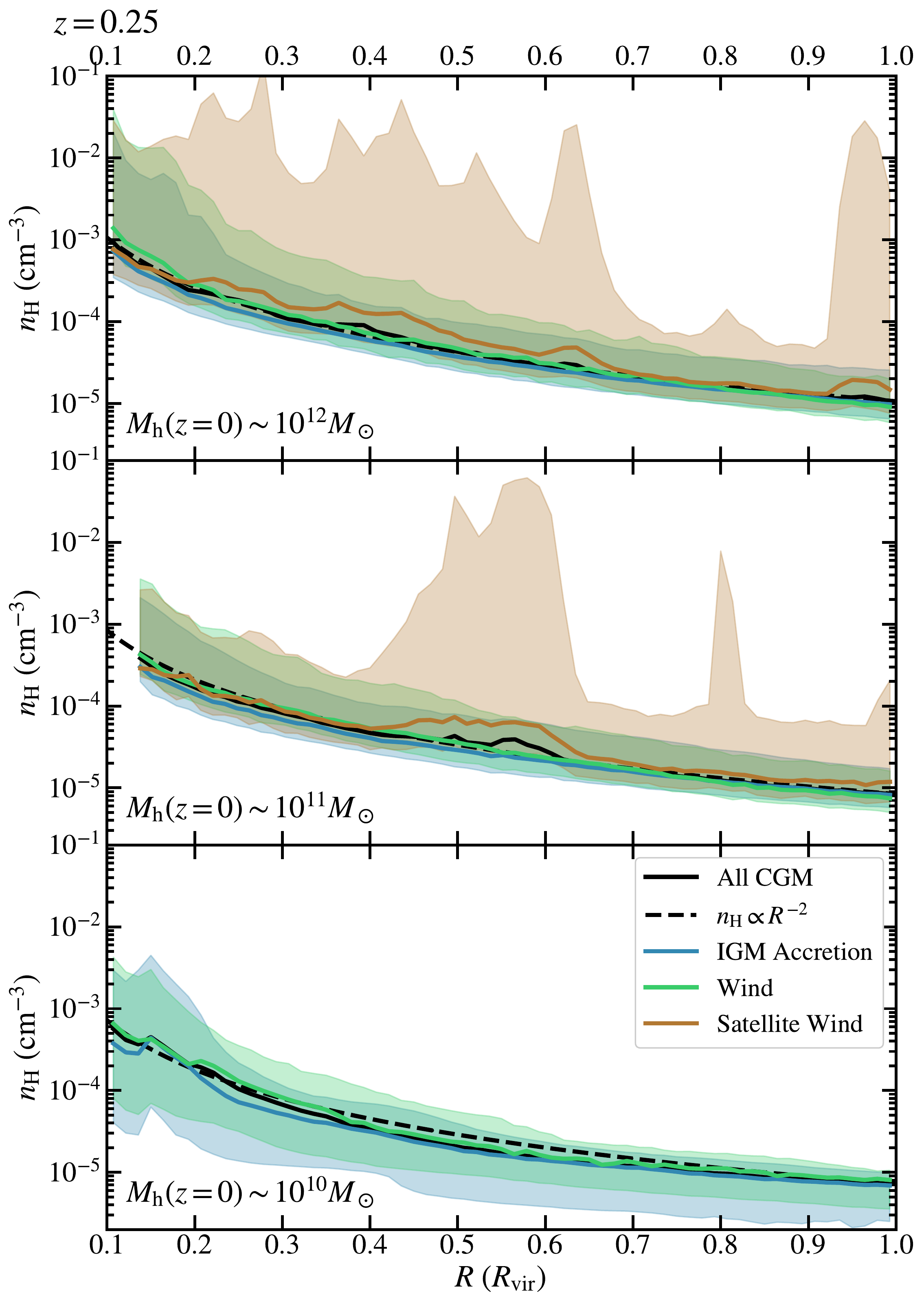}
\end{minipage} \hfill
\begin{minipage}{0.495\textwidth}
\centering
\includegraphics[width=\textwidth]{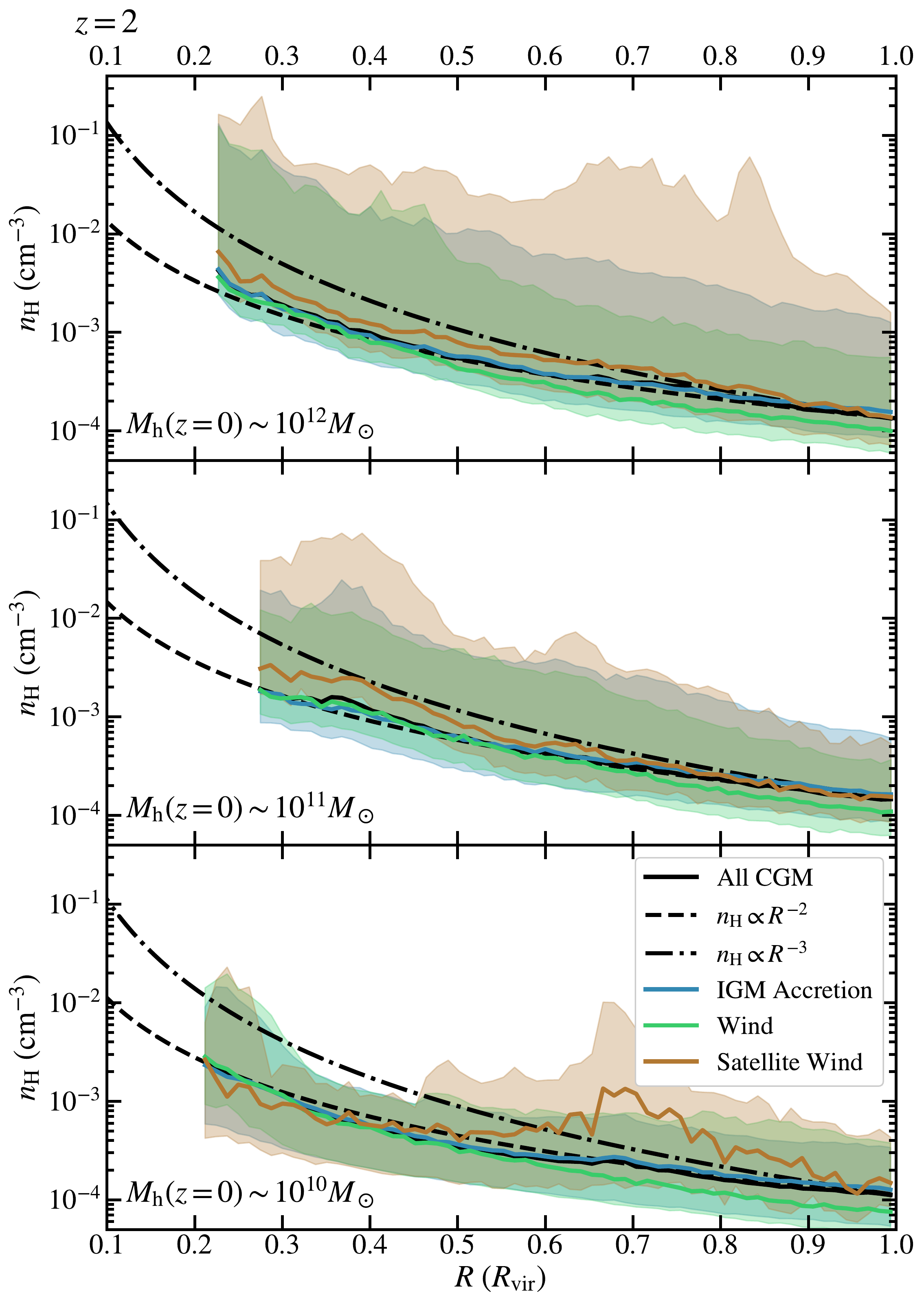}
\end{minipage} \hfill
\caption{
Density profiles of CGM gas for different origins and mass ranges at $z=0.25$ (left) and $z=2$ (right).
The solid lines are the volume-weighted mean hydrogen number densities and the shaded regions are the 16th and 84th percentile densities for all particles.
The values at a given radius are calculated using data from all halos in that mass bin.
For reference, we plot power-law density profiles (dashed and dashed-dotted lines), normalized to match the median CGM density at $R_{\rm vir}$.
The volume-weighted density profile depends on redshift and halo mass but is close to $n_{\rm H} \propto R^{-2}$, i.e. approximately constant total mass per radial bin.
Close to the main galaxy (in the case of wind) or close to the satellite galaxies (in the case of satellite wind) densities are enhanced relative to this mean density profile.
}
\label{fig:CGM_den_profile}
\end{figure*}

It is also interesting to examine the radial distributions of CGM of different origins. 
Figure~\ref{fig:CGM_mass_frac_profile} shows the fraction of CGM mass at a given radius contributed by different origins at $z=0.25$ and $z=2$.
The solid curves indicate the median mass fraction in a radial bin across all simulations in that halo mass range, while shaded regions indicate the 16th to 84th percentiles.
This is complemented by Figure~\ref{fig:CGM_den_profile}, which shows the density profiles of CGM gas at $z=0.25$ and $z=2$.
The mean density profiles are calculated by finding all the particles at a given $R/R_{\rm vir}$ across all halos in that mass bin and calculating the volume-weighted mean densities (solid lines) and the 16th and 84th percentile densities across those particles (shaded regions). 
We chose to plot the volume-weighted mean density because this allows one to recover the total gas mass in any radial bin by multiplying by the corresponding volume. 
To avoid cluttering the figure we do not plot the 16th and 84th percentile interval for the overall CGM category. 
We do not include trends for satellite ISM because in most cases its contribution to the total CGM mass is small, and its radial distribution closely tracks the locations of satellites. 
We also do not plot satellite wind profiles for the $10^{10} M_\odot$ progenitors because many of these halos do not have enough satellite wind to produce robust profiles. 
At small radii, some of the radial bins contain few halos because $R<R_{\rm CGM, inner}$ for some of the halos. 
To avoid artifacts in the profiles due to small number statistics (e.g., sharp jumps), we only show the profiles for radial bins containing 3 simulations more.

IGM accretion provides $\sim 60-80\%$ of the mass at $R\sim R_{\rm vir}$ but only $\sim 20-60\%$ of the mass at the inner edge of the halo ($0.1R_{\rm vir}<R<R_{\rm vir}$). 
Wind, on the other, hand provides $\gtrsim 40\%$ of the mass throughout the inner halo, except for the $10^{12} M_\odot$ progenitors at $z=0.25$. 
This is consistent with winds becoming weaker at low redshift in massive halos in the FIRE simulations \cite[e.g.,][]{Muratov2015}, as well as the fact that most wind mass recycles efficiently (e.g., \citealt{Angles-Alcazar2017}). 
Satellite wind associated either with a subhalo or integrated into the well-mixed halo makes up an approximately constant function of mass across radii, indicating that subhalos can deposit satellite wind at a variety of radii.

We compare our density profiles to power-law profiles $n_{\rm H } = n_{\rm H} (R_{\rm vir})  (R/R_{\rm vir})^{-\gamma}$, where $\gamma =2,3$ and $n_{\rm H} \mid_{R=R_{\rm vir}}$ is the volume-weighted mean density at $R_{\rm vir}$ for all CGM gas in that mass bin.
At both $z=0.25$ and $z=2$, the volume-weighted mean density profile of all gas in the CGM of $10^{11}$ and $10^{12} M_\odot$ progenitor  follows a $n_{\rm H} \propto R^{-2}$ profile, and therefore each radial bin contains approximately the same mass.
Because each radial bin contains approximately the same total gas mass but IGM accretion provides more mass in the outer halo, the majority of IGM accretion is located at larger distances from the main galaxy.
The inverse is true for wind. 
In addition, while the effect is subtle, the wind density profile tends to fall-off more steeply than the IGM accretion profile. 
As discussed previously, some fraction of satellite wind is associated with subhalos (which can produce spikes in density profiles owing to the small number of simulations analyzed) and another fraction is part of the well-mixed halo.

\subsubsection{Origin by Polar Angle}
\label{sec:origin_by_angle}

\begin{figure*}
\includegraphics[width=\textwidth]{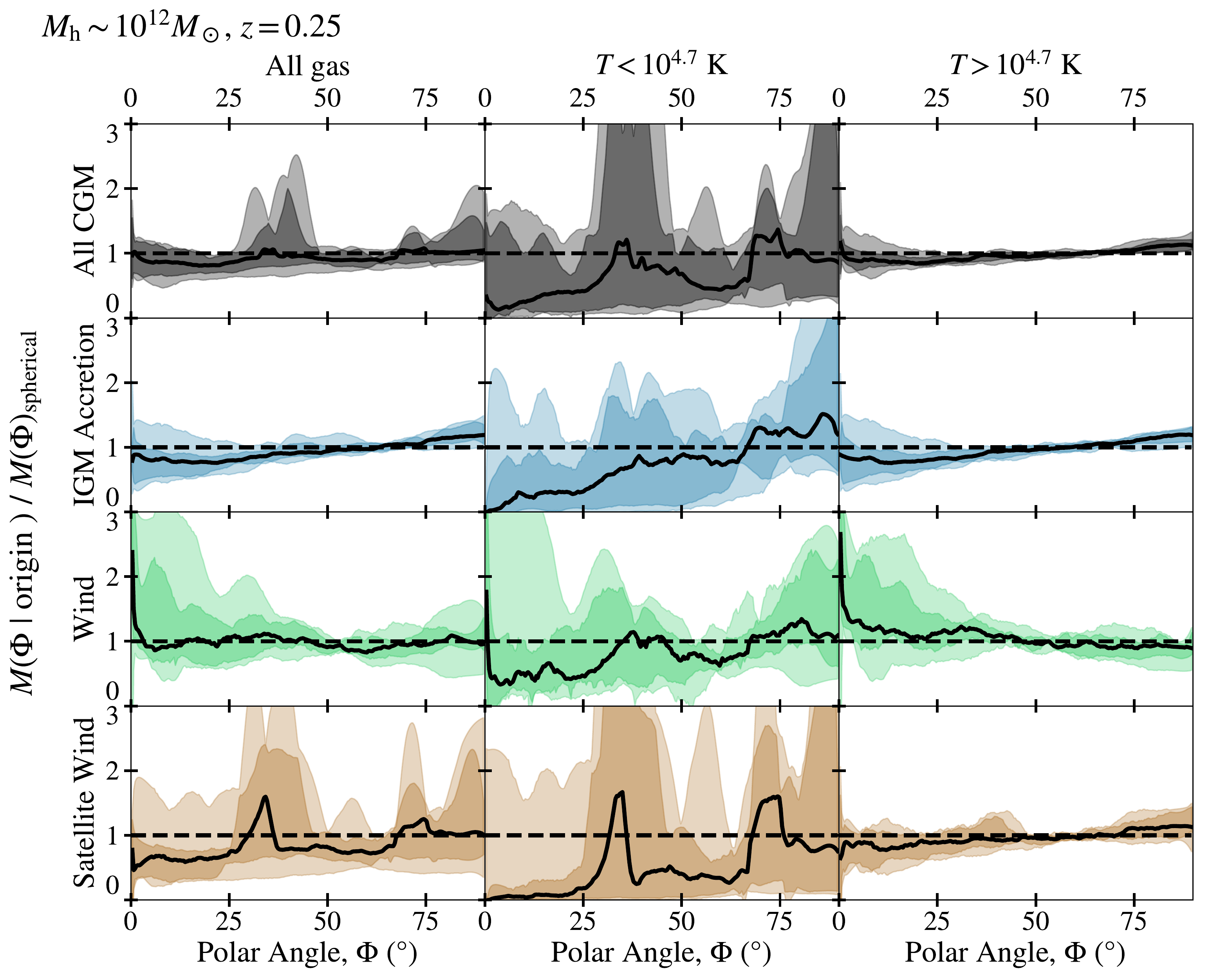}
\caption{
Fraction of CGM mass per unit polar angle, $\Phi$, divided by the fraction of mass that would be found at that angle for a spherical mass distribution.
Results are shown for the CGMs of $M_{\rm h} \sim 10^{12} M_\odot$ halos at $z=0.25$, with the results for different origins shown in different rows and the results for different temperature cuts shown in different columns.
The inner shaded region covers the 16th to 84th percentiles between simulations, while the outer shaded region covers the full extent covered by all simulations in the sample.
Across all origins, cool gas is found preferentially along the galaxy plane but with strong halo-to-halo variability, while hot gas is more spherically distributed.
}
\label{fig:CGM_polar_dists_m12s_snum465}
\end{figure*}

In this section we investigate how the distribution of CGM gas depends on polar angle $\Phi$, which we define as the angle between the position vector of a gas element (relative to the center of the halo) and the angular momentum vector of the stellar disk (the polar direction). 
This is related to the azimuthal angle relative to the semi-minor axis of disk galaxies used in some observational studies \citep[e.g.,][]{Bordoloi2011,Bordoloi2014,Bouche2012,Kacprzak2015}, but for our simulation analysis we evaluate it in three-dimensional space rather than as a projection on the sky. 
We note that studying angular dependences in 3D rather than in 2D should, if anything, enhance possible signals since projecting on the sky introduces a degree of smearing.

We perform this analysis only for the $M_{\rm h} \sim 10^{12}$ M$_\odot$ halos at $z=0.25$, because this is the subset of our simulations for which clear stellar disks are present in all cases. 
The morphologies of these systems are discussed in detail in \cite{Garrison-kimmel2017}. 
The $z=2$ progenitors, as well as the lower-mass galaxies,  are generally in the ``bursty'' star formation regime, corresponding to disturbed gas morphologies and more irregular stellar morphologies \citep[e.g.,][]{Hopkins2014,El-Badry2015,Sparre2017,Faucher-Giguere2017,El-Badry2018}.
We assume that galaxies are statistically symmetric about the disk plane and plot results only for $\Phi < 90\degr$, e.g. we treat $\Phi = 135\degr$ as $45\degr$.

To quantify possible trends with respect to polar angle, we analyze the CGM mass of different origins per unit polar angle relative to the fraction of mass for a spherical mass distribution.
The results are shown in Figure~\ref{fig:CGM_polar_dists_m12s_snum465} for two temperature bins: $T<10^{4.7} K$ (cold/cool) and $T>10^{4.7} K$ (warm/hot).

On average more cold/cool gas is found along the galaxy plane than would be expected from a spherically-symmetric distribution. However, there is a large variance between different halos.
This excess of cold/cool gas around the disk plane is consistent with the ``cold flow disks'' found in previous simulation analyses~\citep[e.g.][]{Keres2009b,2013ApJ...769...74S, Stewart2016}.
On the other hand, warm/hot gas is more spherically distributed. 
Because $\gtrsim 70\%$ of the gas mass in these halos is at warm/hot temperature (Appendix~\ref{sec:mass_by_phase}), the overall mass in the halo is substantially spherical.

The polar angle distribution as a function of origin reveals a number of interesting features.
As expected, cool satellite wind is found primarily close to satellites themselves (the positions of satellites correspond to strong ``peaks'' in the polar angle distributions). 
The preference of cold/cool wind gas to be found along the galaxy plane in some halos reflects an extended disk of recycled material in those halos.
Also of note, while warm/hot wind is distributed close to spherically in the median, a subset of halos have enhanced warm/hot wind mass along the galaxy axis. 
This may reflect the preferential expansion of warm/hot wind normal to the disk plane.

\subsection{Metallicities of CGM Gas of Different Origins}
\label{sec:metallicity}

\begin{figure*}
\centering
\begin{minipage}{0.495\textwidth}
\centering
\includegraphics[width=\textwidth]{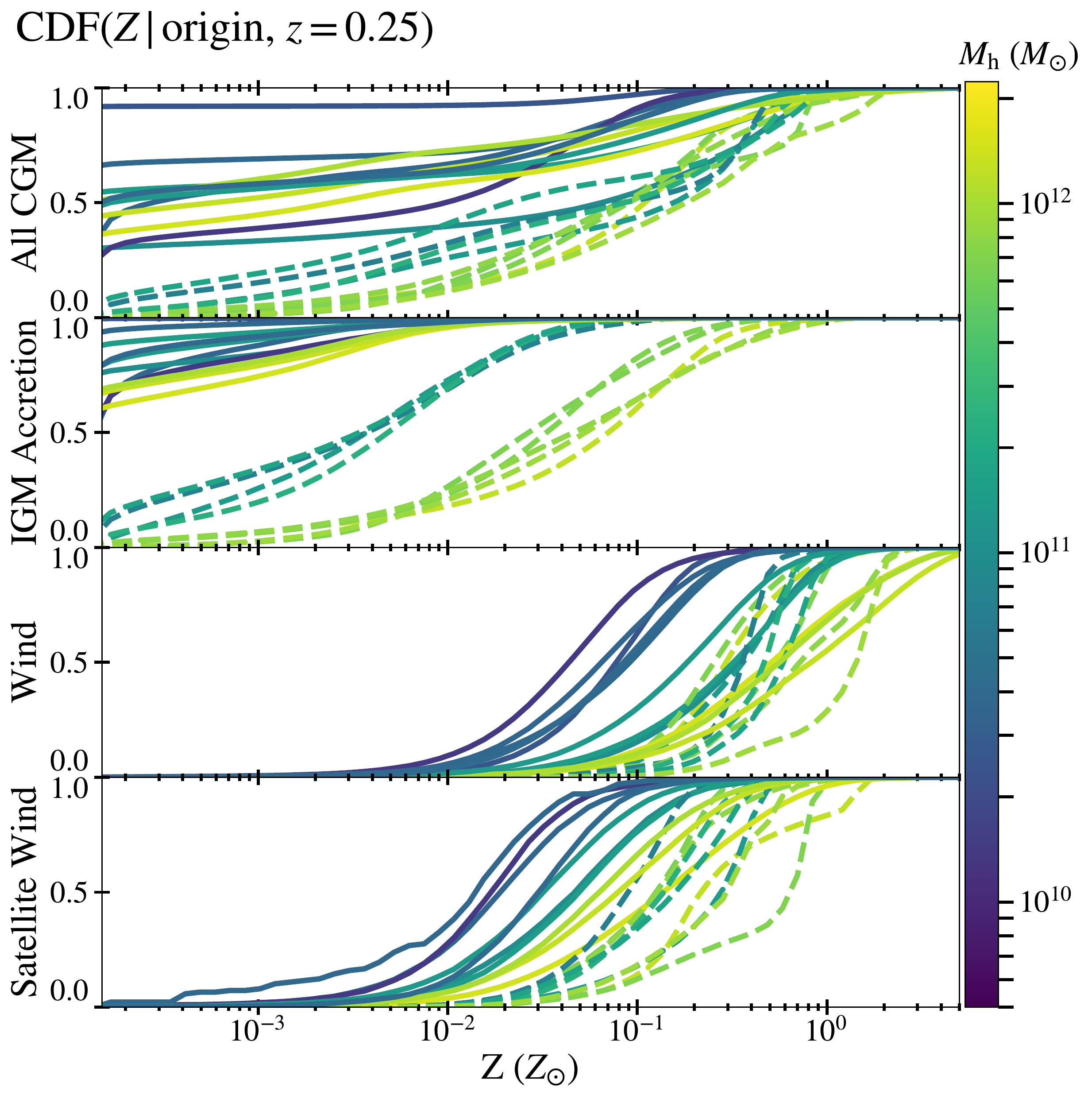}
\end{minipage} \hfill
\begin{minipage}{0.495\textwidth}
\centering
\includegraphics[width=\textwidth]{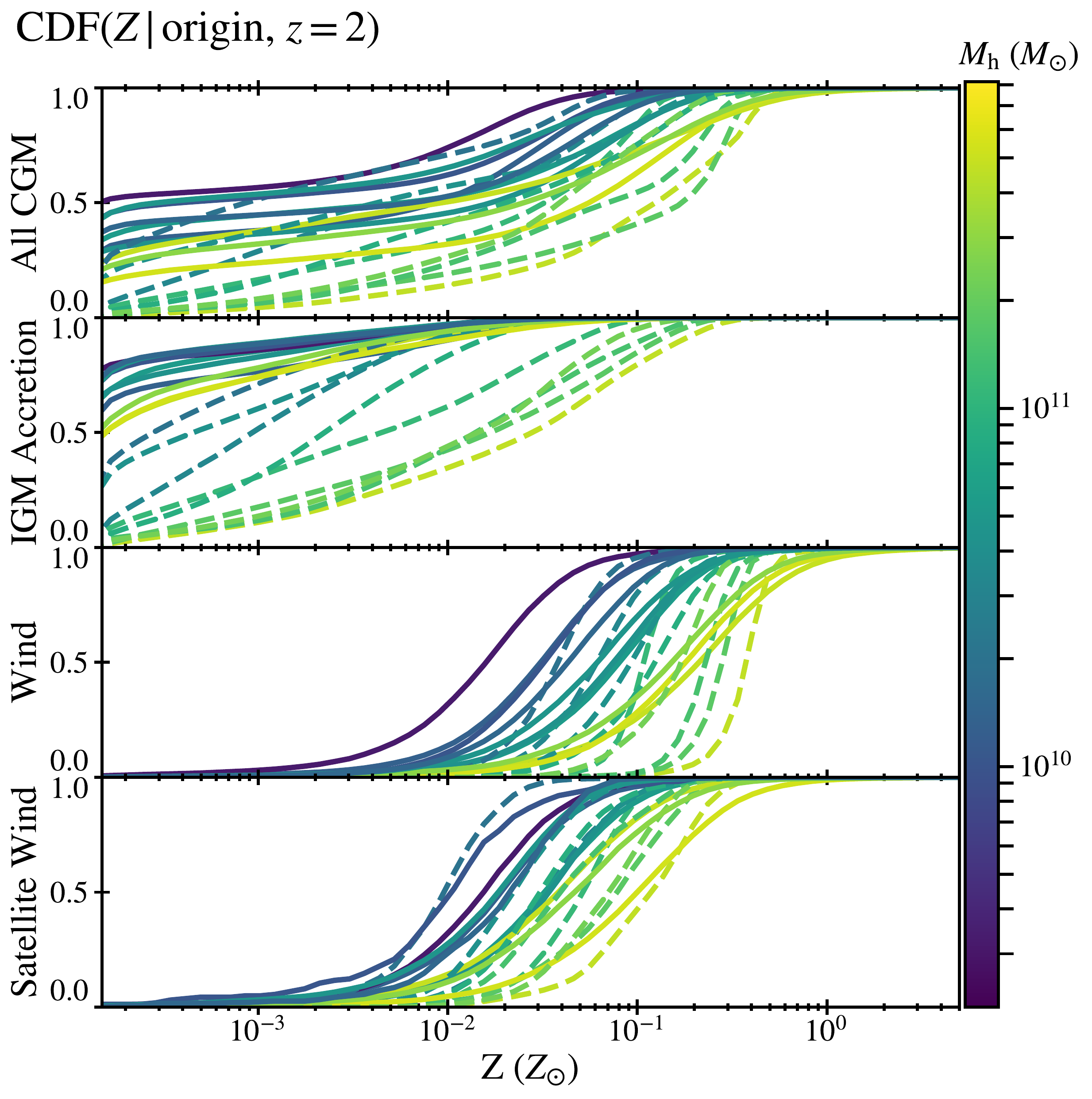}
\end{minipage} \hfill
\caption{
\textbf{Left:}
Cumulative distributions of metallicity for CGM gas at $z=0.25$, classified according to gas particle origin.
Each curve represents the metallicity distribution for a single halo, and the color of the line corresponds to that halo's halo mass at $z=0.25$.
Dashed lines indicate simulations that include subgrid metal diffusion.
The metallicity of IGM accretion is systematically lower than the metallicity of winds. 
The metallicity distribution of IGM accretion is systematically lower than the metallicity distributions of winds (typically by $\gtrsim 1$ dex in the median), but absolute metallicities depend significantly on whether subgrid metal diffusion is included or not. 
The metallicity distribution IGM accretion is most increased by subgrid metal diffusion.
\textbf{Right:}
Same as left, but at $z=2$.
}
\label{fig:CGM_metallicity}
\end{figure*}

\begin{figure*}
\includegraphics[width=\textwidth]{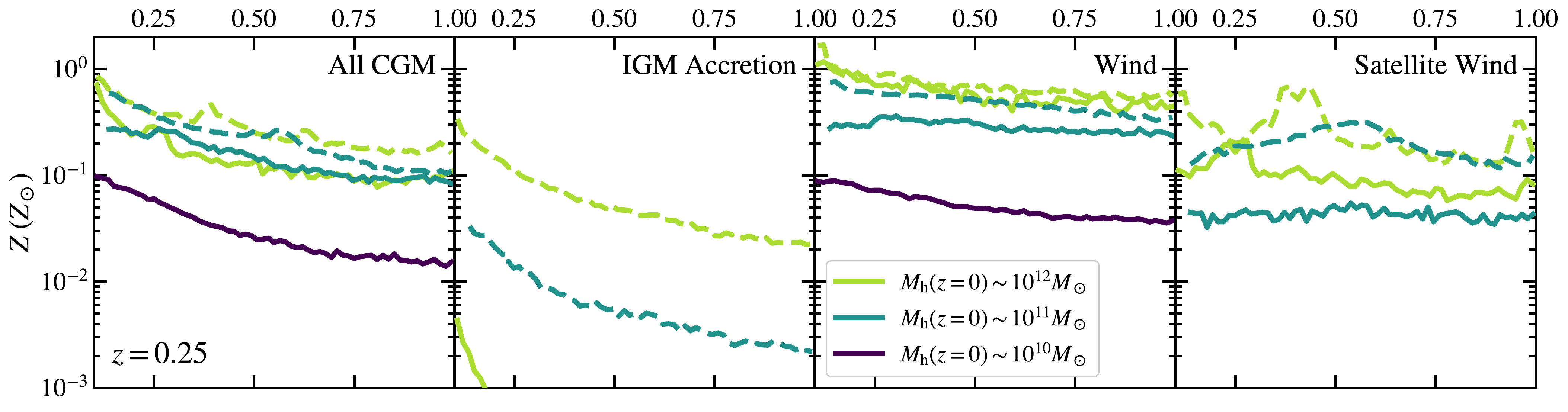}
\includegraphics[width=\textwidth]{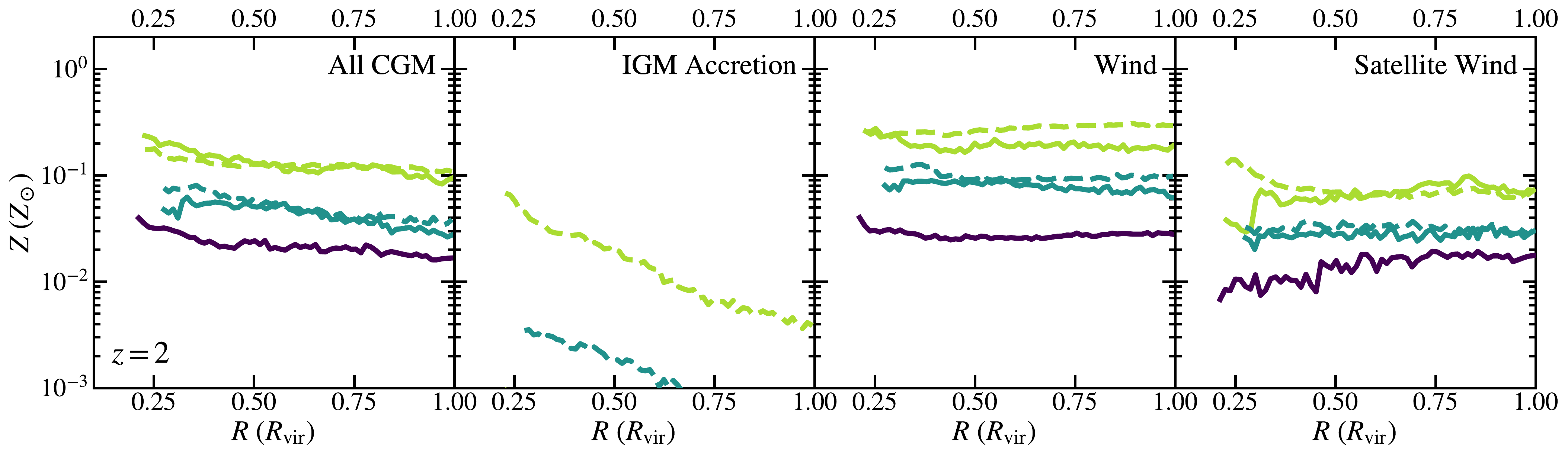}
\caption{
Metallicity as a function of radius for the CGM at $z=0.25$ (top) and $z=2$ (bottom).
The yellow-green, teal, and navy lines correspond to the metallicities for $M_{\rm h}(z=0) \sim 10^{(12,11,10)} M_\odot$ halos, respectively. 
Solid (dashed) lines are for the simulations without (with) metallicity diffusion. 
We plot the mass-weighted mean metallicity when considering all the gas in the CGM (first column), and the median metallicities when considering gas from different origins.
We limit the y-axis to $Z \geq 10^{-3} Z_\odot$ to focus on higher metallicity gas, but the majority of IGM accretion is at $Z\sim 10^{-4} Z_\odot$ (Figure~\ref{fig:CGM_metallicity}).
The decrease in the overall CGM metallicity with increasing radius is driven primarily by the increasing contribution to the total mass from low-metallicity IGM accretion.
}
\label{fig:CGM_met_profiles}
\end{figure*}

Gas metallicity is a potentially useful diagnostic for observationally differentiating IGM accretion from winds ~\citep[e.g.][]{Lehner2013,Quiret2016,Hafen2016}. 
Observations of quasar absorption systems suggest that CGM metallicities span $\gtrsim 2$ dex~\citep[e.g.][]{Wotta2016,Prochaska2017}, though the width of the observationally-inferred metallicity distribution depends on modeling assumptions (e.g. \cite{Stern2016} find a $0.3$ dex metallicity dispersion  for the CGM of COS-Halos galaxies using multi-density rather than single-density ionization modeling). 

Figure~\ref{fig:CGM_metallicity} shows the metallicity distributions at $z=0.25$ and $z=2$ for different CGM origins identified using particle tracking. 
Each curve shows the cumulative metallicity distribution for one main halo (i.e. the fraction of particles with that metallicity or below), and is colored by halo mass.
Simulations that include subgrid turbulent metal diffusion are shown as dashed lines. 
Since simulations without subgrid metal diffusion tend to produce incorrect distributions of stellar metallicities \citep[][]{Escala2018}, we regard the simulations including subgrid metal diffusion as our best model predictions for the CGM.
However, the subgrid prescription for metal diffusion is not guaranteed to always produce accurate predictions and its results can depend on the assumed diffusion coefficient. 
It is therefore informative to also consider CGM metallicity distributions for the simulations that do not include subgrid metal diffusion as an indication of the uncertainties in the predicted metallicities. 
In \S \ref{sec:origin_by_halo_mass} above, we discuss in more detail the interpretation of metallicities in simulations with and without subgrid metal diffusion. 
 
When no subgrid metal diffusion is included, the majority of gas classified as IGM accretion is at the simulation metallicity floor of $Z=10^{-4}~Z_\odot$ (we assume a solar metal mass fraction $Z_\odot = 0.0134$; \citealt{Asplund2010}). 
Even without subgrid metal diffusion, IGM accretion can be lightly enriched by halo stars or brief interactions with galaxies (interactions too short to affect the classification). 
Nevertheless, very little IGM accretion has $Z>10^{-2}Z_\odot$ in the simulations without subgrid metal diffusion.
Winds in general are much more enriched than IGM accretion. 
In general, wind metallicities increase with halo mass \citep[as expected from the mass-metallicity relation and the scaling of wind metallicity with ISM metallicity, e.g.][]{Ma2015, Muratov2016}. 
Moreover, the metallicities of satellite wind are in general lower than those of central galaxies in the same main halo by $\sim0.5 - 0.8$ dex.

With subgrid metal diffusion included, the results change dramatically for IGM accretion.
The median metallicity is no longer at the metallicity floor, but is $Z\sim\metMWmdNEPmedlow Z_\odot$ for the CGM of Milky-Way mass galaxies at $z=0.25$ and $Z\sim\metMWmdNEPmedhigh Z_\odot$ for the progenitor halos at $z=2$.
We note that increasing the metallicity of an IGM accretion particle from the metallicity floor at $Z = 10^{-4} Z_\odot$ to $Z\sim\metMWmdNEPmedlow Z_\odot$ can be done by diffusing only $\sim10\%$ of the metals from a wind particle with metallicity $\approx 0.4 Z_\odot$ (metals can diffuse from many surrounding gas particles).
The metallicity distributions of winds, both from central and satellite galaxies, are also in general enhanced by subgrid metal diffusion, especially at $z=0.25$. 
Metallicity distributions tend to become narrower by $\sim 0.5$ dex when subgrid metal diffusion is included.
This is because in the metal diffusion case the metallicity of nearby particles can equilibrate closer to an average value, an effect that is also found in predicted stellar metallicities \citep[][]{Escala2018}. 

Figure~\ref{fig:CGM_met_profiles} shows metallicity statistics as a function of radius.
The left column shows mass-weighted mean metallicity for all CGM gas, while the right three columns show the median metallicity for different origins.
As before, solid (dashed) curves correspond to simulations without (with) subgrid metal diffusion.
From the inner edge of the CGM to $R_{\rm vir}$, the mean CGM metallicity decreases by a factor $\sim 3.5-7$ at $z=0.25$ and $\sim 2-3$ at $z=2$. 
This is driven primarily by the fact that the CGM mass at larger radii is dominated by low-metallicity IGM accretion (see Figure~\ref{fig:CGM_mass_frac_profile}).

\section{Discussion}
\label{sec:discussion}

\subsection{Comparison with Previous Work}
\subsubsection{Particle Tracking Results}

We first compare our results with some previous particle tracking analyses that quantified the origins of the CGM. 

\cite{Ford2014} (hereafter \fordp) also classified mass in the CGM using a particle-tracking analysis of cosmological SPH simulations.
Beyond differences in the simulations (e.g., the simulations analyzed by \fordp~were full cosmological volumes that included many more galaxies but at lower resolution), the analysis of \fordp ~used a different classification scheme.
In our analysis, we use a definition of ISM that depends on both density and distance to a galaxy center, while \fordp~select ISM as all gas with $n_{\rm H} \ge 0.13$ cm$^{-3}$.
Note that the modeling and resolution employed in the \fordp~simulation sample are very different from our simulations (e.g., the simulations analyzed by \fordp~employed a subgrid, pressurized model for the ISM, whereas our simulations in principle resolve the multiphase ISM shaped by gravity, cooling, and stellar feedback). 
Therefore, density cuts in the two sets of simulations do not in general correspond one-to-one. 
By using only a density cut \fordp's wind definition does not distinguish between wind from central galaxies and from satellites.
On the other hand, \fordp~divide up their wind based upon how long ago it was launched, and whether or not it eventually accretes onto a galaxy. 
For simulated halos with $M_{\rm h}(z=0.25) > 10^{11.5} M_\odot$,  \fordp~find that IGM accretion provides 67\% of the CGM mass, with winds (either from the central galaxy or other galaxies) providing the remaining 33\%. 
This is in broad agreement with our results, although we find that the total wind mass is slightly higher (Figure~\ref{fig:CGM_mass_frac_vs_Mh_CGM} shows that IGM accretion provides $\sim$\fracnepMWmeanlow\% across our halo mass range). 
For halos with $M_{\rm h}(z=0.25) < 10^{11.5} M_\odot$, \fordp~find that IGM accretion provides 57\% of the CGM mass and wind accounts for the remaining 43\%, also in broad agreement with our results.

\cite{Nelson2013} analyzed gas accretion onto $10^{10} < M_{\rm h} / M_\odot < 10^{12}$ halos at $z=2$ in simulations evolved with the AREPO moving-mesh code, but without galactic winds.
The resulting gas accretion rates are dominated by IGM accretion (referred to by \citeauthor{Nelson2013} as ``smooth accretion''), with a secondary contribution from merging subhalos (``clumpy accretion''). 
Without wind to expel gas from galaxies, they found that gas stripped from other halos contributed only a small fraction ($\lesssim10\%$) of the gas accretion onto halos.

\cite{Suresh2018} studied the origin of cool gas in the CGM of a $M_{\rm h} \sim 10^{12} M_\odot$ halo at $z \sim 2$ using AREPO cosmological zoom-in simulations with enhanced resolution in the CGM, resulting in median CGM mass resolution of $2200\,M_\odot$ (compared to the typical, uniform $\sim7070~M_\odot$ mass resolution in similar halos in this work).
In their simulation that includes a prescription for galactic winds, \citeauthor{Suresh2018} find that $\sim 70\%$ of the cool gas has been processed by the central galaxy, with much smaller contributions from IGM accretion and satellite material. 
This confirms the general importance of winds in building up the CGM. 
We caution that in this work we do not study in depth the relationship between CGM origin and gas temperature, and that \citeauthor{Suresh2018} studied a more massive halo at $z=2$ than any we analyze (at $z=2$, our most massive halo has mass $M_{\rm h} \approx 5 \times 10^{11} M_\odot$). 
Therefore we cannot compare our results in detail with \citeauthor{Suresh2018}

\cite{Sokolowska2017} used four zoom-in SPH simulations to study the origin of hot halos around Milky Way-mass galaxies.
Their analysis suggests that at $z \gtrsim 2$ the hot halo has significant contributions from feedback-driven winds, with IGM accretion contributing more at lower redshifts, consistent with our analysis.

\cite{Fernandez2012} used a cosmological zoom-in hydrodynamic simulation evolved with an adaptive mesh refinement (AMR) code to study the origin of cool/cold gas (selected based on HI content) in the halo of a single MW-mass galaxy at $z<0.5$.
\citeauthor{Fernandez2012} did not explicitly track the flow of mass in the grid simulation, but instead analyzed a sequence of >100 projection maps and used metallicity to distinguish between IGM accretion and other origins.
Similar to our study, they find that gas removed from satellites can be a significant, but not dominant, source of mass, and filamentary IGM accretion accounts for $\sim$25\%-75\% of the cool/cold halo gas. 
Even though the AMR code used by \cite{Fernandez2012} should tend to over-predict gas diffusion below the resolution scale, they find that IGM accretion is metal-poor compared to satellite wind, consistent with our results.
Note, however, that \citeauthor{Fernandez2012} focus on cool/cold gas mass, while our particle tracking analysis includes all mass in the CGM (in Appendix \ref{sec:mass_by_phase} we quantify CGM mass fractions in our simulations for different temperature bins).

Because of differences in analysis methodology, none of the works discussed here allow for a truly one-to-one comparison.
To make future comparisons more conclusive, it would be beneficial for different groups to adopt consistent classifications. 
The different classifications developed in this work represent one way forward.

While we focus on the origins of the CGM in this paper, we note that there is extensive literature on the origin of gas and stars in galaxies~\citep[e.g.][]{Oppenheimer2010,VandeVoort2011,Nelson2015,VandeVoort2016a}.
A useful comparison of the present results is with \cite{Angles-Alcazar2017}, who studied the origin of baryons of central galaxies in FIRE-2 simulations.
\citeauthor{Angles-Alcazar2017} found that $\sim 30-60\%$ of the stellar mass in the central galaxies of $M_{\rm h} \sim 10^{9} - 10^{12} M_\odot$ at $z=0$ halos formed from wind recycling, consistent with the large fraction of CGM mass originating from wind in the inner CGM found in our study (Figure~\ref{fig:CGM_mass_frac_profile}).
Similarly, \citeauthor{Angles-Alcazar2017} found that $\sim 20-40\%$ of the stellar mass of central galaxies in $10^{12} M_\odot$ halos at $z=0$ originates as intergalactic transfer (i.e., satellite wind that accretes onto the central galaxy), which is consistent with the significant contributions of satellite wind to the CGM mass that we find here. 
In general, however, the origins of the CGM can differ greatly from those of the central galaxy.

\subsubsection{The Metal Content of Halos}
\label{sec:discussion_metal_content}

Recently, \cite{Christensen2018} analyzed metals in galaxy halos from a suite of zoom-in SPH simulations evolved with a different set of physics modules. 
Over the galaxy stellar mass range $M_\star \sim 10^{7} - 10^{11} M_\odot$, they find total halo metal masses spanning $\sim 30\%-80\%$ of the estimated metal budget at $z=0.25$ and $\sim 60\%$ at $z=2$.
As we find for our simulations (see Figure~\ref{fig:metal_mass_budget}), \cite{Christensen2018} find that the majority of metals inside halos are outside galaxies at $z=2$, but that this fraction decreases by $z=0.25$, up to a factor of $\sim2$ for $L^{\star}$ progenitors in their simulations.
\cite{Christensen2018}'s total retained halo metal masses are on average lower than those found in our simulations, peaking at $\sim80\%$.
In addition, the retained metals scale differently with galaxy/halo mass: at $z=2$ their total retained halo metal masses \textit{decrease} from $\sim80\%$ for $M_\star \sim 10^6 M_\odot$ galaxies to $\sim 60\%$ for $M_\star \sim 10^{10} M_\odot$ galaxies.
These differences may arise from different feedback models, which can result in different efficiencies at ejecting metals from galaxies and their halos, as discussed by \citeauthor{Christensen2018} when comparing to previous results from the FIRE project.

\cite{Peeples2014}, hereafter \peeplesp, inferred the metal masses contained in $M_\star \sim 10^{(9-11.5)} M_\odot$ galaxies at $z\sim0$ by combining observations with semi-analytic modeling. 
As shown in the top panel of Figure~\ref{fig:metal_mass_budget}, in our simulations most of the metals produced by $10^{12} M_\odot$ progenitors by $ z=0.25$ are retained in the galaxy.
This result is contrary to the result of \peeplesp~that most of the metals produced by the stars are not observed in the galaxy.
In particular, our  value $M_{\rm Z} / ( y_{\rm box} M_\star ) = M_{\rm Z}/M_{\rm Z, budget}$ for $10^{12} M_\odot$ progenitors at $z=0.25$ is $\approx \fretdivfretPeeples \times$ the analogous quantity inferred by \peeplesp~for similar-mass galaxies. 
This large difference arises from two different factors: in our simulations, the total metal mass $M_{\rm Z}$ is $\approx \metmassdivmetmassPeeples \times$ larger than inferred by \peeplesp, while the metal budget $M_{\rm Z,budget}$ in our simulations is only $\metbudgdivmetbudgPeeples \times$ that inferred by \peeplesp~for these galaxies. 
The differences in $M_{\rm Z}$ are primarily differences in the metal mass contained in stars, as stars contribute the majority of metals for these galaxies.
\peeplesp~use the stellar metallicity measurements of~\cite{Gallazzi2005} to calculate the stellar metal masses, and apply a correction to convert from B-band weighted stellar metallicities to mass-weighted stellar metallicities. 
This correction lowers the metallicities calculated by \peeplesp~by a factor of $\approx \zstardivzstarcorrected$ for galaxies in the mass range considered.
Furthermore, the stellar metal masses calculated by \peeplesp~depend on the solar metallicity, which they assumed to be $Z_\odot = 0.0153$ \citep[][]{Caffau2011}. 
As discussed by \cite{Muratov2016}, the difference in $M_{\rm Z, budget}$ is a result of different type II SNe yields and stellar mass loss rates in FIRE vs. \peeplesp~\citep[this issue is also discussed by][]{Christensen2018}.
Note that \cite{Muratov2016} analyzed the FIRE-1 simulations, which use the SNe II yields of \cite{Woosley1995}, while our FIRE-2 simulations use the yields of~\cite{Nomoto2006}.
However, the IMF-averaged yields are within $\sim10\%$ with the exception of Mg and Ne~\citep{Hopkins2017}, and our the total halo metal budgets agree well with the corresponding FIRE-1 halo metal budgets (see Appendix~\ref{sec:supplementary_material}). 
If taken at face value, the factor $\fretdivfretPeeples$ difference in galaxy the ``metal retention fraction'' relative to \peeplesp~would suggest that low-redshift, $\sim L^\star$ FIRE-2 galaxies remove from the galaxies $\sim 3 \times$ too few metals compared to analogous real galaxies. 
However, we caution that the substantial uncertainties in inferred metal masses and assumed metal budgets preclude us from drawing a robust conclusion regarding the significance of this discrepancy.

\subsection{Implications for Observations}
We now discuss some implications of our simulation analysis for the interpretation of CGM observations. 

First, our analysis suggests some broad expectations for the structures probed by CGM observations in different regimes. 
The particle trajectories in Figures \ref{fig:pathlines_CGM_snum465}-\ref{fig:pathlines_CGM_snum172} along with the CGM mass fractions vs. radius in Figure \ref{fig:CGM_mass_frac_profile} indicate that sight lines with impact parameter $<R_{\rm vir}$ will almost always intersect CGM gas representing a mix of different origins. 
These origins include IGM accretion, wind from the central galaxy, and wind from satellites. 
Our results indicate that this issue will especially complicate the interpretation of low-redshift observations, as $z \sim 0.25$ halos are predicted to be more throughly mixed than $z\sim 2$ halos. 
With regards to detecting galactic winds, our analysis confirms results previously obtained using FIRE-1 simulations. 
Namely, winds from $\sim L^{\star}$ galaxies should be easier to detect in the CGM of high-redshift star-forming galaxies (i.e., Lyman break galaxies) as galactic winds become much weaker for $\sim L^{\star}$ galaxies at $z \lesssim 1$ \citep[e.g.,][]{Muratov2015, Angles-Alcazar2017}. 
Figure \ref{fig:CGM_mass_frac_profile} shows that this is also the case for the FIRE-2 simulations analyzed in this paper: the CGM mass fraction from the winds of central galaxies for $M_{\rm h} \sim 10^{12}$ M$_{\odot}$ progenitors is much larger at $\sim R_{\rm vir}$ at $z=2$ than at $z=0.25$.

Next, we consider specific observational diagnostics that have been discussed for identifying the origin of CGM gas in observations. 

Some observations of absorption equivalent width as a function of azimuthal angle relative to the galaxy semi-minor axis suggest a bimodal or otherwise aspherical distribution, which could be due to different physical processes~\citep[e.g.,][]{Bordoloi2011,Bouche2012,Bordoloi2014,Kacprzak2015,Ho2016}.
In addition, the ongoing Project AMIGA will constrain the geometry of the CGM of M31 with nearly 50 sight lines~\citep[][]{2015ApJ...804...79L, Howk2017}. 
As discussed in \S\ref{sec:origin_by_angle}, we find that while warm/hot ($T > 10^{4.7}$ K) gas is distributed approximately spherically around low-redshift $\sim L^{\star}$ galaxies, cold/cool gas ($T < 10^{4.7}$ K) appears preferentially located near disk plane of galaxies (albeit with large halo-to-halo variations).
In order to determine the observational signatures expected from the distributions found in the simulations, it would be beneficial to produce mock absorption spectra, as well as to analyze a larger simulation sample to better understand the expected scatter.

The metallicity of a CGM absorber is often discussed as a diagnostic of its physical nature, with high-column but metal-poor systems often associated with fresh IGM accretion \citep[e.g.,][]{2011ApJ...743..207R, 2011Sci...334.1245F,Fumagalli2016a}. 
In \cite{Hafen2016} we studied the FIRE-1 simulations and found that metallicity is in general a poor diagnostic of inflows and outflows as defined purely based on radial kinematics relative to central galaxies. 
This is in part because the prevalence of wind recycling implies that metal-rich wind gas often later falls back onto the galaxy from which it was ejected. 
By classifying IGM accretion and wind gas in the CGM not based on instantaneous radial kinematics but instead using particle tracking, we can revisit the question of metallicity as a diagnostic of CGM gas origin. 
For example, the metallicity distributions in Figure \ref{fig:CGM_metallicity} indicate that at $z=0.25$ most CGM gas with $Z \lesssim 10^{-2}Z_\odot$ is associated with IGM accretion, in the case of simulations without subgrid metal diffusion. 
However, similar to what was found by \cite{Hafen2016}, more metal-rich gas can arise either in IGM accretion or in winds, if metal diffusion in the CGM is accurately captured by our subgrid prescription.

\cite{Lehner2013}, \cite{Wotta2016}, and the COS CGM Compendium \citep{Lehner2018,Wotta2018,Lehner2019} analyzed Lyman Limit Systems at $z \lesssim 1$ and reported evidence for a bimodal metallicity distribution of cosmologically-selected partial Lyman limit systems at $0.45 < z < 1$, with a dip in the distribution at $Z\sim10^{-1}Z_\odot$ (in contrast to \citealt{Fumagalli2016}, who report a broadly unimodal distribution at $z=2.5-3.5$).
As another result from \cite{Hafen2016}, we found that for similarly-selected absorbers the CGM across a wide range of halo masses contributed to the observed metallicity distribution, which produced a distribution without a statistically significant metallicity bi-modality (although our modest sample of zoom-in simulations introduced significant noise in the analysis). 
The lack of a clear metallicity dip appears in conflict with the analysis of the above observational studies~\citep[see also][]{Rahmati2018}.
Our particle tracking results suggest a similar conclusion: while different origins have different metallicities, the halo mass range induces a $\gtrsim 1$ dex spread in metallicities for a given origin (see Figure~\ref{fig:CGM_metallicity}).
This provides additional evidence that any analysis that aims to compare to the metallicity distribution of cosmologically-selected CGM absorbers must account for the full mass distribution of halos probed by random sight lines.

Recently, a number of observational studies have estimated the metallicity of the low-redshift CGM surrounding $\sim L^{\star}$ galaxies \citep[e.g.,][]{Stern2016,Prochaska2017,Bregman2018}.
Idealized analytic models of these systems have also been constructed and used to constrain the metallicity necessary to explain observations \citep[][]{Faerman2017,McQuinn2017,Mathews2017,Stern2018,Voit2018a}, especially the high OVI absorption columns measured out to $\sim R_{\rm vir}$ \citep{Tumlinson2011, Johnson2015}. 
The consensus from these different analyses is that the observations require a relatively high CGM metallicity $\gtrsim 0.3 Z_{\odot}$.
The mass-weighted mean CGM metallicity of our simulated $L^\star$ galaxies with subgrid metal diffusion included at $z=0.25$ is in the range $\sim0.2-0.3$ Z$_{\odot}$ at $R\approx (0.5-1)R_{\rm vir}$ (Figure \ref{fig:CGM_met_profiles}). 
This is marginally consistent with, but on the low end of, the observational inferences mentioned above. 
We caution that metallicities in the simulations are subject to uncertainties in the treatment of subgrid metal diffusion. 
Moreover, forward modeling using synthetic absorption spectra are ultimately needed to accurately compare simulation predictions to observations. 
In particular, metallicity is usually not a direct observable but instead relies on model-dependent inferences based on observed ions; assumptions about observed systems can be avoided by comparing direct observables to synthetic observations produced from the simulations. 
Several comparisons of FIRE-2 predictions to observed metal absorption systems are forthcoming (Hummels et al. in prep; Li et al. in prep; Dong et al. in prep).

\section{Conclusions}
\label{sec:conclusions}

We used FIRE-2 cosmological zoom-in simulations to study the properties and origins of the CGM over the redshift interval $z\sim0-2$ for the progenitors of halos with $M_{\rm h}(z=0) \sim 10^{10} - 10^{12} M_\odot$.
Using the galaxy growth particle tracking analysis of \cite{Angles-Alcazar2017} as a starting point, we tracked the trajectories of CGM gas elements throughout the full duration of each simulation (i.e. their pathlines).
Using the particle histories, we classified each gas particle as having one of the following (most recent) origins: IGM accretion, wind from the central galaxy, satellite wind, and satellite ISM.
Our results can be summarized as follows:

\begin{enumerate}
\item The baryonic mass fraction within $R_{\rm vir}$ on average increases with halo mass.
At $z=0.25$, it is $\sim$70\% of the cosmic baryon budget for $M_{\rm h}\sim 10^{12}$ M$_{\odot}$ halos (Figure~\ref{fig:mass_budget}).
The CGM mass is $\sim$\fCGMMWmeanlow-\fCGMdwarfmeanlow\% of the baryonic mass retained within the virial radius of the halo baryon; on average, the halo mass fraction in the CGM increases with decreasing halo mass.
\item A significant fraction ($\gtrsim 20$\%) of the metals produced by stars within halos escape the virial radius in most halos analyzed (Figure~\ref{fig:metal_mass_budget}). 
The main exception are $M_{\rm h}\sim 10^{12}$ M$_{\odot}$ at $z\sim0.25$. 
In general, the metal mass fraction ``lost'' from halos increases with decreasing halo mass. 
In low-redshift $\sim L^{\star}$ halos, most of the metals are located in the central galaxy, consistent with efficient recycling of wind ejecta. 
\item Across all halo masses and redshifts considered, IGM accretion provides $\sim 60-80\%$ of the CGM gas mass at $R \sim R_{\rm vir}$, but $\sim 40-80 \%$ of the gas in the inner CGM ($0.1R_{\rm vir}<R<0.5R_{\rm vir}$) has recycled through the central galaxy and is classified as wind (Figure~\ref{fig:CGM_mass_frac_profile}).
We include as IGM accretion unprocessed gas associated with the CGM of infalling galaxies.
Wind from the central galaxy is regularly recycled on time scales ranging from tens of Myrs to Gyrs (e.g. Figure~\ref{fig:r_vs_time_m12i_CGM_snum465}). 
The CGM mass is, as a whole, distributed with approximately constant mass per $\Delta R$, corresponding to a volume-weighted average density profile $n_{\rm H} \propto R^{-2}$ (Figure~\ref{fig:CGM_den_profile}). 
\item Wind and gas otherwise removed from satellites is not only a major contributer to the growth of central galaxies~\citep{Angles-Alcazar2017}, but can also contribute as much (or more) mass to the CGM as wind from the central galaxy for $M_{\rm h} \sim 10^{12} M_\odot$ halos at $z\sim0.25$ (Figure~\ref{fig:CGM_mass_frac_vs_Mh_CGM}). 
Satellite wind also provides up to $\sim$20\% of the CGM metal mass, with the remaining CGM metal mass dominated by wind from the central galaxy (Figure~\ref{fig:CGM_metal_mass_frac_vs_Mh_CGM}). 
\item For $M_{\rm h} \sim 10^{12} M_\odot$ halos at $z=0.25$, which host well-defined stellar disks, we find tentative evidence for cold/cool ($T < 10^{4.7}$ K) gas to be found preferentially near the galaxy disk plane, but with strong halo-to-halo variations (Figure~\ref{fig:CGM_polar_dists_m12s_snum465}).
\item The metallicity distribution of IGM accretion gas particles is systematically lower than the metallicity distributions of winds (typically by $\gtrsim 1$ dex in the median), although actual values of gas metallicity in the CGM and IGM depend significantly on the treatment of subgrid metal diffusion (Figure~\ref{fig:CGM_metallicity}).
The mean wind metallicity increases with increasing halo mass.
The overall CGM metallicity decreases with increasing radius, which is driven by an increasing mass fraction contributed by low-metallicity IGM accretion and a decreasing mass fraction contributed by more enriched winds (Figure~\ref{fig:CGM_met_profiles}).

\end{enumerate}
Overall, our analysis reveals that the CGM is a complex mix of diffuse IGM accretion, CGM and ISM stripped from infalling satellites, and winds from central and satellite galaxies.
We find, for example, that a significant fraction of CGM mass can come from satellite winds. 
Thus, future analyses (including semi-analytic models) should not only consider IGM accretion and wind from central galaxies, but also satellites and the larger-scale flows in which they are embedded (e.g., infalling filaments). 
On average, the CGM becomes more thoroughly mixed with decreasing redshift. 
As a result, sight lines through galaxy halos will in general probe gas of different physical origins, and this complexity should be taken into account when interpreting CGM observations. 
To provide additional intuition regarding the structure and origins of the CGM, we have produced an interactive 3D visualization of the $z=0.25$ CGM mass distribution, classified according to origin, for one of our $L^*$ galaxies, as well as a visualization of the trajectories of gas particles of different origins (see footnote \ref{foot:firefly}).
Data in electronic form for several of the figures in this paper are also available online.\footnote{\url{https://github.com/zhafen/CGM_origin_analysis}}

There are a number of directions in which our analysis can be expanded. 
Esmerian et al. (in prep.) use the particle tracking analysis described in this work to investigate in more detail the hot gaseous halos of $L^*$ galaxies. 
While we considered the origins of mass in the CGM in this analysis, a natural follow-up is to investigate the fates of CGM gas. 
How much CGM gas stays in the CGM, is ejected into the IGM, or accretes onto the central galaxy?
How does this correlate with the origin of gas elements? 
The simulations analyzed in this paper were evolved with ideal hydrodynamics. 
Additional physics, especially cosmic rays, can potentially significantly modify circumgalactic gas flows~\citep[e.g.][]{Chan2018a,Su2018a,Hopkins2019}, so it will be important to repeat a similar particle tracking analysis on simulations implementing different physics.
Finally, an ultimate goal that our particle tracking analysis and its extensions will support is the development of diagnostics that can be used to observationally distinguish CGM gas with different origins and fates.

\section*{Acknowledgements}

We thank
Andrey Kravtsov,
Thorsten Naab,
Sarah Wellons,
Hsiao-Wen Chen,
Michele Fumagalli,
John Stocke,
Jason Tumlinson,
Nicolas Lehner,
and
Rachel Somerville
for useful discussions.
We thank our referee, Dylan Nelson, for a comprehensive response that greatly improved the quality of this work.
We thank Alex Gurvich for help integrating \textsc{Firefly} into our analysis.
We thank Peter Behroozi for the making the UniverseMachine Early Data Release available~\citep{Behroozi2018}.
Zach Hafen was supported by the National Science Foundation under grant DGE-0948017.
CAFG was supported by NSF through grants AST-1412836, AST-1517491, AST-1715216, and CAREER award AST-1652522, by NASA through grants NNX15AB22G and 17-ATP17-0067, by STScI through grants HST-GO-14681.011, HST-GO-14268.022-A, and HST-AR-14293.001-A, and by a Cottrell Scholar Award from the Research Corporation for Science Advancement.
DAA acknowledges support by a Flatiron Fellowship.
The Flatiron Institute is supported by the Simons Foundation.
JS is supported as a CIERA Fellow by the CIERA Postdoctoral Fellowship Program (Center for Interdisciplinary Exploration and Research in Astrophysics, Northwestern University).
DK and TKC were supported by NSF grant AST-1715101 and by a Cottrell Scholar Award from the Research Corporation for Science Advancement.
Support for CBH was provided by NASA through Hubble Space Telescope (HST) theory grants HST-AR-13917 with additional funding from the NSF Astronomy and Astrophysics Postdoctoral Fellowship program.
CE acknowledges support from NSF grant AST-1714658.
Support for SGK and PFH was provided by an Alfred P. Sloan Research Fellowship, NSF Collaborative Research Grant \#1715847 and CAREER grant \#1455342, and NASA grants NNX15AT06G, JPL 1589742, 17-ATP17-0214.
KE was supported by an NSF Graduate Research Fellowship.
AW was supported by NASA through ATP grant 80NSSC18K1097, and grants HST-GO-14734 and HST-AR-15057 from STScI. 
NM acknowledges the support of the Natural Sciences and Engineering Research Council of Canada (NSERC).
This research was undertaken, in part, thanks to funding from the Canada Research Chairs program.
Numerical calculations were run on
the Quest computing cluster at Northwestern University, 
the Wheeler computing cluster at Caltech, 
XSEDE allocations TG-AST130039 and TG-AST120025;
Blue Waters PRAC allocation NSF.1713353, 
and NASA HEC allocation SMD-16-7592.



\bibliographystyle{mnras}
\bibliography{Mendeley,references}



\appendix

\section{Phases of CGM Gas}
\label{sec:mass_by_phase}

\begin{figure}
\includegraphics[width=\columnwidth]{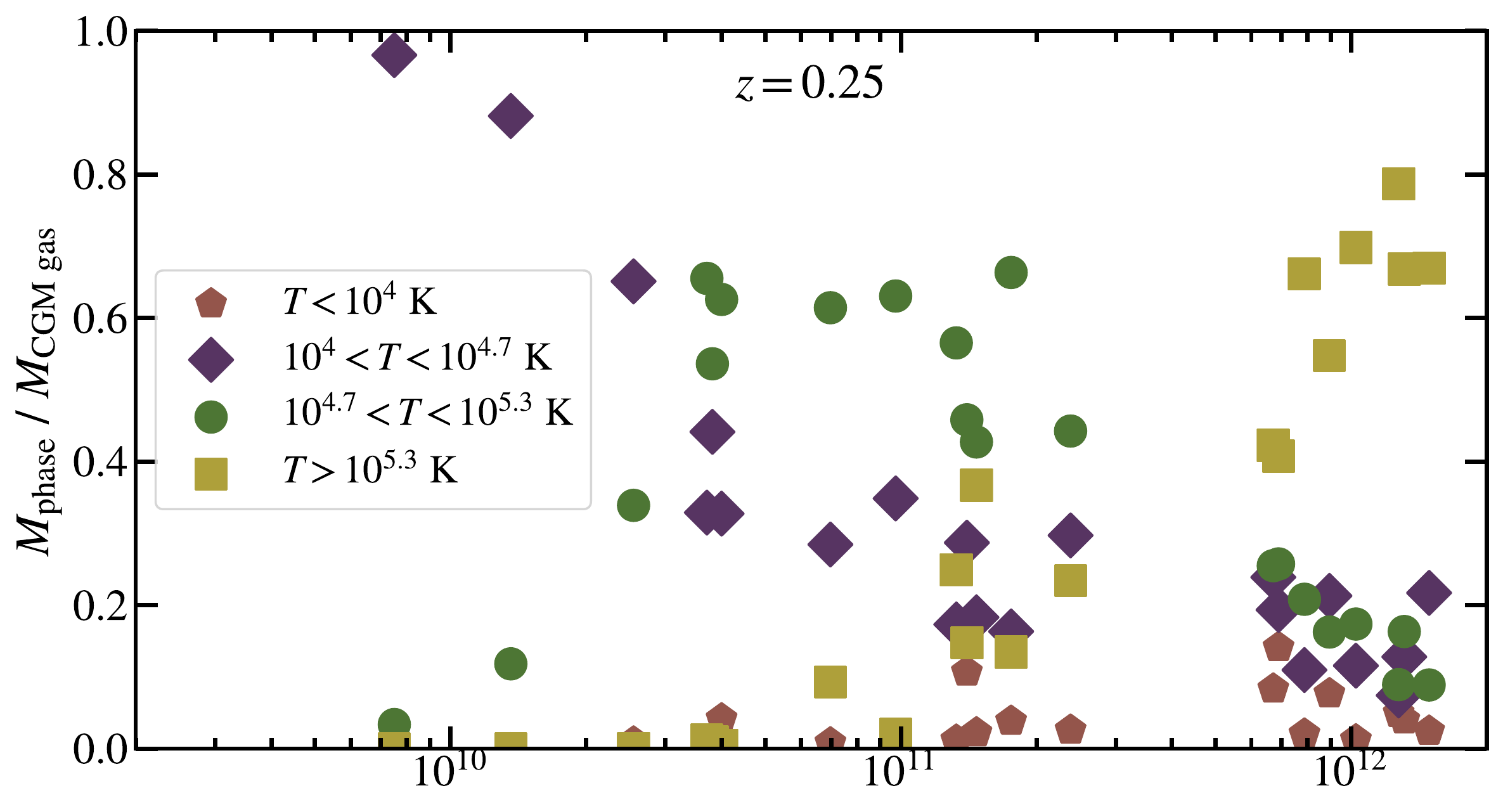}
\includegraphics[width=\columnwidth]{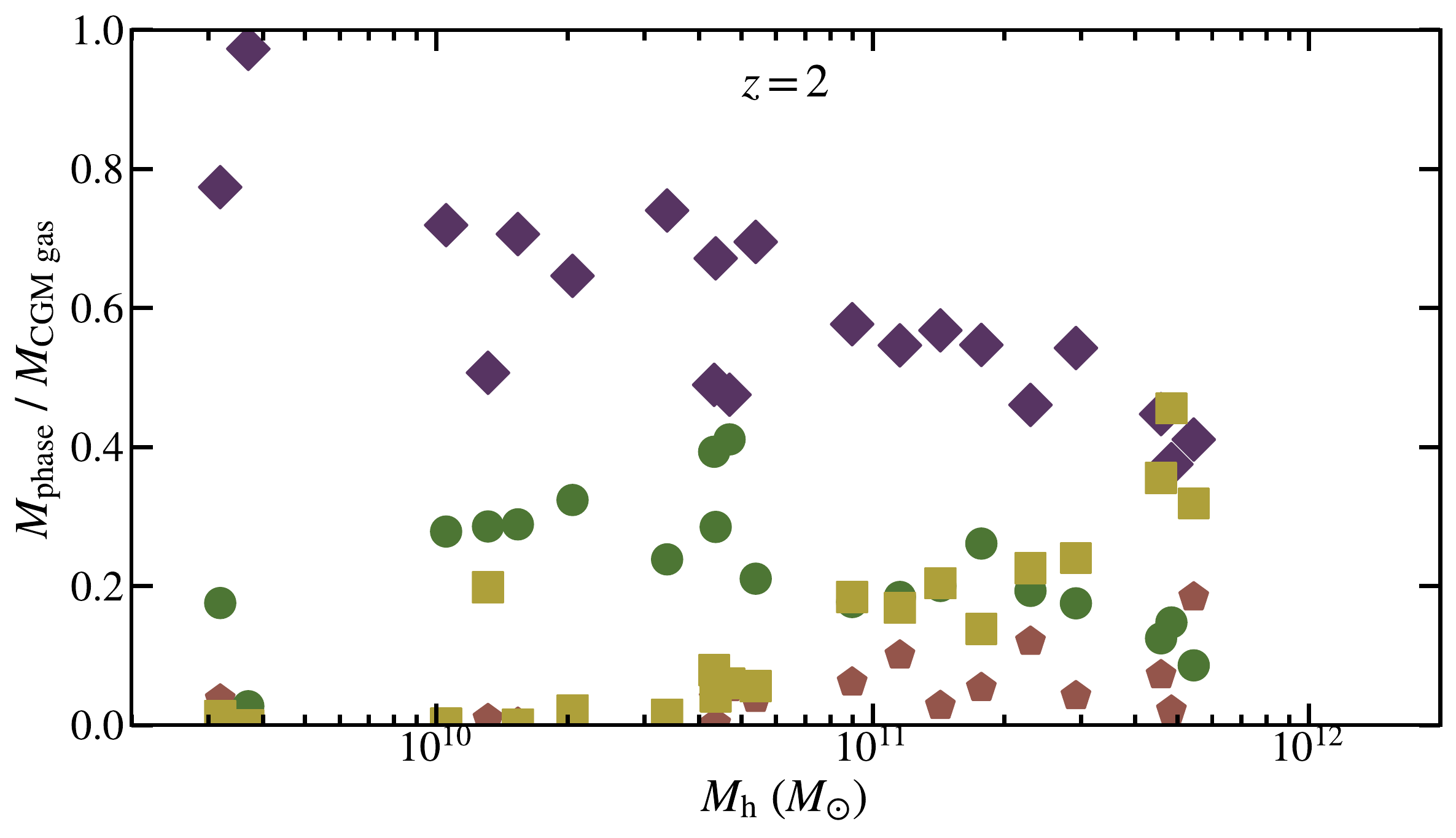}
\caption{
CGM gas mass in the cold (red pentagons; $T < 10^4$ K), cool (purple diamonds; $10^4$ K $< T < 10^{4.7}$ K), warm (green circles; $10^{4.7}$ K $< T < 10^{5.3}$ K), and hot (yellow squares; $10^{5.3}$ K $< T$) phases, as fractions of the total CGM mass.
At $z=0.25$ (top), the warm-hot halo provides the majority of the CGM mass for $M_{\rm h}\sim 10^{11}-10^{12}$ M$_{\odot}$ halos. 
On average, the CGM contains a larger mass fraction in cool gas at $z=2$. 
}
\label{fig:phase_cgm_mass_budget}
\end{figure}

As discussed throughout the paper, CGM gas from different origins exists in coherent structures (typically of cooler gas) or in a more well-mixed and diffuse halo (typically hotter gas). 
In this section we quantify the CGM mass fractions in different temperature bins. 
This provides additional context for the results of this paper regarding the CGM composition, as well as metrics that can be used to compare our simulations with others.
Figure~\ref{fig:phase_cgm_mass_budget} shows the mass in the CGM divided according to  cold (red pentagons; $T < 10^4$ K), cool (purple diamonds; $10^4$ K $< T < 10^{4.7}$ K), warm (green circles; $10^{4.7}$ K $< T < 10^{5.3}$ K), and hot (yellow squares; $10^{5.3}$ K $< T$) phases. 
The temperature cuts match those used in the FIRE-1 analysis of \cite{Muratov2015}.

\section{Supplementary Material}
\label{sec:supplementary_material}

This section contains plots which expand on the results presented in the main text.

\begin{figure}
\includegraphics[width=\columnwidth]{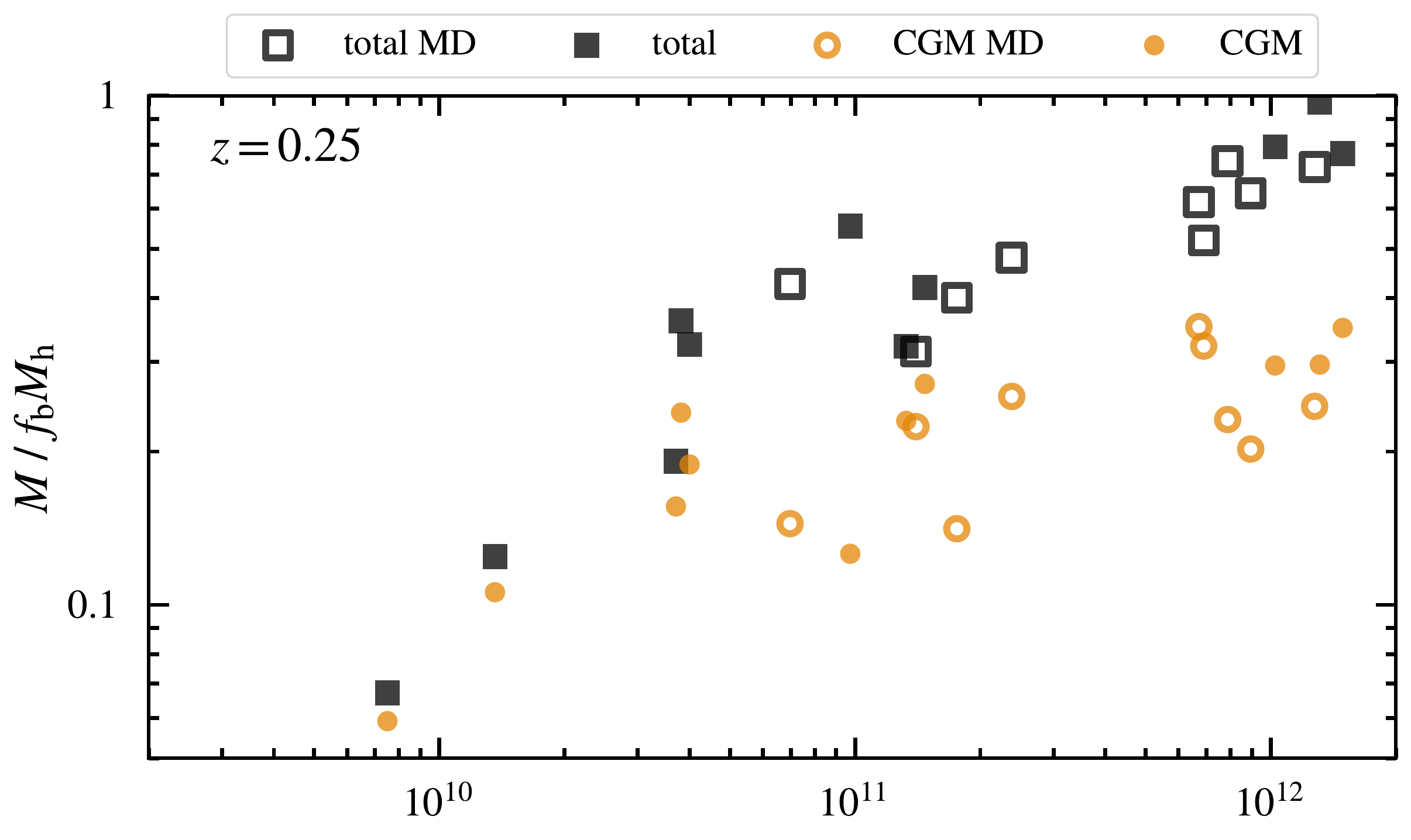}
\includegraphics[width=\columnwidth]{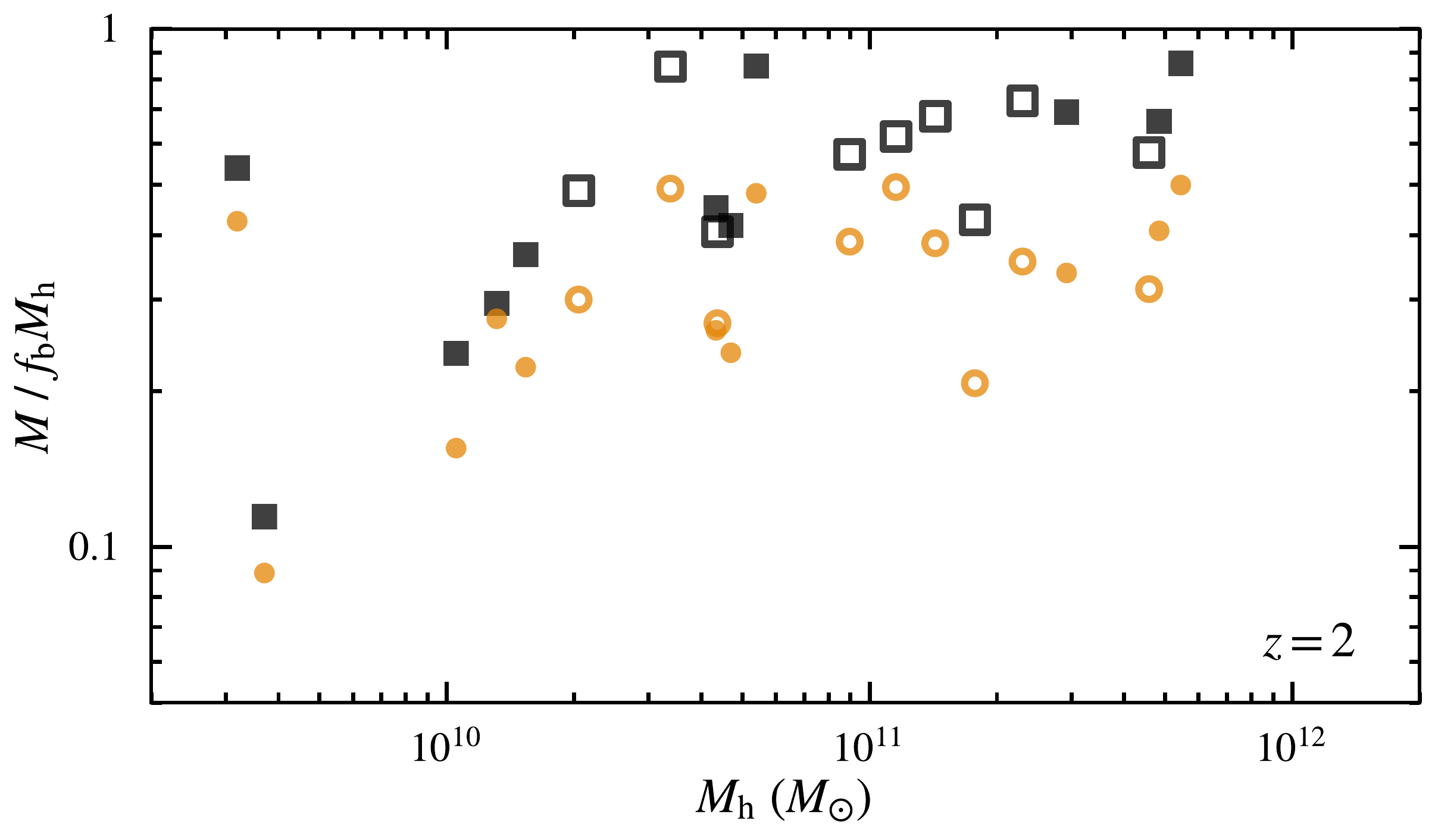}
\caption{
Similar to Figure~\ref{fig:mass_budget}, but the simulations with and without subgrid metal diffusion represented as closed or open symbols, respectively.
We only plot the total halo baryon mass and CGM gas mass, as these are the focus of our analysis.
Simulations with and without subgrid metal diffusion have consistent halo baryon masses and CGM gas masses.
}
\label{fig:mass_budget_md_comp}
\end{figure}

\begin{figure}
\includegraphics[width=\columnwidth]{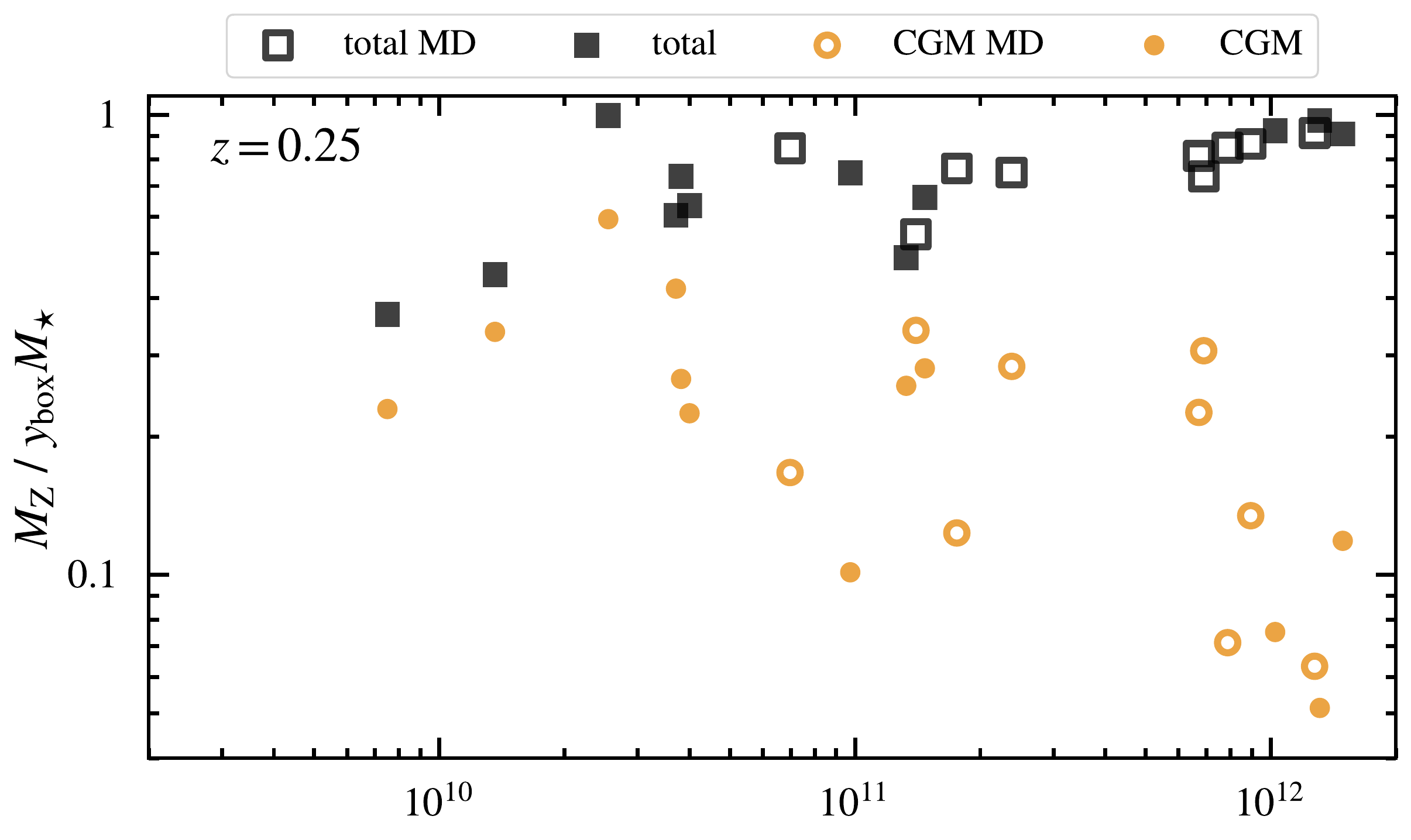}
\includegraphics[width=\columnwidth]{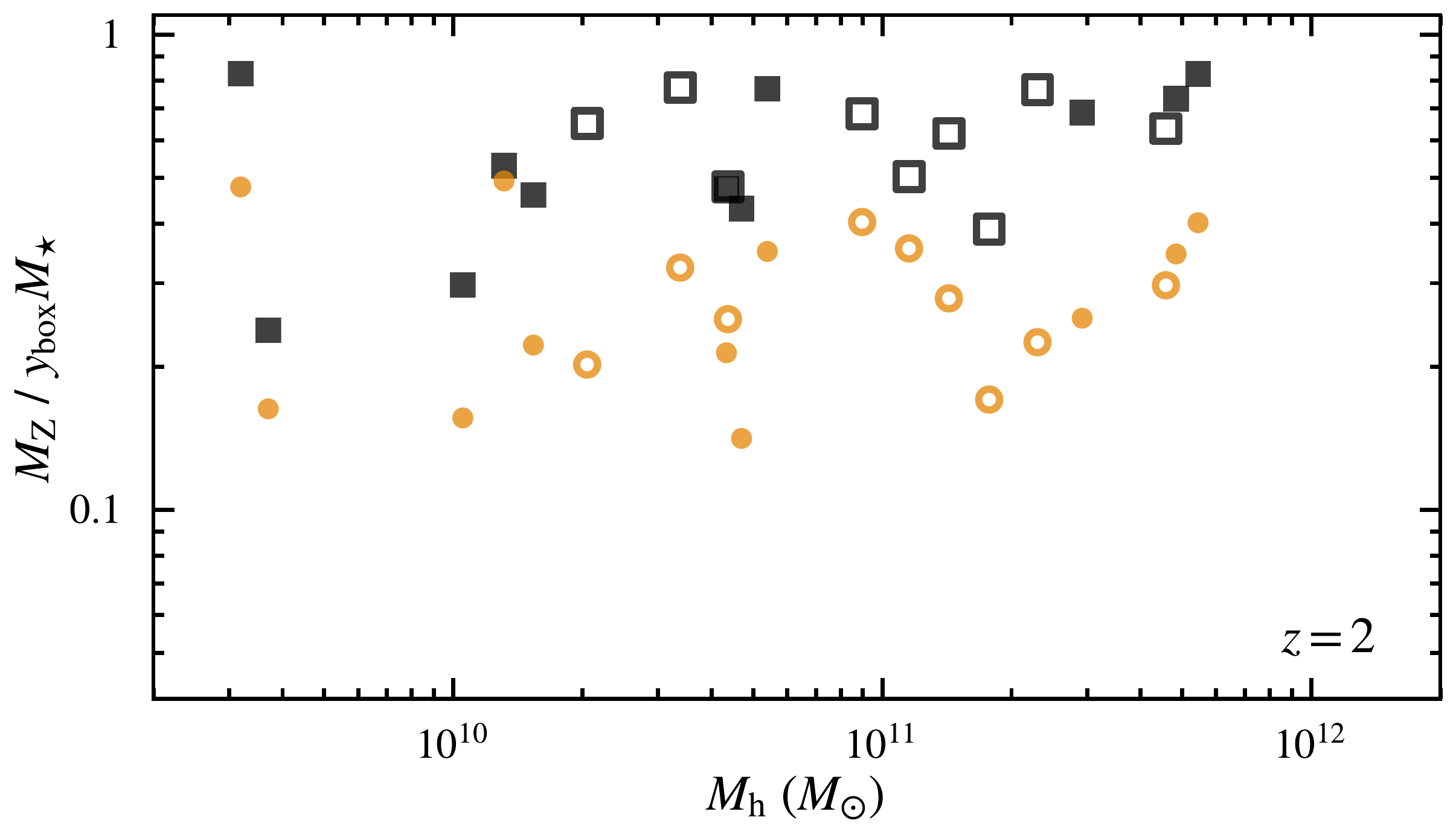}
\caption{
Similar to Figure~\ref{fig:metal_mass_budget}, but the simulations with and without subgrid metal diffusion represented as closed or open symbols, respectively.
Simulations with and without subgrid metal diffusion have consistent halo metal masses and CGM gas metal masses.
}
\label{fig:metal_mass_budget_md_comp}
\end{figure}

Figure~\ref{fig:mass_budget_md_comp} compares simulations with and without subgrid metal diffusion for the total halo baryon mass and the total CGM mass.
Similarly, Figure \ref{fig:metal_mass_budget_md_comp} compares the simulations with and without subgrid metal diffusion for the total halo metal mass and the total CGM gas metal mass, normalized by stellar mass. 
Note that our simulation sample does not include initial conditions evolved both with and without subgrid metal diffusion, so every simulation displayed is from independent initial conditions.
These comparisons indicate no significant systematic differences for these quantities between the simulations with and without subgrid metal diffusion.
This justifies the approach in the main text of combining simulations with and without subgrid metal diffusion for most of the analysis.

\begin{figure}
\includegraphics[width=\columnwidth]{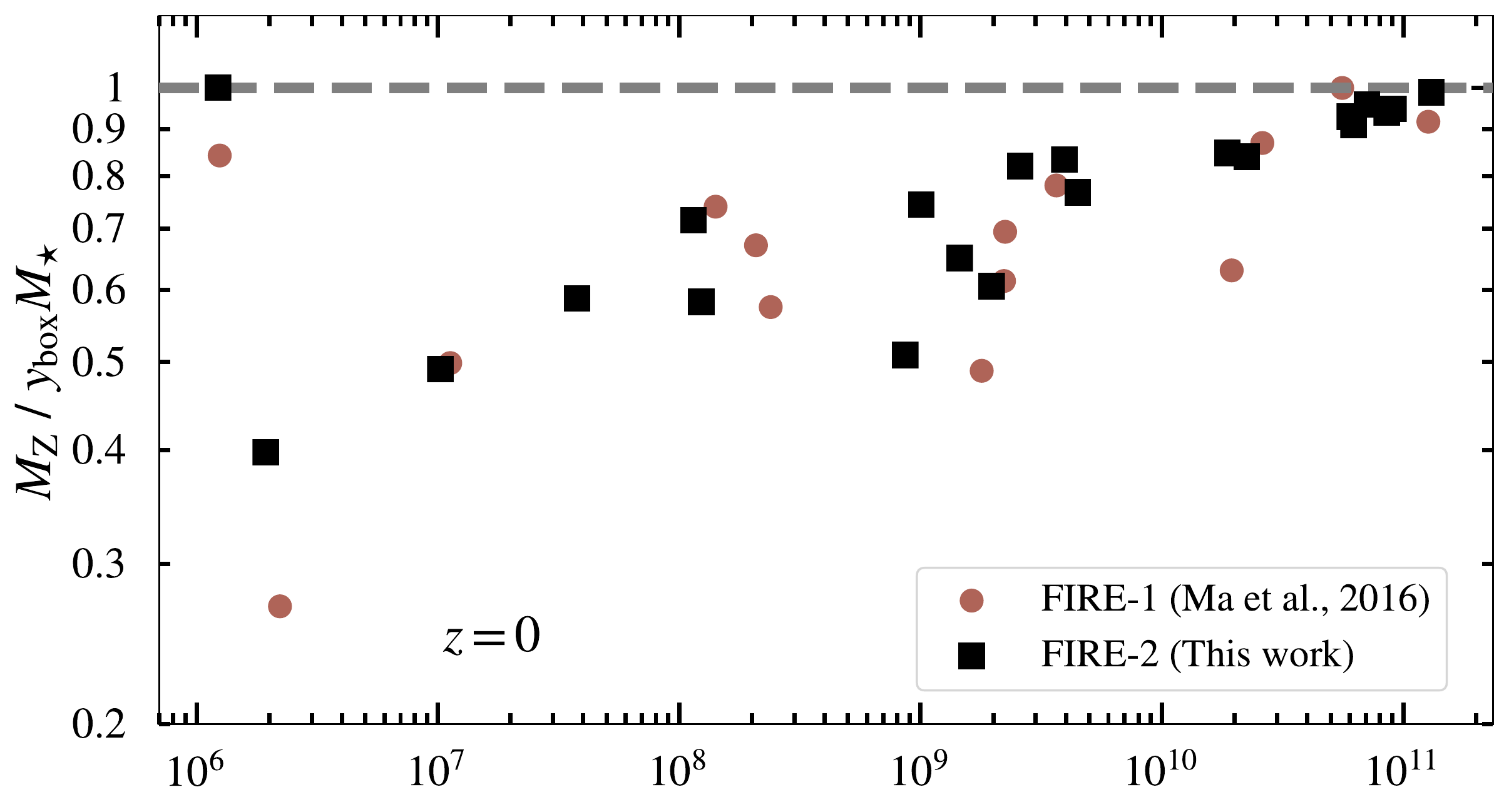}
\includegraphics[width=\columnwidth]{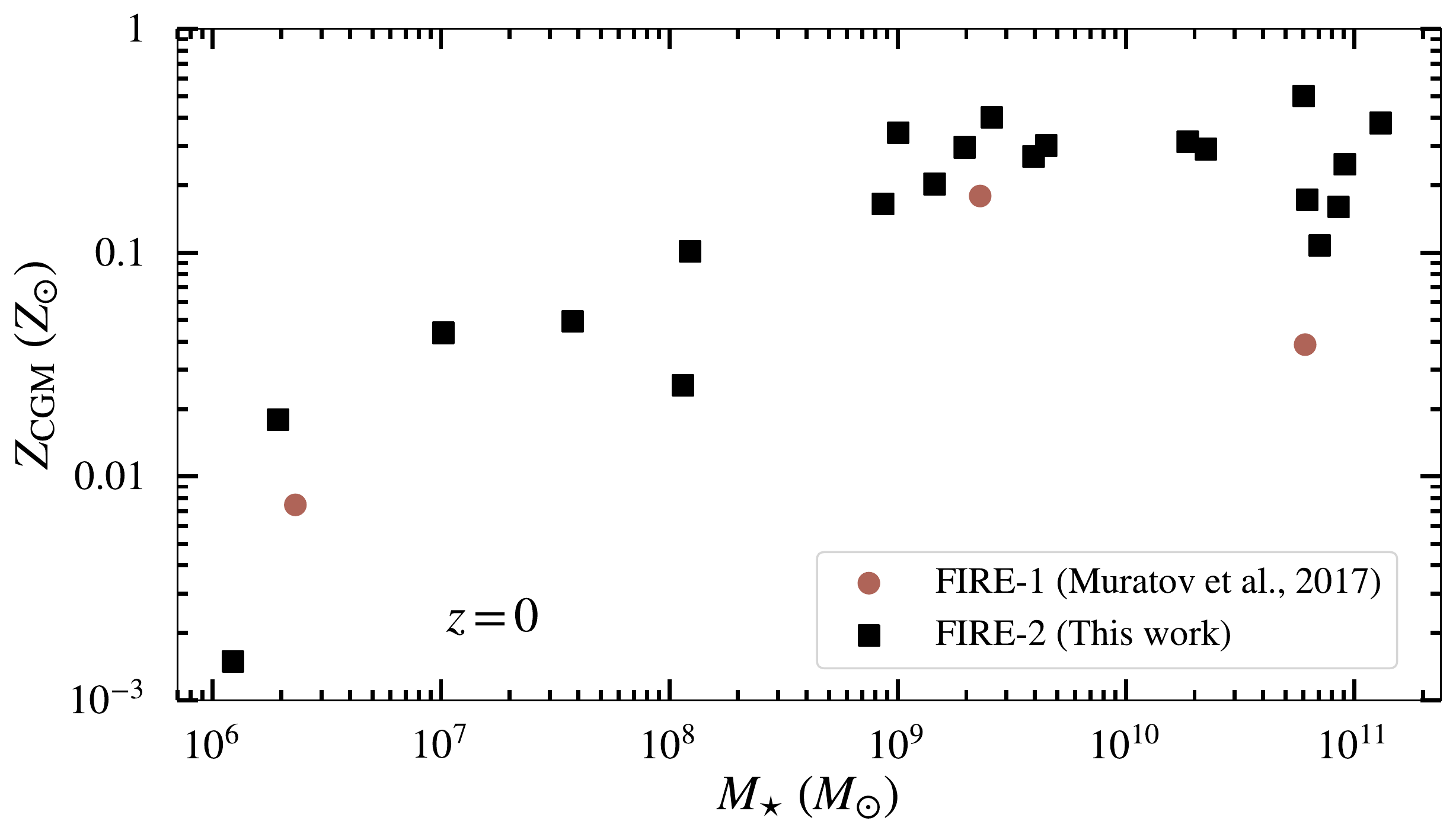}
\caption{
Mass fraction of produced metals that are retained in the halo (top) and mean CGM metallicity (bottom) as a function of stellar mass at $z=0$. 
Results from a sample of FIRE-2 simulations (this work) are plotted as black squares, and results from a sample of FIRE-1 simulations are plotted as red circles \citep{Ma2015,Muratov2016}. 
These results indicate no significant systematic difference between FIRE-1 and FIRE-2 simulations for the total metal mass fraction retained in the halo as a function of stellar mass. 
However, other related quantities such as CGM metallicities can differ more significantly for certain galaxies, e.g. the FIRE-1 data point at $M_{\star} \sim 10^{11}$ M$_{\odot}$ in the bottom panel, corresponding to the {\tt m12i} initial conditions.
}
\label{fig:metal_mass_budget_comp}
\end{figure}

As a complement to Figure~\ref{fig:metal_mass_budget}, Figure~\ref{fig:metal_mass_budget_comp} shows the total metal mass retained in the entire main halo ($<R_{\rm vir}$) and the mean CGM metallicity (defined as the total CGM metal mass divided by the total CGM gas mass) for each of our simulations at $z=0$ as a function of stellar mass of the main central galaxy. 
We compare to results obtained using P-SPH simulations and the FIRE-1 implementation of the galaxy formation physics published by~\cite{Ma2015} and~\cite{Muratov2016}.
The results from this paper were obtained using similar simulations but evolved with the MFM hydro solver and the FIRE-2 version of the galaxy formation physics (see \S\ref{sec:simulations}). 
The figure reveals no systematic differences between halo metal retention fractions in FIRE-1 vs. FIRE-2. 
However, there can be more significant differences between FIRE-1 and FIRE-2 CGM metallicities. 
In particular, the bottom panel of Figure \ref{fig:metal_mass_budget} shows that at $z=0.25$ the FIRE-1 version of \texttt{m12i}~\citep{Hopkins2014} has a mean CGM metallicity $\sim 6\times$ lower than the mean CGM metallicity in our FIRE-2 $\sim 10^{12}$ M$_\odot$ halos. 
This is the result of the FIRE-1 CGM metal mass in \texttt{m12i} being $\sim 4 \times$ lower than in the FIRE-2 version, while the total CGM gas mass in the FIRE-1 version is $\sim 1.5 \times$ higher than in the FIRE-2 counterpart. 
More analysis is needed to quantify how systematic the differences are between FIRE-1 and FIRE-2.


\bsp	
\label{lastpage}
\end{document}